\documentclass[twoside,12pt]{article}
\usepackage{epsfig}

\def\Journal#1#2#3#4{{#1} {#2} (#4) #3 }

\def\NPA{{\em Nucl. Phys.} A}

\def\NPB{{\em Nucl. Phys.} B}

\def\PLB{{\em Phys. Lett.} B}

\def\PRL{\em Phys. Rev. Lett.}
\def\PREV{\em Phys. Rev.}
\def\PREP{\em Phys. Rep.}

\def\PRD{{\em Phys. Rev.} D}
\def\PRC{{\em Phys. Rev.} C}

\def \ketv #1>{\mbox{$|{#1}\rangle$}} 

\newcommand{\bq}{\mbox{\boldmath $q$}}

\newcommand{\bk}{\mbox{\boldmath $k$}}
\newcommand{\br}{\mbox{\boldmath $r$}}
\newcommand{\bx}{\mbox{\boldmath $x$}}
\newcommand{\by}{\mbox{\boldmath $y$}}
\newcommand{\bz}{\mbox{\boldmath $z$}}

\newcommand{\bw}{\mbox{\boldmath $w$}}

\newcommand{\brs}{\mbox{\scriptsize \boldmath $r$}}
\newcommand{\bxs}{\mbox{\scriptsize \boldmath $x$}}
\newcommand{\bys}{\mbox{\scriptsize \boldmath $y$}}

\newcommand{\bks}{\mbox{\scriptsize \boldmath $k$}}

\newcommand{\be}{\begin{equation}}
\newcommand{\ee}{\end{equation}}
\newcommand{\bea}{\begin{eqnarray}}
\newcommand{\eea}{\end{eqnarray}}
\newcommand{\nn}{\nonumber}
\newcommand{\Pu}{p_{\uparrow}}

\newcommand{\Nu}{n_{\uparrow}}
\newcommand{\Nd}{n_{\downarrow}}

\begin{document}

\title{ \vspace{1cm} Hadron interactions in lattice QCD}
\author{S.\ Aoki$^{1,2}$ \\
for HAL QCD Collaboration\\
\\
$^1$Graduate School of Pure and Applied Sciences, University of Tsukuba, \\ Tsukuba 305-8571, Japan\\
$^2$Center for Computational Sciences, University of Tsukuba, \\ Tsukuba, 305-8577, Japan}
\maketitle
\begin{abstract} 
Progress on the potential method, recently proposed to investigate hadron interactions in lattice QCD, is reviewed.  The strategy to extract the potential  in lattice QCD is explained in detail.
The method is applied to extract  $NN$ potentials, hyperon potentials and the meson-baryon potentials. A theoretical investigation is made to understand the origin of  the repulsive core using the operator product expansion.  Some recent extensions of the method are also discussed.
\end{abstract}

\tableofcontents
\section{Introduction}
Quarks and gluons, the fundamental degrees of freedom of Quantum Chromodynamics  (QCD)
never appear in Nature as asymptotic one-particle states. This phenomena is called confinement.
Instead,  only their bound states, i.e., mesons and baryons, are observed in experiment 
and these hadronic states are the asymptotic states whose interactions can be parametrized by  the S-matrix. 
An ultimate goal for theoretical studies of the strong interaction, therefore, is to extract the properties of the hadronic S-matrix from QCD.

For a description of hadronic interactions,  the nuclear force is one of the most fundamental quantities in nuclear physics.  The origin of the nuclear force, however, is still one of the major unsolved problems in strong interaction physics even after the establishment of QCD.
Although the nuclear force is not well understood theoretically, a large number of proton-proton and neutron-proton scattering data as well as deuteron properties have been experimentally accumulated and summarized as the nucleon-nucleon ($NN$) potential, which is designed to  reproduce these experimental properties through the Schr\"odinger equation for the $NN$ wave function. Once the potential is determined, it can be used to study systems with more than 2 nucleons by using various many-body techniques.

Phenomenological $NN$ potentials which can fit the $NN$ data precisely (e.g. more than 2000 data with $\chi^2/{\rm dof}\simeq 1$ ) at $T_{\rm lab} < 300 $ MeV are called the high-precision $NN$ potentials. Those in coordinate space, some of which are shown in Fig.~\ref{fig:NNpotential}, reflect characteristic features of the $NN$ interaction at different length scales\cite{Taketani1967,Hoshizaki1968,Brown1976,Machleidt1989,Machleidt2001}:
\begin{figure}[tb]
\begin{center}
\includegraphics[width=0.5\textwidth]{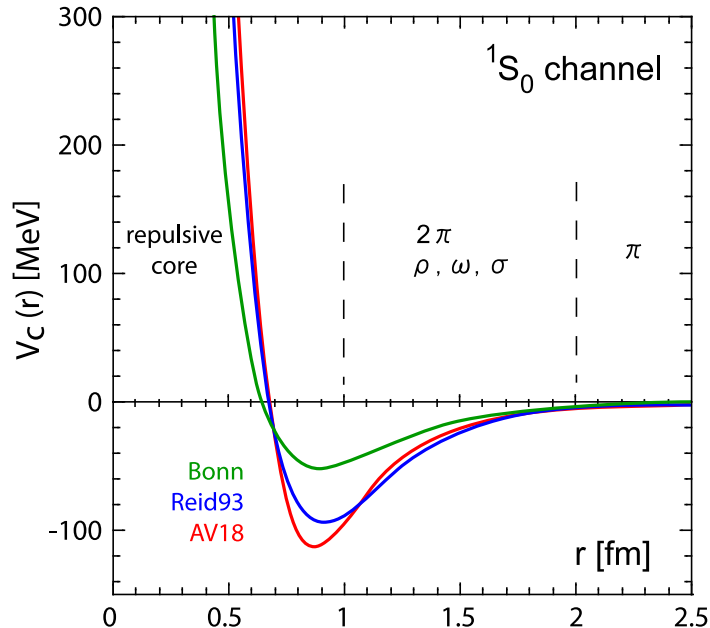}
\caption{Three examples of the modern $NN$ potential in $^1S_0$ (spin-singlet and $S$-wave) channel: CD-Bonn\protect\cite{Machleidt2001a}, Reid93\protect\cite{Stoks:1994wp} and Argonne $v_{18}$\protect\cite{Wiringa:1994wb}. Taken from Ref.~\protect\cite{Ishii:2006ec}.}
\label{fig:NNpotential}
\end{center}
\end{figure}
\begin{enumerate}
  \item[(i)]
  The long range part of the nuclear force  (at relative distance 
 $r >  2$  fm) 
 is dominated by one-pion exchange introduced by Yukawa\cite{Yukawa1935}.
  Because of the pion's Nambu-Goldstone character, it
  couples to the spin-isospin density of the nucleon and hence
  leads to a strong spin-isospin dependent force, namely the tensor force.
 \item[(ii)]  The medium range part ($1\ {\rm fm} < r < 2$ fm) receives 
  significant contributions from the exchange of  
  two-pions ($\pi\pi$) and/or heavy mesons ($\rho$, $\omega$, and $\sigma$).
  In particular, the spin-isospin independent attraction
 of about 50 -- 100 MeV in this region plays an essential role
  for the binding of atomic nuclei.
\item[(iii)]  The short  range part ($r < 1$ fm) is best described by
  a strong repulsive core as originally introduced by Jastrow \cite{Jastrow1951}.
  Such a short range repulsion is important for  
  the stability of atomic nuclei against collapse, 
 for determining the maximum mass of neutron stars, and for
 igniting the Type II supernova explosions \cite{Tamagaki1993,Heiselberg2000,Lattimer2000}.
\end{enumerate}   

It is then a challenge for the theoretical particle and nuclear physics communities to extract these properties of the $NN$ interaction from first principle non-perturbative QCD calculations, in particular lattice QCD simulations. 
A theoretical framework suitable for such a purpose was first proposed by L\"{u}scher\cite{Luscher:1990ux}. For two hadrons in a finite box with the size $L \times L \times L$ with periodic boundary conditions,  an exact relation between  the energy spectra in the box
and the elastic scattering phase shift at these energies was derived. If the range of the hadron interaction $R$  is sufficiently smaller than the size of the box $R<L/2$, the behavior of the 
two-particle Nambu-Bethe-Salpeter (NBS) wave function $\psi ({\bf r})$ in the interval $R < \vert {\bf r} \vert < L/2 $ has sufficient information to relate the phase shift and the two-particle spectrum.  L\"{u}scher's method bypasses the difficulty to treat the real-time scattering process on the Euclidean lattice.  Furthermore, it utilizes the finiteness of the lattice box effectively to extract the information of the on-shell scattering matrix and the phase shift.  

A closely related but an alternative approach to the $NN$ interactions from lattice QCD has been proposed recently\cite{Ishii:2006ec,Aoki:2008hh,Aoki:2009ji}.  
The starting point is the same NBS wave function  $\psi (\br)$ as discussed in Ref.~\cite{Luscher:1990ux}.  Instead of looking at the wave function outside the range of the interaction, the authors consider the internal region $ |\br | < R$ and define an energy-independent  non-local potential $U(\br, \br')$ from $\psi (\br)$ so that  it  obeys the Schr\"{o}dinger type equation in a finite box.
Since $U(\br, \br')$ for strong interactions is localized in its spatial coordinates due to confinement of quarks and gluons, the potential receives only weak finite volume effect in a large box. Therefore, once $U$ is determined and is appropriately extrapolated to  $L \rightarrow \infty$, one may simply use the Schr\"{o}dinger equation in infinite space to calculate the scattering phase shifts and bound state spectra to compare  with  experimental data.   A further advantage of utilizing the potential is that it is a smooth function of the quark masses so that it is relatively easy to handle on the lattice. This is in sharp contrast to the scattering length which shows a singular behavior in the quark mass corresponding to the formation of the $NN$ bound state.

Since the recent progress for the study of the $NN$ interaction by the first method has already been reviewed in Ref.\cite{Beane:2010em}, the recent progress for the second method is mainly considered in this review.
In Sec.~\ref{sec:strategy}, the strategy  of Ref.~\cite{Ishii:2006ec,Aoki:2008hh,Aoki:2009ji} to define the $NN$ potential in QCD is explained in detail, and the lattice formulation is introduced
in Sec.~\ref{sec:lattice}.
Results of lattice QCD calculations for $NN$ potentials are given in both quenched and full QCD
in Sec.~\ref{sec:NNpotential}. Central potentials at the leading order of the velocity expansion is shown to reproduce qualitative features of the $NN$ potential such as the repulsion at short distance and the attraction at medium to long distances. 
The tensor potential, which exists also at leading order,  is extracted.  Contrary to the case of the central potentials, it does not have a repulsive core.  Higher order contributions in the velocity expansion are also investigated and shown to be small at low energy and low orbital angular momentum $L$.   
In Sec.~\ref{sec:hyperon}, the method to extract the potential is applied to the hyperon-nucleon interactions such as $N\Xi$ and $N\Lambda$ systems. Interactions between octet baryons in general are also investigated in the flavor SU(3) limit, where up, down and strange quark masses are all equal. In Sec.~\ref{sec:OPE},  
we also consider a recent attempt to understand the origin of the repulsive core in the $NN$ potential.  Using the operator product expansion and renormalization group analysis in QCD, the potential derived from the NBS wave function in Sec.~\ref{sec:NNpotential} is shown to have a repulsive core, whose functional form is also theoretically predicted.  
In Sec.~\ref{sec:extension}, two extensions of the potential method are considered, together with explicit applications of these extensions to hadron interactions. One is the extension of the potential method to inelastic scattering, in order to investigate the $\Lambda\Lambda$ system,
while the other is the extraction of the potential from the time dependent NBS wave function in lattice QCD. With the latter method, the existence of the $H$-dibaryon is investigated in the flavor SU(3) limit.
In Sec.~\ref{sec:others}, applications of the method to the three nucleon force,  meson-baryon potentials and the potential in 2-color QCD are considered. 
Brief concluding remarks are given in Sec.~\ref{sec:conclusion}.

\section{Strategy to extract potential in QCD}
\label{sec:strategy}
\subsection{Nambu-Bethe-Salpeter (NBS) wave function and its asymptotic behavior}
A key quantity to extract the potential from QCD is the equal time Nambu-Bethe-Salpeter wave function, defined by
\bea
\varphi^W({\bf x}) e^{-Wt}&=& \langle 0 \vert T\{N({\bf r} + {\bf x},t) N({\bf r}, t)\} \vert 2N,W, s_1 s_2\rangle,
\eea
where $\vert 2N, W, s_1s_2\rangle$ is an eigen-state of QCD for two nucleons with total energy $W=2\sqrt{{\bf k}^2+m_N^2}$ and  the total three-momentum ${\bf P}=0$, whose helicities are denoted by $s_1$, $s_2$.  Here the local nucleon operator is given by
\bea
N_{\alpha}(x) &\equiv& \left(\begin{array}{c}
p_\alpha (x) \\
n_\alpha (x) \\
\end{array}\right) =
\varepsilon^{abc} \left( u_a (x) C\gamma_5 d_b(x) \right) q_{\alpha}(x), \quad
q(x)=\left(\begin{array}{c}
u (x) \\
d (x) \\
\end{array}\right),
\eea
where $x=({\bf x},t)$, the color indices are denoted by $a,b,c$,  and $\alpha$ is the spinor index.
The charge conjugation matrix in the spinor space is given by $C=\gamma_2\gamma_4$, and
$p,n$ denotes proton and neutron operators while $u,d$ denote up and down quark operators.
Note that $\varphi^W$ implicitly has two pairs of spinor-flavor indices which come from $N_\alpha ({\bf r} + {\bf x},t) N_\beta({\bf r}, t)$ and  two helicity indices $s_1$ and $s_2$.

The most remarkable property of the above NBS wave function is explained as follows.
If the $W$ is smaller than the threshold energy for one-pion production ({\it i.e.} $W < 2 m_N + m_\pi$ ),  then its asymptotic behavior for  large $\vert{\bf x} \vert$ can be 
evaluated\cite{Ishizuka2009a,Aoki:2009ji}. The helicity component in the spin singlet channel ($S=0$) is given by $
\vert s_1 s_2 \rangle = \frac{1}{\sqrt{2}}\left(\vert +\frac{1}{2},+\frac{1}{2}\rangle +\vert -\frac{1}{2},-\frac{1}{2}\rangle\right)
$, where the relative $+$ sign is our convention.
For this case, we have
\bea
\varphi^W({\bf r})_{S=0}  &\simeq  & \sum_{l,l_z} Z^{l,l_z} (S=0)Y_{l l_z}(\Omega_{\bf r}) \frac{\sin( k r - l\pi/2 + \delta_{l0}(k))}{k r}  e^{i\delta_{l0}(k)} 
\label{eq:asympt_singlet}
\eea
where $r=\vert{\bf r}\vert$, $k=\vert{\bf k}\vert$, and 
$\delta_{lS}(k)$ is the $NN$ scattering phase shift  in QCD with the total angular momentum $l$ and the total spin $S$, which is  determined by the unitarity of  S-matrix in QCD below the inelastic threshold\cite{Aoki:2009ji}.  Here $Y_{lm}(\Omega_{\bf r})$ is the spherical harmonic function with the solid angle $\Omega_{\bf r}$ of ${\bf r}$. The coefficient $Z^{ll_z}(0)$ has spinor components $\alpha,\beta$ and is given by
\bea
Z^{ll_z}_{\alpha\beta}(0) &=& Z D^l_{l_z0}(\Omega_{\bf k}) U_{\alpha \hat\alpha}(\nabla) U_{\beta \hat\beta}(-\nabla) \chi_{\hat\alpha \hat\beta} (0,0)
\eea
where $Z$ is the wave function renormalization for the nucleon operator $N(x)$, $D^l_{m\lambda}$ is the Wigner $D$-matrix, and $U_{\alpha \hat\alpha}(\nabla)$ and $U_{\beta \hat\beta}(-\nabla)$ are the $4\times 2$ matrices acting on the $2\times 2$ matrix $\chi_{\hat\alpha\hat\beta}(S,S_z)$. Explicitly we have
\bea
U(\nabla) &=& \sqrt{W+m_N}\left( 
\begin{array}{cc}
I_{2\times 2}, &  \displaystyle\frac{-i\sigma\cdot \nabla}{W+ m_N} \\
\end{array}\right) \\
\chi (0,0)&=&\frac{1}{\sqrt{2}} i\sigma_2, \quad \chi(1,0) =\frac{1}{\sqrt{2}}\sigma_1,\quad
\chi(1,\pm 1) = \frac{1}{2}(I_{2\times 2}\pm \sigma_3).
\eea
For the spin triplet channel ($S=1$), the asymptotic behavior of $\varphi^W$ is more involved but schematically written as
\be
\varphi^W({\bf r})_{S=1} \propto  \sum Y_{l l_z}(\Omega_{\bf r}) \frac{\sin( k r - l\pi/2 + \delta_{l1}(k))}{k r}  e^{i\delta_{l1}(k)} 
\label{eq:asympt_triplet} .
\ee 
An explicit form of the asymptotic behavior for the NBS wave function in the triplet channel
is given in appendix~\ref{app:NBS}. 

The asymptotic behaviors in eqs.~(\ref{eq:asympt_singlet}) and (\ref{eq:asympt_triplet}) tell us that the NBS wave function at large separation $r$ describes the scattering wave of the quantum mechanics whose phase shift agrees with the phase of the S-matrix in QCD. 
Therefore
the NBS wave function satisfies the free Schr\"odinger equation at large $r$ as
\be
\left[\frac{k^2}{2\mu} - H_0\right] \varphi^W({\bf r} )\simeq 0, \qquad H_0 =\frac{-\nabla^2}{2\mu} 
\ee
at $W < 2m_N + m_\pi$, where $ \mu = m_N/2$ is the reduced mass of the $NN$ system.  
Note that these properties hold 
without using the non-relativistic approximation/expansion. In particular, only the upper  components of the spinor indices for the NBS wave function ($\alpha=1,2$ and $\beta =1,2$) are enough to reproduce all $NN$ scattering phase shifts $\delta_{lS}(k)$ with $l=0,1,2,3,\cdots$ and $S=0,1$. From these properties,  (the upper spinor components of)  the NBS wave function can be regarded as the "wave function" of the $NN$ system at $W < 2m_N + m_\pi$.

It is also noted that the equal-time constraint for the NBS wave function here is not a restriction to the extraction of physical observables such as the scattering phase shift, as evident from the fact that all informations on the scattering phase shifts are encoded in the asymptotic behavior of the equal-time NBS wave function. Moreover, the equal-time NBS wave function with non-zero total momentum (${\bf P}\not=0$) is equivalent to the NBS wave function in the center of mass frame with non-zero time separation.  In this case the 4-dimensional distance between two nucleon operators is always space-like.

\subsection{Non-local potential from the NBS wave function}

Since the NBS wave function satisfies the free Schr\"odinger equation at large $r$, one can define short-ranged non-local potential  by
\bea
\left[E_k- H_0\right] \varphi^W_{\alpha\beta}({\bf x} ) &=&
\int U_{\alpha\beta;\gamma\delta}({\bf x}, {\bf y})  \varphi^W_{\gamma\delta}({\bf y} )d^3y, \qquad
E_k=\frac{k^2}{2\mu}.
\label{eq:schroedinger}
\eea
It is noted that  the spinor indices $\alpha,\beta,\gamma,\delta$ here run from 1 to 2, since 
all $NN$ scattering phase shifts can be reproduced from the NBS wave function with $\alpha,\beta\in \{1,2\}$ as discussed in the previous subsection.
Therefore $U_{\alpha\beta;\gamma\delta}$ has $4\times 4$ components, which can be determined from $4$ components of $\varphi_{\alpha\beta}^W$ for 4 different combinations of $(s_1, s_2)$. 
Note that since the NBS wave function $\varphi^W$ is multiplicatively renormalized, the potential $U({\bf x},{\bf y})$ is finite and does not depend on the particular renormalization scheme .   

The non-local function $U({\bf x},{\bf y})$ is shown to be energy-independent as follows.
Let $V_{\rm th}$ be the space spanned by the wave function with $W\le W_{\rm th} \equiv 2m_N+m_\pi$: $V_{\rm th}=\{\varphi^W\vert W\le W_{\rm th}\}$, and
the projection operator to $V_{\rm th}$ is defined as
\bea
P^{W_{\rm th}}({\bf x},{\bf y}) &=&\int_{W_{1,2}\le W_{\rm th} }\rho(W_1) dW_1\, \rho(W_2) dW_2\, \varphi^{W_1}({\bf x}) N^{-1}(W_1,W_2)\varphi^{W_2}({\bf y})^\dagger \nonumber \\
&\equiv& \int_{W_1\le W_{\rm th} }\rho(W_1) dW_1\, P(W_1;{\bf x},{\bf y}) 
\eea
where $\rho(W)$ is the density of states at energy $W$, and $N^{-1}(W_1,W_2)$ is the inverse 
of  the hermitian operator $N(W_1,W_2)$ defined by
\be
N(W_1,W_2) = \int  \varphi^{W_1}({\bf r})^\dagger \varphi^{W_2}({\bf r})\, d^3 r,
\ee 
so that
\be
\int \rho(W) dW\, N(W_1,W) N^{-1}(W,W_2) =\frac{1}{\rho(W_1)}\delta(W_1-W_2).
\ee
The non-local potential is then defined by
\bea
U^{W_{\rm th}}({\bf x},{\bf y}) &=&  \int_{W_{1,2}\le W_{\rm th} }\rho(W_1) dW_1\, \rho(W_2) dW_2\,  \left[E_k - H_0\right]\varphi^{W_1}({\bf x}) N^{-1}(W_1,W_2)\varphi^{W_2}({\bf y})^\dagger \nonumber \\
&=&  \int_{W_1\le W_{\rm th} }\rho(W_1) dW_1\,   \left[E_k - H_0\right] P(W_1;{\bf x},{\bf y}).
\label{eq:non-local}
\eea
It is easy to see that the above non-local potential satisfies eq.(\ref{eq:schroedinger})  at 
$W\le W_{\rm th}$ as follows.
\bea
\int U({\bf x},{\bf y})^{W_{\rm th}}\varphi^W({\bf y})\, d^3y &=&
\int_{W_1\le W_{\rm th} }\rho(W_1) dW_1\,  \left[E_k - H_0\right]\varphi^{W_1}({\bf x}) 
\frac{1}{\rho(W)}\delta(W_1-W)\nonumber \\
&=& \theta(W_{\rm th}-W)\, \left[E_k - H_0\right]\varphi^{W }({\bf x}) .
\eea
It should be noted that the non-local potential $U({\bf x},{\bf y})$ which satisfied 
eq.(~\ref{eq:schroedinger})  at  $W\le W_{\rm th}$  is not unique: For example, we can add the following term 
\be
\int_{W>W_{\rm th}} \rho(W)\, dW\, f_W({\bf x}) P(W;{\bf x},{\bf y})
\ee
with arbitrary functions $f_W({\bf x})$ to the non-local potential $U({\bf x},{\bf y})$ without affecting eq.(\ref{eq:schroedinger})  at  $W\le W_{\rm th}$.
The non-local potential  $U({\bf x},{\bf y})$ in eq. (\ref{eq:non-local}) is energy independent by construction. 

Alternatively we can define a different non-local potential by
\be
U^{\infty}({\bf x},{\bf y}) =  \int_0^\infty \rho(W) dW\,   \left[E_k - H_0\right] P(W;{\bf x},{\bf y}),
\ee 
which satisfies eq. (\ref{eq:schroedinger}) for all $W$. This potential, however, becomes long-ranged, due to the presence of inelastic contributions above $W_{\rm th}$.
The extension of this method to non-elastic cases will be discussed in Sec.~\ref{sec:extension}. 

In Ref.~\cite{Beane:2010em}, it is claimed that the NBS wave function satisfies the Schr\"odinger equation with a non-local and energy-dependent potential. In general this is true but, as shown here, there is a scheme which makes the non-local potential energy-independent.  This is sufficient for the strategy considered in this report.
 
\subsection{Velocity expansion of the non-local potential}
In principle, if one knows all NBS wave functions $\varphi^W$, the non-local potential $U$ can be constructed according to eq. (\ref{eq:non-local}). In practice, however, one can obtain only a few of them corresponding to the ground state as well as a few low lying excited states in lattice QCD simulations. Therefore, for practical applications, it is convenient to expand the non-local potential  in terms of the velocity(derivative) with local functions as
\be
U({\bf x},{\bf y}) = V({\bf x},\nabla) \delta^3({\bf x}-{\bf y}) .
\ee 
At the lowest few orders we have
\bea
V({\bf r},\nabla) &=&\underbrace{V_0(r) + V_\sigma(r) \vec\sigma_1\cdot\vec\sigma_2 + V_T(r) S_{12}}_{\rm LO} + \underbrace{V_{\rm LS} (r){\bf L}\cdot{\bf S}}_{\rm NLO} + O(\nabla^2),
\label{eq:velocity_exp}
\eea
where $r=\vert{\bf r}\vert$, $\vec\sigma_i$ is the Pauli-matrix acting on the spin index of the $i$-th nucleon, ${\bf S}=(\vec\sigma_1+\vec\sigma_2)/2$ is the total spin, ${\bf L} = {\bf r}\times {\bf p}$ is the angular momentum, and
\be
S_{12} = 3 \frac{({\bf r}\cdot \vec\sigma_1) ({\bf r}\cdot \vec\sigma_2) }{r^2} -\vec\sigma_1\cdot\vec\sigma_2
\ee
is the tensor operator. Each coefficient function is further decomposed into its flavor components as
\be
V_X(r) = V_X^0(r) + V_X^\tau(r) \vec\tau_1\cdot \vec\tau_2, \quad
X=0, \sigma, T, {\rm LS},\cdots,
\ee
where $\vec \tau_i$ is the Pauli-matrix acting on the flavor index of the $i$-th nucleon. 
The form of the velocity expansion (\ref{eq:velocity_exp}) agrees with the form determined by symmetries\cite{okubo1958}.

For the leading order of the velocity expansion, the local potential is given by
\be
V^{\rm LO}({\bf r}) = V_0(r) +  V_\sigma(r) \vec\sigma_1\cdot\vec\sigma_2 + V_T(r) S_{12},
\ee
which can be obtained from the NBS wave function at one value of $W$. Since $S_{12}=0$ for the spin single state, for example, we have
\be
V_c(r, S=0)\equiv V_c(r) - 3V_\sigma(r) = \frac{\left[E_k -H_0\right]\varphi^W({\bf r})}{\varphi^W({\bf r})} .
\ee

\subsection{Remarks}
There are several remarks on the extraction of the potential from QCD described in the previous sections.

First of all, it should be noted that the potential itself is not a physical observable, and it is therefore not unique.  In particular the potential depends on the choice of the nucleon operator to define the NBS wave function. The local nucleon operator is one choice here but other definitions are equally possible, though the local operator is a convenient choice for the reduction formula of composite particles such as the nucleon\cite{Nishijima,Zimmermann,Haag}. A choice of the nucleon operator to define the NBS wave function is considered to be a "scheme" to define the potential. The potential is therefore a scheme dependent quantity, while
physical observables such as the scattering phase shift and the binding energy of the deuteron are of course scheme independent.
  
Is such a scheme-dependent quantity useful ? The answer to this question is probably "yes", since the potential is useful to "understand" or "describe"  the phenomena. For example, the repulsive core best summarizes the behavior of the NN scattering phase shift at larger energy in terms of the short distance behavior of the potential. 
A well-known example for a scheme-dependent but useful quantity is the running coupling in quantum field theory. Although the running coupling is of course scheme dependent, it is useful to understand the deep inelastic scattering data at high energy ({\it i.e.} asymptotic freedom).

Among different schemes, there exist of course good schemes. A good convergence of the perturbative expansion for a certain class of observables is  a reasonable criteria for a good scheme  in the case of the running coupling. In the case of the potential, a good convergence of the velocity expansion is important. In this respect, a completely local and energy-independent potential is the best one, and moreover the inverse scattering method tells us that it must be unique if it exists. We think that such a local and energy-independent potential  exists only if no inelastic scattering appears at all values of energy ({\it i.e.} $W_{\rm th}=\infty$).

There exists a criticism that the repulsive core in the phenomenological potentials is meaningless since the low energy scattering data can be described equally well by other potentials without the repulsive core. In other words, the repulsive core can be removed by a unitary transformation of the wave function without changing the scattering phase shift at low energy. Even though this may be true, the resulting "potential" is highly non-local, and therefore is beyond the concept of a potential.
This criticism thus corresponds to claiming that the asymptotic freedom is meaningless since it can be removed by some non-perturbative scheme of the "coupling", which is however beyond the concept of a coupling. 

For the definition of the potential in QCD from the NBS wave function, the convergence of the velocity expansion can  be checked by examining the energy ($W$) dependence of the lower order  potentials. For example, if we have $\varphi^{W_n}$ for $n=1,2,\cdots N$, we can determine the $N-1$ unknown local functions of the velocity expansion in $N$ different ways.
The variation among $N$ different determinations gives an estimate of the size of the higher order terms neglected. Furthermore one of these higher order terms can be determined from   $\varphi^{W_n}$ for $n=1,2,\cdots N$.
The convergence of the velocity expansion will be  investigated explicitly in Sec.~\ref{sec:NNpotential}.

Let us consider what we have shown so far from a slightly different point of view.
Instead of adopting the point of view that  the "potential" can be defined in QCD,  
the analysis in this section shows that 
the use of quantum mechanics with potentials to describe the NN scattering can be justified  in a more fundamental quantum field theory (QCD) through the NBS wave function, whose asymptotic behavior encodes phases of the S-matrix for the NN scattering. 
If the local approximation for the potential defined from the NBS wave function is good in the low energy region, one can use the local potential combined with quantum mechanics to investigate nuclear physics\footnote{It is more legitimated to show if  three or more body potentials remain  subdominant in the same framework in QCD before using the potential in nuclear physics.   There exists another approach to investigate nuclei, by combining the chiral effective field theory with lattice calculations. (See Ref.\cite{Epelbaum:2010xt} and references therein.) The chiral effective can easily incorporate may-body interactions order by order in the chiral expansion, though there are many free parameters to be determined from experimental data. On the other hand, the potential from lattice QCD does not contain such free parameters.  Therefore it is interesting and important to determine some parameters of the chiral effective theory from the potential in lattice QCD.  } .  

It is now clear that there is no unique definition for the NN potential. Ref.~\cite{Beane:2010em,Detmold:2007wk,Beane:2008ia}, however, criticized that the NBS wave function is not "the correct wave function for two nucleons" and that its relation to the correct wave function is given by
\be
\varphi^W({\bf r}) = Z_{NN}(\vert {\bf r}\vert) \langle 0 \vert T\{N_0({\bf x}+{\bf r},0) N_0({\bf x},0)\}\vert  2N, W,s_1,s_2\rangle + \cdots
\ee 
where $N_0({\bf x},t)$ is "a free-field nucleon operator"  and the ellipses denote "additional contributions from the tower of states of the same global quantum numbers".
Thus $\langle 0 \vert T\{N_0({\bf x}+{\bf r},0) N_0({\bf x},0)\}\vert  2N, W,s_1,s_2\rangle$ is  considered to be "the correct wave function".
In this claim it is not clear what is "a free-field nucleon operator"  in the interacting quantum field theory.   An asymptotic {\it in} or {\it out} field operator may be a candidate. If the asymptotic field is used for $N_0$, however, the potential defined from the wave function identically vanishes for all ${\bf r}$ by construction.
To be more fundamental,  a concept of "the correct wave function" is doubtful. If  some wave function were "correct",  the potential would be uniquely defined from it. This clearly contradicts the fact discussed above that the potential is not an observable and therefore is not unique. 
This argument shows that the criticism of Ref.~\cite{Beane:2010em,Detmold:2007wk,Beane:2008ia} is flawed.

\section{Lattice formulation}
\label{sec:lattice}
In this section, we discuss the procedure to extract the NBS wave function from lattice QCD simulations.
For this purpose, let us consider the correlation function on the lattice defined by
\be
F({\bf r},t-t_0)=\langle 0\vert  T\{N({\bf x}+{\bf r},t ) N({\bf x},t)\} \overline{\cal J}(t_0)\vert 0 \rangle
\label{eq:4-pt}
\ee
where $\overline{\cal J}(t_0)$ is a source operator which creates a two nucleon state of states and its explicit form will be considered later. By inserting a complete set and considering baryon number conservation, we have
\bea
F({\bf r},t-t_0) &=&\langle 0\vert  T\{N({\bf x}+{\bf r},t ) N({\bf x},t)\} \sum_{n,s_1,s_2} \vert 2N, W_n, s_1,s_2\rangle \langle 2N, W_n,s_1,s_2 \vert  \overline{\cal J}(t_0)\vert 0 \rangle \nonumber \\
&=& \sum_{n, s_1,s_2} A_{n, s_1,s_2} \varphi^{W_n}({\bf r}) e^{-W_n (t-t_0)}, \quad
A_{n,s_1,s_2} =\langle 2N, W_n,s_1,s_2 \vert  \overline{\cal J}(0)\vert 0 \rangle .
\eea
For a large time separation $(t-t_0)\rightarrow \infty$, we have
\be
\lim_{(t-t_0)\rightarrow\infty} F({\bf r},t-t_0) = A_0 \varphi^{W_0}({\bf r}) e^{-W_0 (t-t_0)}
+ O(e^{- W_{n\not=0} (t-t_0)})
\label{eq:ground}
\ee 
where $W_0$ is assumed to be the lowest energy of NN states.
Since the source dependent term $A_0$ is just a multiplicative constant to the NBS wave function $\varphi^{W_0}({\bf r})$, the potential defined from $\varphi^{W_0}({\bf r})$ in the previous section is manifestly source-independent. Therefore the statement  that  the potential in this scheme is "source-dependent" in Ref.~\cite{Beane:2008dv} is clearly wrong. 

In this extraction of the wave function, the ground state saturation for the correlation function $F$ in eq. (\ref{eq:ground}) is important. In principle, one can achieve this by taking a large $t-t_0$.
In practice, however, $F$ becomes very noisy at large $t-t_0$, so that the extraction of $\varphi^{W_0}$ becomes difficult at large $t-t_0$. Therefore it is crucial to find the region of $t$ where the ground state saturation is approximately satisfied while the signal is still reasonably good.
The  choice of the source operator becomes important in order to have such a good $t$-region. 

\subsection{Choice of source operator}
We can choose the source operator $\bar{\cal J}$ to fix quantum numbers of the state $\vert 2N, W, s_1,s_2\rangle $ such as $(J, J_z)$.  Since a lattice QCD simulation is usually performed on the (finite) hyper-cubic lattice, we consider the cubic transformation group $SO(3,{\bf Z})$ instead of the $SO(3,{\bf R})$ as the symmetry of 3-dimensional space. Therefore the quantum number is classified in terms of the irreducible representation of $SO(3,{\bf Z})$,  denoted by $A_1$, $A_2$, $E$, $T_1$, $T_2$ whose dimensions are $1,1,2,3,3$. 
A relation of irreducible representations between $SO(3,{\bf Z})$ and $SO(3,{\bf R})$ is given in table~\ref{tab:cubic} for $L\le 6$, where $L$ represents the angular momentum  for the  irreducible representation in $SO(3,{\bf R})$. For example,  the source operator $\bar{\cal J}(t_0)$ in the $A_1$ representation with positive parity generates states with $L=0,4,6,\cdots$ at $t=t_0$, while 
the operator in the $T_1$ representation with negative parity produces states with $L=1,3,5,\cdots$.  For two nucleons, the total spin $S$ becomes $1/2\otimes 1/2 = 1 \oplus 0$, which corresponds to $T_1$($S=1$) and $A_1$($S=0$) of the $SO(3,{\bf Z})$.
Therefore, the total representation $J$ for a two nucleon system is determined by the product $R_1\otimes R_2$, where $R_1=A_1,A_2,E,T_1,T_2$ for the orbital "angular momentum" while $R_2=A_1, T_1$ for the total spin. In table~\ref{tab:product}, the product $R_1\otimes R_2$ is decomposed into the direct sum of irreducible representations. 

\begin{table}
\begin{center}
\caption{The number of each representation of  $SO(3,{\bf Z})$ which appears in the angular momentum $L$ representation of $SO(3,{\bf R})$. $P=(-1)^L$ represents an eigenvalue under parity transformation.}
\label{tab:cubic}
\vspace{0.3cm}
\begin{tabular}{|cc|ccccc|}
\hline
$L$ & $P$ & $A_1$ & $A_2$ & $E$ & $T_1$ & $T_2$ \\
\hline
0 (S)&  $+$ & 1 & 0 &0 &0 & 0 \\
1 (P)&  $-$  & 0 & 0 &0 & 1 & 0\\
2 (D)&  $+$ & 0 & 0 &1 & 0 & 1 \\
3 (F)&  $-$  & 0 & 1 &0 & 1 & 1\\
4 (G)&  $+$ & 1 & 0 &1 & 1 & 1 \\
5 (H)&  $-$  & 0 & 0 &1 & 2 & 1\\
6 (I) &  $+$ & 1 & 1 &1 & 1 & 2 \\
\hline
\end{tabular}
\end{center}
\end{table}     

\begin{table}
\begin{center}
\caption{The decomposition for a product of two irreducible representations, $R_1\otimes R_2$, into irreducible representations in $SO(3,{\bf Z})$.  Note that $R_1\otimes R_2= R_2\otimes R_1$ by definition. }
\label{tab:product}
\vspace{0.3cm}
\begin{tabular}{|c||ccccc|}
\hline
  & $A_1$ & $A_2$ & $E$ & $T_1$ & $T_2$ \\
\hline
\hline
$A_1$ & $A_1$ & $A_2$ & $E$ & $T_1$ & $T_2$ \\
$A_2$   & $A_2$  & $A_1$ & $E$ & $T_2$ &  $T_1$\\
$E$       &  $E$ & $E$  &$A_1\oplus A_2\oplus E$ & $T_1\oplus T_2$ & $T_1\oplus T_2$ \\
$T_1$    & $T_1$ & $T_2$& $T_1\oplus T_2$ & $A_1\oplus E \oplus T_1\oplus T_2$ & $A_2\oplus E \oplus T_1\oplus T_2$\\
$T_2$  & $T_2$ & $T_1$  &$T_1\oplus T_2$& $A_2\oplus E \oplus T_1\oplus T_2$ &$A_1\oplus E \oplus T_1\oplus T_2$  \\
\hline
\end{tabular}
\end{center}
\end{table}

Most of  the results in this report use the wall source at $t=t_0$ defined by
\be
{\cal J}^{\rm wall}(t_0)_{\alpha\beta, fg} = N^{\rm wall}_{\alpha,f} (t_0) N^{\rm wall}_{\beta,g} (t_0)
\ee
where $\alpha,\beta=1,2$ are upper component spinor indices while $f,g$ are flavor indices.
Here $N^{\rm wall}(t_0)$ is obtained by replacing the local quark field $q(x)$ of $N(x)$ by the wall quark field,
\be
q^{\rm wall}(t_0) \equiv \sum_{\bf x} q({\bf x},t_0) 
\ee
with the Coulomb gauge fixing only at $t=t_0$. Note that this gauge-dependence of the source operator disappears for the potential.
All states created by the wall source have zero total momentum. Among them 
the state with zero relative momentum has the largest magnitude.
The most important reason to employ the wall source  is that the ground state saturation for the potential at long distance is better achieved for the wall source than other sources such as the smeared source.

By construction, the source operator $\bar{\cal J}^{\rm wall}(t_0)$ has zero orbital angular momentum at $t=t_0$, which corresponds to the $A_1$ representation with positive parity.
By the spin projection operator $P^{(S)}$, e.g.  $P^{(S=0)}=\sigma_2$ and $P^{(S=1,S_z=0)}=\sigma_1$,  we fix the $J$ of  the source as
\be
{\cal J}(t_0;J^{P=+},I) = P^{(S)}_{\beta\alpha} {\cal J}^{\rm wall}(t_0)_{\alpha\beta,fg} 
\ee
where $P=\pm $ is the parity and $I=1,0$ is the total isospin of the system. Since the  nucleon is a fermion, an exchange of two nucleon operators in the source should give a minus sign. This fact fixes the total isospin once the total spin is given: $(S,I)=(0,1)$ or $(1,0)$.
(Note that $S,I=0$ are antisymmetric while $S,I=1$ are symmetric under the exchange.)
Since $A_1^+\otimes A_1(S=0) = A_1^+$  and $A_1^+\otimes T_1(S=1) = T_1^+$, the state with either $(J^P,I) =(A_1^+, 1)$ for the spin-singlet or $(J^P,I) =(T_1^+, 0)$ for the spin-triplet is created at $t=t_0$ by the corresponding source operator.  The NBS wave function extracted at $t > t_0$ has the same  quantum numbers $(J^P,I)$  as they are conserved under QCD interactions. 
In addition the total spin $S$ is conserved at $t > t_0$ for the two-nucleon system with equal up and down quark masses: Under the exchange of the two particles, the constraint $(-1)^{S+1+I+1}P=-1$ must be satisfied due to the fermionic nature of nucleon, while
the parity $P$ and the isospin $I$ are conserved in this system. Therefore $S$ is conserved.
It is  noted, however, that $L$ is not conserved in general. While the state with $(J^P,I) =(A_1^+, 1)$  always has $L=A_1^+$ even at $t > t_0$, the one with  $(J^P,I) =(T_1^+, 0)$  has both $L=A_1^+$ and $L=E^+$ components\footnote{This can be seen from Table~\ref{tab:product} for $R_2=T_1$(spin-triplet), which also tells us  existences of $L= T_1^+$ and $L=T_2^+$ components in addition. These extra components are expected to be small since they appear as a consequence of the violation of $SO(3,{\bf R})$ on the hyper-cubic lattice.} at $t> t_0$, which corresponds to $L=0$ and $L=2$ in $SO(3,{\bf R})$, respectively.
Note that  $J$ or $L$ in this report is used to represent the total or orbital  quantum number of $SO(3,{\bf Z})$ as well as $SO(3,{\bf R})$, depending on the context. 

The orbital angular momentum $L$ of the NBS wave function can be fixed to a particular value by the projection operator $P^{(L)}$ as
 \be
 \varphi^W({\bf r}; J^P,I, L,S) = P^{(L)} P^{(S)} \varphi^W({\bf r}; J^P,I) 
 \ee
 where $\varphi^W({\bf r}; J^P,I) $ is extracted from
 \be
 F({\bf r},t-t_0;J^P,I) \simeq A(J^P,I) \varphi^{W}({\bf r}; J^P,I)e^{-W(t-t_0)} ,\quad
 A(J^P,I) =\langle 2N,W  \vert \bar{\cal J}(0;J^P,I)\vert 0\rangle
 \ee
 for large $t-t_0$. Here we also apply the total spin projection operator $P^{(S)}$ but this is redundant since the total spin $S$, already fixed by the source, is conserved as mentioned above.
The projection operator $P^{(L)}$ to an arbitrary function $\phi({\bf r})$ is defined in general by
\be
P^{(L)} \phi({\bf r}) \equiv \frac{d_L}{24}\sum_{g\in SO(3,{\bf Z})} \chi^L(g) \phi(g^{-1}\cdot{\bf r})
\ee
for $L=A_1,A_2,E,T_1,T_2$, where $\chi^L$ denotes the character for the representation $L$ of the cubic group $SO(3,{\bf Z})$, $g$ is one of 24 elements in $SO(3,{\bf Z})$ and $d_L$ is the dimension of $L$.

\subsection{Leading order potential: spin-singlet case}
We present the procedure to determine potentials at the reading order(LO):
\be
V^{\rm LO}({\bf r} ) = V_0(r) +V_\sigma(r)(\vec\sigma_1\cdot\vec\sigma_2)+ V_T(r) S_{12}.
\ee
Since $S_{12}=0$ and $\vec\sigma_1\cdot\vec\sigma_2= - 3$ for the spin-singlet case, 
the LO  central potential in the case of $(J^P,I)=(A_1^+,1)$ state is extracted as
\be
V_C(r)^{(S,I)=(0,1)} \equiv V^{I=1}_0 (r)-3 V^{I=1}_\sigma (r)
= \frac{\left[E_k- H_0\right]\varphi^W({\bf r}; A_1^+,I=1,A_1,S=0)}{\varphi^W({\bf r}; A_1^+,I=1,A_1,S=0)},
\ee
where $V_X^{I=1} = V_X^0+V_X^\tau$ in isospin space.
The potential $V_C({\bf r})^{(S,I)=(0,1)}$ in the above
is often referred to as the central potential for the $^1S_0$ state, where the notation $^{2S+1}L_J$ represents the orbital angular momentum $L$ (see table~\ref{tab:cubic}), the total spin $S$ and the total angular momentum $J$ of ${\bf J}={\bf L}+{\bf S}$. It is noted, however, that in the leading order of the velocity expansion, the potential does not depend on the quantum number of the state $J=L=A_1$. Moreover the $A_1$ state may contain $L=4,6,\cdots$ components other than $L=0$, though the $L=0$ component may dominate in the state. Therefore it is more precise to refer to $V_C({\bf r})^{(S,I)=(0,1)}$ as the spin-singlet (isospin-triplet) central potential determined from the state with $J=L=A_1$. A possible difference of spin-singlet central potentials between this determination and others such as the one determined from $J=L=E$ gives an estimate for contributions from higher order terms in the velocity expansion. 

\subsection{Leading order potential: spin-triplet case}
Both the tensor potential $V_T$ and central potential $V_C$ appear at the LO
for the spin-triplet case.  Let us consider the determination from the $(J^P,I)=(T_1^+,0)$ state. The 
Schr\"odinger equation for this state becomes
\be
\left[H_0+V_C(r)^{(S,I)=(1,0)} + V_T(r) S_{12}\right]\varphi^W({\bf r};J^P=T_1^+,I=0) = E_k \varphi^W({\bf r};J^P=T_1^+,I=0)
\ee 
where the spin-triplet central potential is given by
\be
V_C(r)^{(S,I)=(1,0)} \equiv  V_0^{I=0}(r) + V_\sigma^{I=0}(r), \qquad
V_X^{I=0} = V_X^0-3V_X^\tau .
\ee
The projections to $A_1$ and $E$ components read
\bea
{\cal P}\varphi^W_{\alpha\beta} &\equiv& P^{(A_1)}\varphi^W_{\alpha\beta} ({\bf r}; J^P=T_1^+,I=0)
\label{eq:projA1}
\\
{\cal Q}\varphi^W_{\alpha\beta} &\equiv& P^{(E)}\varphi^W_{\alpha\beta} ({\bf r}; J^P=T_1^+,I=0)
\simeq (1-P^{(A_1)}) \varphi^W_{\alpha\beta} ({\bf r}; J^P=T_1^+,I=0).
\label{eq:projE}
\eea
The last quantity in eq. (\ref{eq:projE})  is an approximation of the first line and a difference comes from $T_1$ and $T_2$ components, which are expected to be small.  
This approximate representation for ${\cal Q}$ is often employed in numerical simulations.

Using these projections, $V_C$ and $V_T$ can be extracted as
\bea
V_C(r)^{(1,0)} &=& E_k -\frac{1}{\Delta({\bf r})}\left( [{\cal Q} S_{12}\varphi^W]_{\alpha\beta}({\bf r})
H_0[{\cal P}\varphi^W]_{\alpha\beta}({\bf r}) - [{\cal P} S_{12}\varphi^W]_{\alpha\beta}({\bf r})
H_0[{\cal Q}\varphi^W]_{\alpha\beta}({\bf r})
\right) \label{eq:central}\\
V_T(r) &=& \frac{1}{\Delta({\bf r})}\left( [{\cal Q} \varphi^W]_{\alpha\beta}({\bf r})
H_0[{\cal P}\varphi^W]_{\alpha\beta}({\bf r}) - [{\cal P} \varphi^W]_{\alpha\beta}({\bf r})
H_0[{\cal Q}\varphi^W]_{\alpha\beta}({\bf r})
\right) \label{eq:tensor}
\\
\Delta({\bf r}) &\equiv& [{\cal Q} S_{12}\varphi^W]_{\alpha\beta}({\bf r})
[{\cal P}\varphi^W]_{\alpha\beta}({\bf r}) - [{\cal P} S_{12}\varphi^W]_{\alpha\beta}({\bf r})
[{\cal Q}\varphi^W]_{\alpha\beta}({\bf r}) .
\eea
In numerical simulations, $(\alpha,\beta)=(2,1)$ is mainly employed. 

If one neglects $V_T$ by putting $V_T=0$ in the above, one obtains the effective central potential for the spin-triplet (isospin-singlet) state as
\be
V_C^{\rm eff}(r)^{(1,0)} =  \frac{\left[E_k- H_0\right] {\cal P}\varphi^W_{\alpha\beta}({\bf r})  }{{\cal P}\varphi^W_{\alpha\beta}({\bf r}) } . 
\label{eq:effective_central}
\ee
The difference between $V_C$ and $V_C^{\rm eff}$ is $O(V_T^2)$ in the second order perturbation for small $V_T$.  

\subsection{A comparison with the finite volume method in lattice QCD}
In this subsection we briefly compare the potential method with the direct extraction of the phase shift via the finite volume method in lattice QCD. 

First of all, by construction, the potential method gives the correct phase shift at $k =\sqrt{W^2/4-m_N^2}$ where $W$ is the total energy of the state from which the NBS wave function is defined, while phase shifts at other values of $k$ are approximated ones obtained in the velocity expansion of the non-local potential. 

Secondly, the finite size correction to the potential is expected to be small. Indeed the finite volume method for the extraction of the phase shift in lattice QCD assumes that there is no finite size correction to the potential as long as the volume is large enough so that the interaction range of the potential is smaller than  half of the lattice extension, $L/2$.  Under this condition, there exists an asymptotic region in the periodic box where the scattering wave satisfies the free Schr\"odinger equation with a specific value of the energy, from which one can determine the phase shift at  certain values of $k$ in the infinite volume. This is L\"uscher's finite volume formula for the phase shift\cite{Luscher:1990ux}.  

Thirdly, we also expect that the quark mass dependence of the potential is much milder than that of physical observables such as the scattering length. While the scattering length is almost zero at the heavy quark mass region, it diverges when the bound state is formed at the lighter quark mass region.   
In this situation, the scattering length varies from zero to infinity as the quark mass changes\cite{Kuramashi:1995sc}.
Such a drastic change of  the scattering length can easily be realized by a small change to the shape of the potential as a function of the quark mass.

Let us assume that higher order terms in the velocity expansion give negligible contributions at low energy so that the leading order local potential well reproduces the scattering phase shift.
In this situation, some problems of the finite size method can be avoided by using 
the potential method.  To extract the phase shift in the finite size method in lattice QCD, one has to assume that one particular angular momentum gives a dominant contribution among possible angular momenta in a given representation of the cubic group. For example, although a state in the  $A_1$ representation   contains not only an $L=0$ contribution but also $L=4,6,\cdots$ contributions, one usually assumes that the $L=0$ contribution dominates so that the energy shift in the finite volume is related to the scattering phase for the $L=0$ state.  In the case of the potential, on the other hand, such an assumption is unnecessary. One can determine the local potential in the velocity expansion from the $A_1$ state without specifying the dominant angular momentum. Once the potential is obtained, one can calculate the scattering phase shift for an arbitrary  $L$ by solving the Schr\"odinger equation in the infinite volume with the extracted potential.
Furthermore, one can check the assumption made for the finite size method by comparing sizes of the scattering phases among different $L$'s.
 
\begin{figure}[tb]
\begin{center}
\includegraphics[width=0.33\textwidth,angle=270]{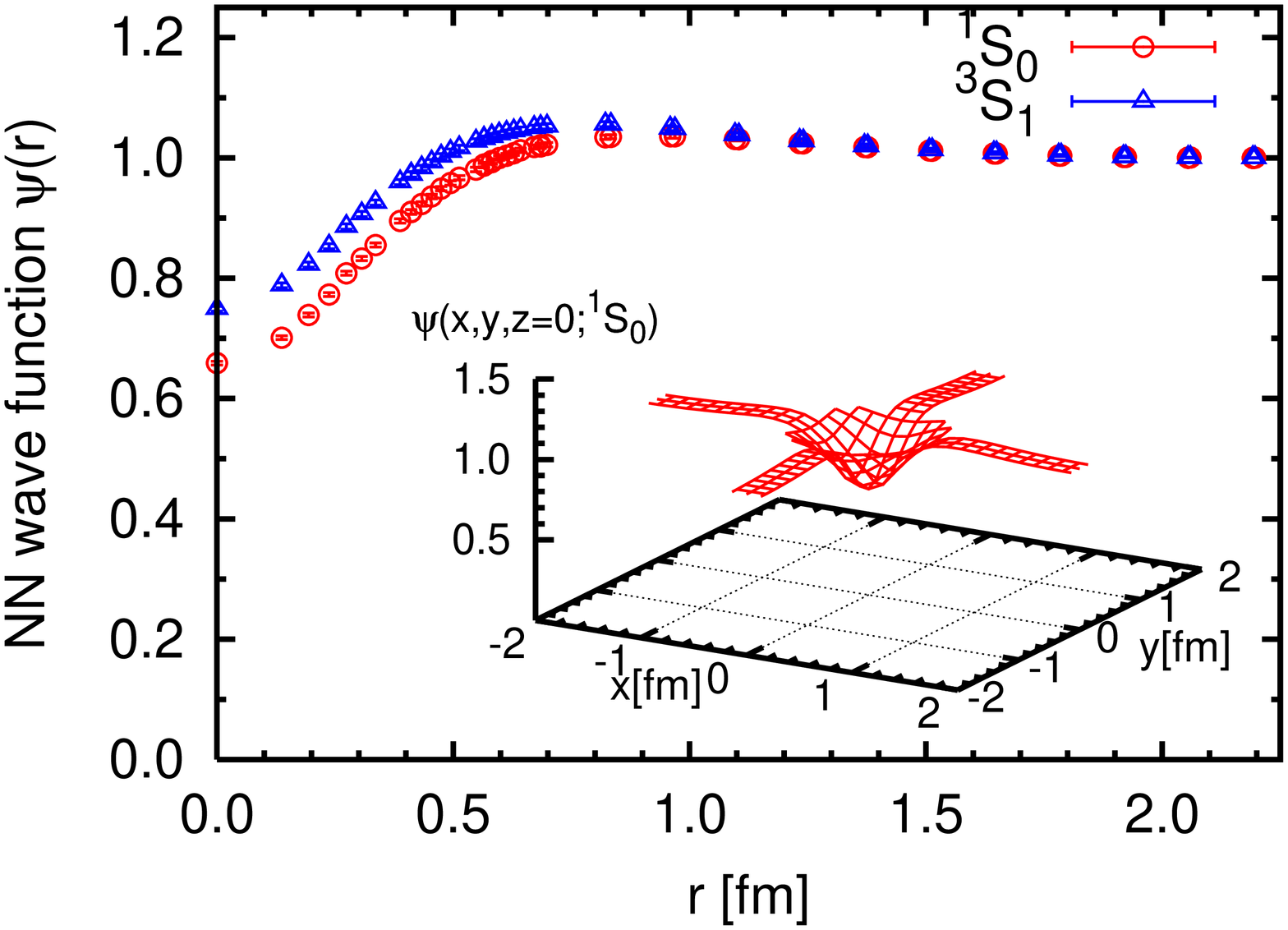}
\includegraphics[width=0.33\textwidth,angle=270]{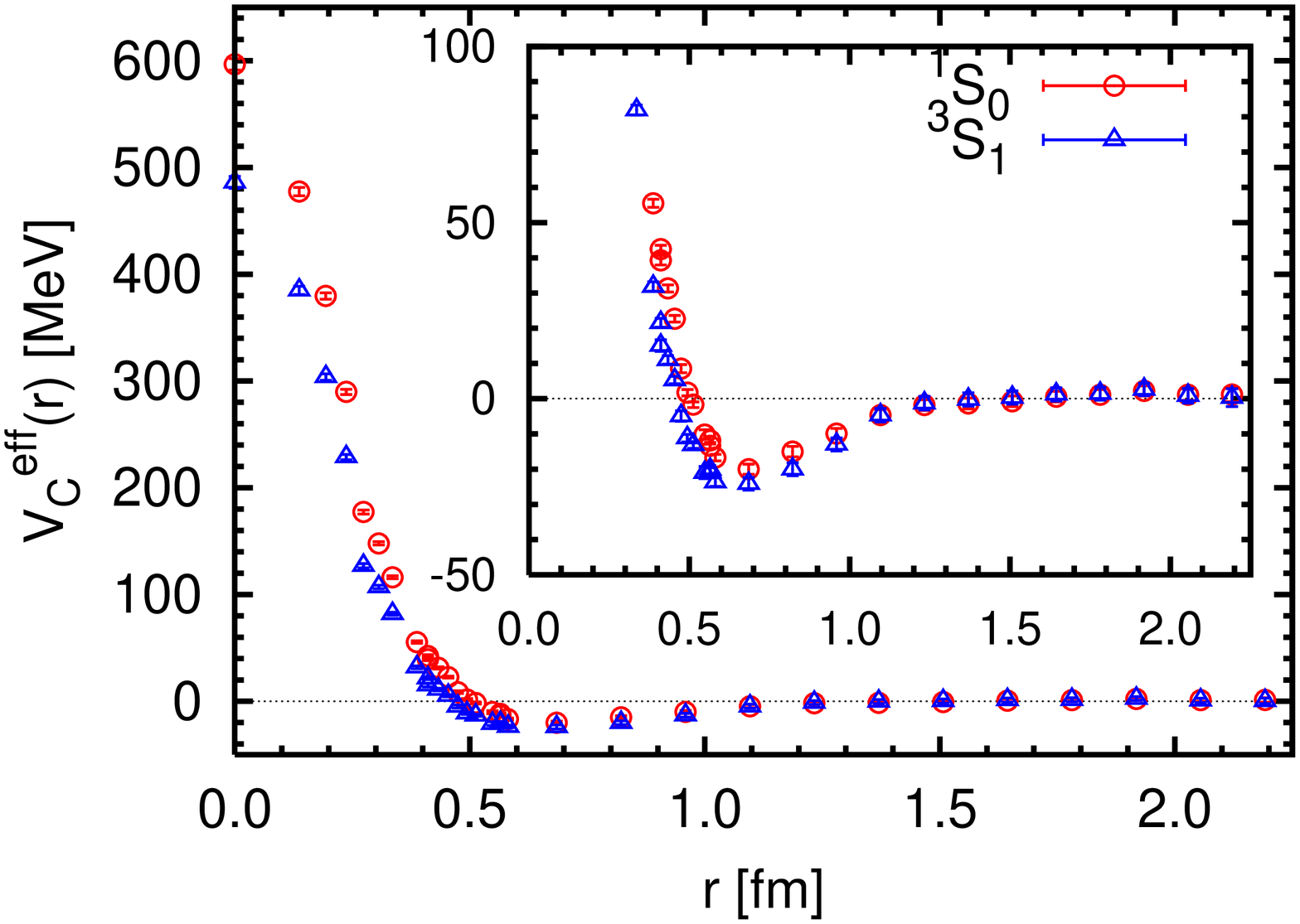}
\caption{(Left)The NN wave function for the spin-singlet and spin-triplet channels in the orbital $A_1^+$ representation at $m_\pi\simeq 529$ MeV and $a\simeq 0.137$ fm in quenched QCD. The insert is a three-dimensional plot of the spin-singlet wave function $\varphi^W(x,y,z=0)$. 
(Right)  The NN (effective) central potential for the spin-singlet (spin-triplet) channel determined from the orbital $A_1^+$ wave function. Both figures are taken from Ref.~\protect\cite{Aoki:2009ji}. }
\label{fig:wave_potential}
\end{center}
\end{figure}

\section{Results for nuclear potentials from lattice QCD}
\label{sec:NNpotential}.

\subsection{Quenched QCD results for (effective) central potentials}
Let us show results in the quenched QCD, where creations and annihilations of virtual quark-antiquark pairs  are all neglected. For the simulations, the standard plaquette gauge action is employed on a 32$^4$ lattice at the bare gauge coupling constant $\beta=6/g^2=5.7$, which corresponds to the lattice spacing $a\simeq 0.137$ fm ($1/a=1.44(2)$ GeV), determined from the $\rho$ meson mass in the chiral limit, and the physical size of the lattice $L\simeq 4.4$ fm\cite{Ishii:2006ec}. As for the quark action, the standard Wilson fermion action is employed at three different values of the quark mass corresponding to the pion mass $m_\pi \simeq  731, 529, 380$ MeV and the nucleon mass $m_N \simeq 1560,1330,1200$ MeV, respectively. 

Fig.~\ref{fig:wave_potential}(Left) shows the NBS wave functions for the spin-singlet and the spin-triplet channels in the orbital $A_1$ representation at $m_\pi \simeq 529$ MeV. These wave functions are normalized to be 1 at the largest spatial point $r\simeq 2.2$ fm.

\begin{figure}[tb]
\begin{center}
\includegraphics[width=6.5cm,angle=270]{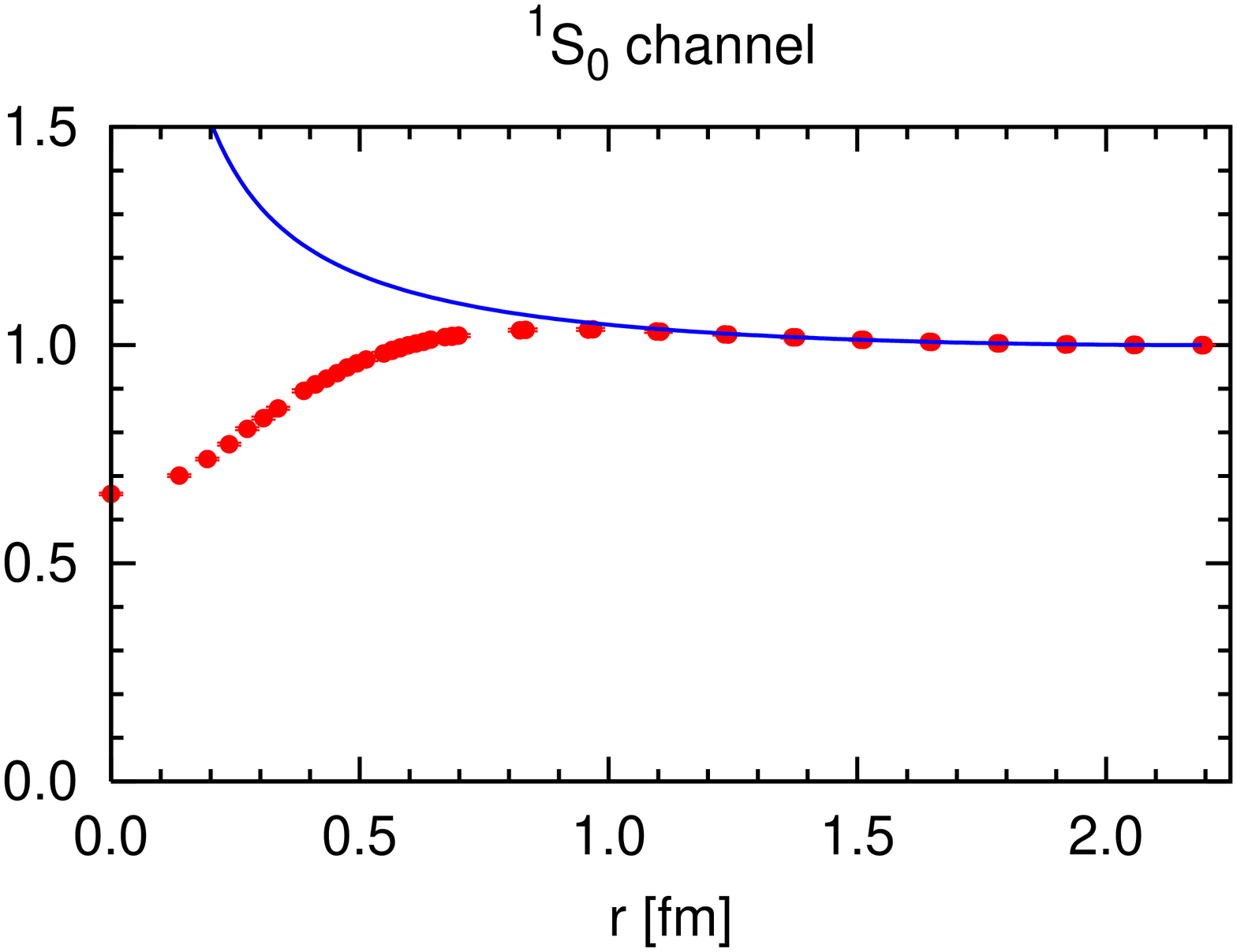}
\includegraphics[width=6.5cm,angle=270]{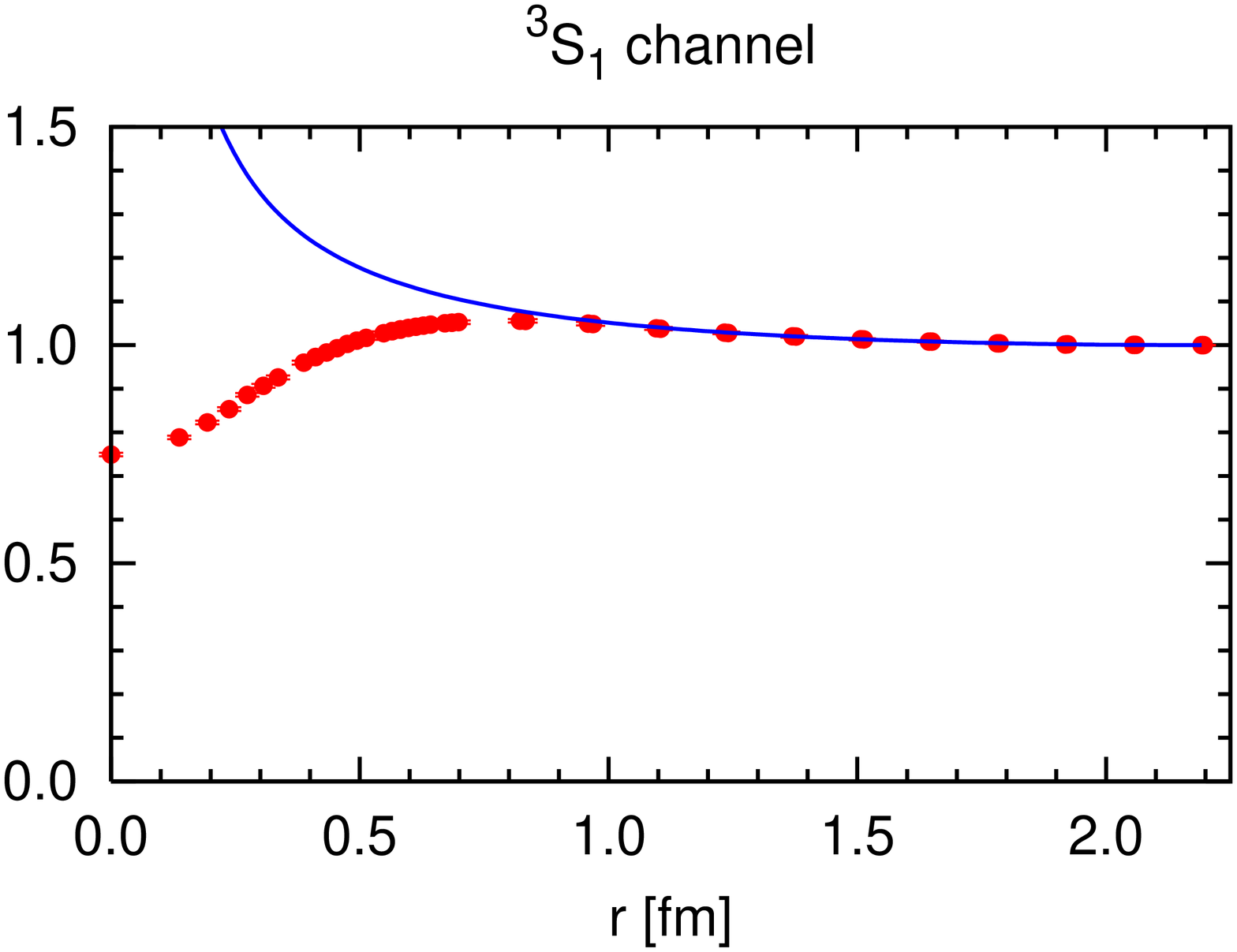}
\caption{(Left) The fit of the $NN$ wave functions at $m_\pi\simeq 529$ MeV for the spin-singlet channel in the orbital $A_1^+$ representation using the Green's function in the fit range $11\le r/a \le 15$. (Right) A similar fit for the spin-triplet channel. Taken from Ref.~\protect\cite{Aoki:2009ji}.}
\label{fig:wave_fit}
\end{center}
\end{figure}

The central potential in the spin-singlet channel and the effective central potential in the spin-triplet channel reconstructed from the wave functions at $m_\pi\simeq 529$ MeV are shown in Fig.~\ref{fig:wave_potential}(Right). These potentials reproduce the qualitative features of the phenomenological $NN$ potentials, namely  the repulsive core at short distance  surrounded by the attractive well at medium and long distances. From this figure one observes that the interaction range of the potential is smaller than 1.5 fm. Therefore the box size $L\simeq 4.4$ fm is large enough to extract the phase shift  by the finite size method, and furthermore the finite size corrections to the potentials themselves are expected to be small. Labels $^1S_0$ and $^3S_1$
of the potentials in the figure represent the fact that potentials are determined from $A_1$ wave functions, which are dominated by $S$  wave components.

\begin{figure}[tb]
\begin{center}
\includegraphics[width=0.35\textwidth,angle=270]{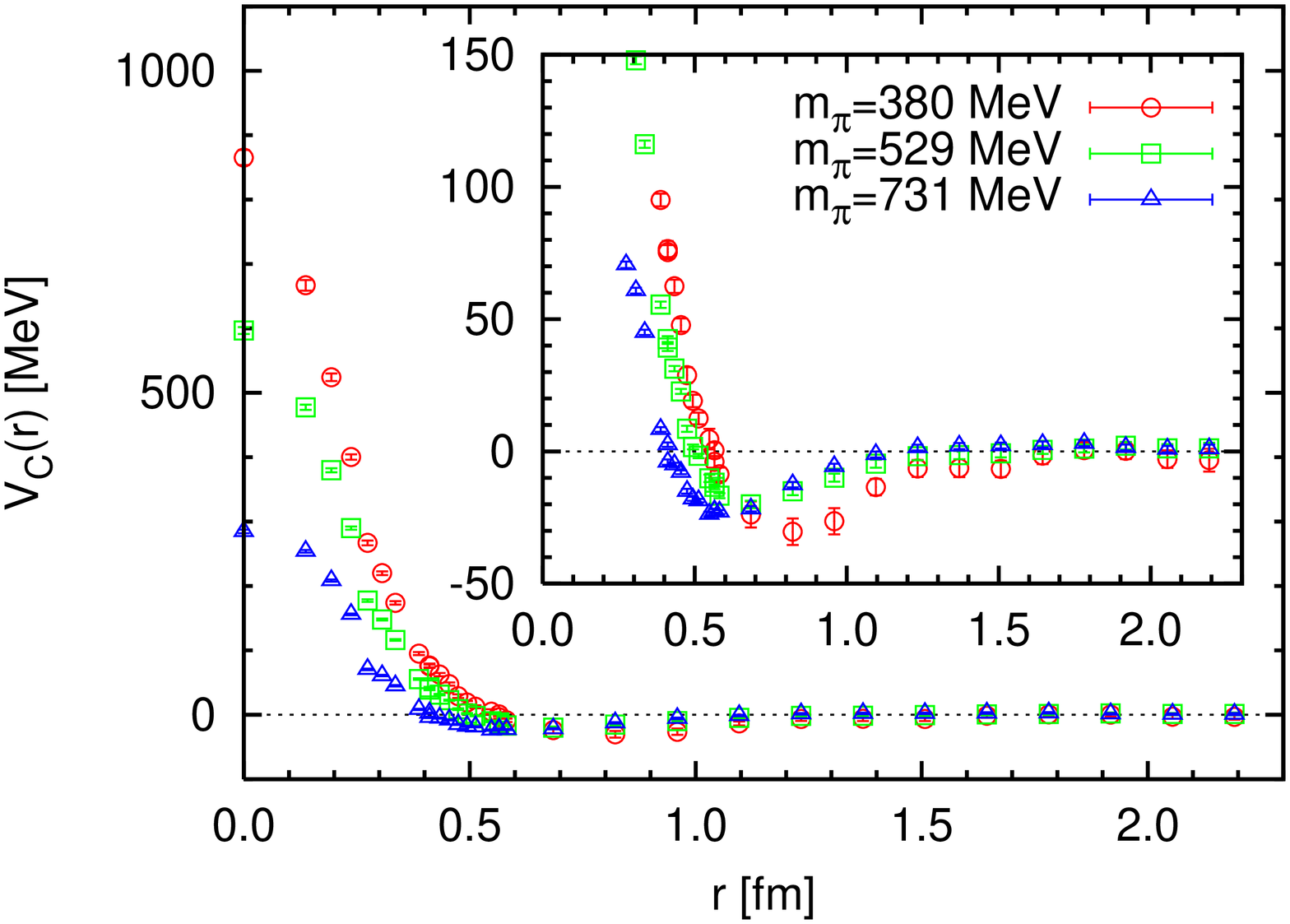}
\caption{The central potentials for the spin-singlet channel from the orbital $A_1^+$ representation at three different pion masses in quenched QCD. Taken from Ref.~\protect\cite{Aoki:2009ji}.}
\label{fig:mass-dep}
\end{center}
\end{figure}

Instead of calculating the energy shift due to the finite size, one can extract the asymptotic momentum $k$, by fitting the NBS wave function $\varphi({\bf r})$ at large distance with the Green's function $G({\bf r};k^2)$ in a finite and periodic box for
the Helmholtz equation $(\nabla^2 + k^2)G({\bf r};k^2) = -\delta_{\rm lat}({\bf r})$ with $\delta_{\rm lat}({\bf r})$ being the periodic delta-function. Explicitly it is given by
\begin{equation}
G({\bf r};k^2) =\frac{1}{L^3}\sum_{{\bf n}\in{\bf Z}^3}\frac{e^{i(2\pi/L){\bf n\cdot r}}}{(2\pi/L)^2{\bf n^2}-k^2}.
\end{equation}
 The asymptotic momentum $k$ is related to the scattering phase shift $\delta_0(k)$ or the scattering length $a_0$ for the S-states\footnote{ We here assume that the dominant component of the scattering wave in $A_1$ representation has $L=0$.}  
as
\begin{equation}
k \cot \delta_0(k) = \frac{2}{\sqrt{\pi} L}Z_{00}(1;q^2) =\frac{1}{a_0} + O(k^2),
\label{eq:SL}
\end{equation}
where $Z_{00}(1,q^2)$ with $q=\frac{kL}{2\pi}$ is (the analytic continuation of) the generalized zeta-function $Z_{00}(s,q^2) = \displaystyle\frac{1}{\sqrt{4\pi}}\sum_{{\bf n}\in {\bf Z}^3}({\bf n}^2 -q^2)^{-s}$.  Fig.~\ref{fig:wave_fit} shows the fits of the wave functions in the interval $11 a\simeq  1.5$ fm $\le r \le 16 a\simeq 2.2$ fm using the above form at $m_\pi = 529$ MeV.  This leads to the values of the effective energy $E\equiv k^2/m_N$, which can be translated to the scattering length $a_0$ by the L\"uscher's formula (\ref{eq:SL}).
 
In Fig.\ref{fig:mass-dep}, we compare the $NN$ central potentials in the spin-singlet channel for three different pion masses. As the pion mass decreases, the repulsion at short distance and the attraction at medium distance are enhanced simultaneously. 
In table~\ref{tab:EandSL}, we give values of $E$ and the S-wave scattering length $a_0$, which show a net attraction of the NN interactions in both channels at these pion masses, though the absolute magnitudes of  the scattering length $a_0$ are much smaller than the experimental values at the physical pion mass $m_\pi \simeq 140$ MeV: $a_0^{(\rm exp)}=(^1S_0)\sim 20$ fm and  $a_0^{(\rm exp)}=(^3S_1)\sim -5$ fm. 
\begin{table}[bt]
\begin{center}
\caption{Effective center of mass energies $E$ obtained from the asymptotic momenta and the scattering length $a_0$ at different pion masses. Taken from Ref.~\protect\cite{Aoki:2009ji}.}
\label{tab:EandSL}
\vspace{0.3cm}
\begin{tabular}{|l || ll | ll |}
\hline
& \multicolumn{2}{c|}{$E$[MeV] } &  \multicolumn{2}{c|}{$a_0$[fm] } \\
\hline
$m_\pi$[MeV]  & spin-singlet & spin-triplet & $^1S_0$ & $^3S_1$ \\
\hline
\hline
731.1(4) & -0.40(8) & -0.48(10) & 0.12(3) & 0.14(3) \\
529.0(4) & -0.51(9) & -0.56(11) & 0.13(3) & 0.14(3) \\
379.7(9) & -0.68(26) & -0.97(37) & 0.15(7) & 0.23(10) \\
\hline
\end{tabular}
\end{center}
\end{table}  

The above discrepancy is partly caused by the heavier pion masses and  the absence of the dynamical quarks in quenched simulations. 
If we get closer to the physical pion mass in full QCD simulations, there should appear the "unitary region" where the $NN$ scattering length shows  the singularity associated with the formation of the di-nucleon bound state, so that $a_0$ changes sign\cite{Kuramashi:1995sc}. Therefore the $NN$ scattering length must become a non-linear function of the pion mass in this region. 
Unlike the scattering length, on the other hand, the $NN$ potential does not necessarily have singular behavior in the unitary region, as demonstrated in the well-known quantum mechanical examples such as the low-energy scattering between ultracold atoms. To check this in QCD, it is of course important to study the $NN$ potential in the full QCD simulations at lighter pion masses.
 
In addition to the above reasoning, there is a possibility that extracted values of $k^2$ have large systematic uncertainties caused by the contamination of the excited states at large distance for the wave functions. 

These $NN$ scattering lengths extracted from the NBS wave function agree in sign  but are much smaller in magnitude than the previous quenched results from the finite size method in smaller volume\cite{Fukugita:1994na}, while they disagree even in sign with the recent full QCD results form the finite size method (See Ref.\cite{Beane:2010em} and references therein.).

\begin{figure}[tb]
\begin{center}
\includegraphics[width=0.32\textwidth,angle=270]{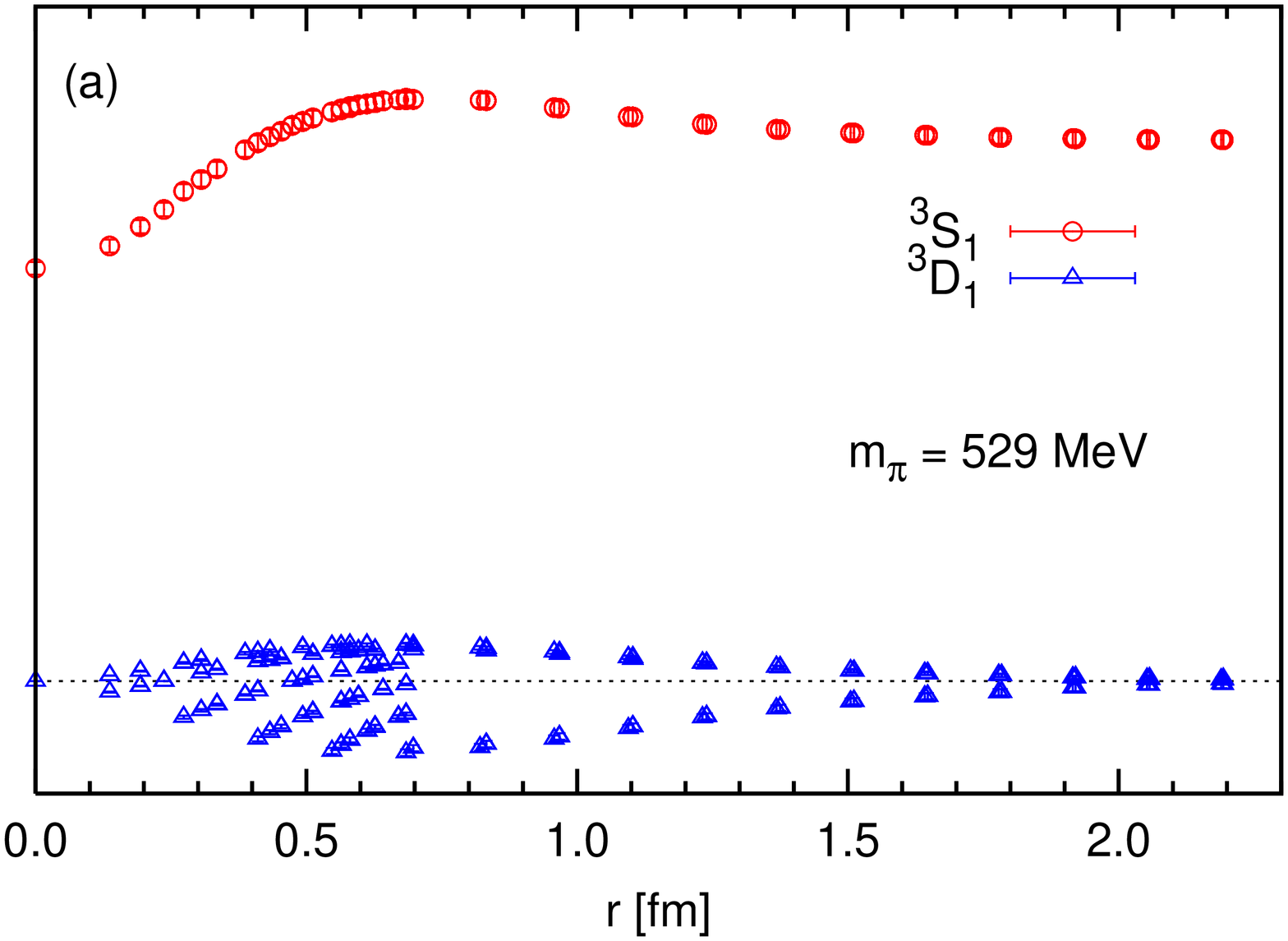}
\includegraphics[width=0.32\textwidth,angle=270]{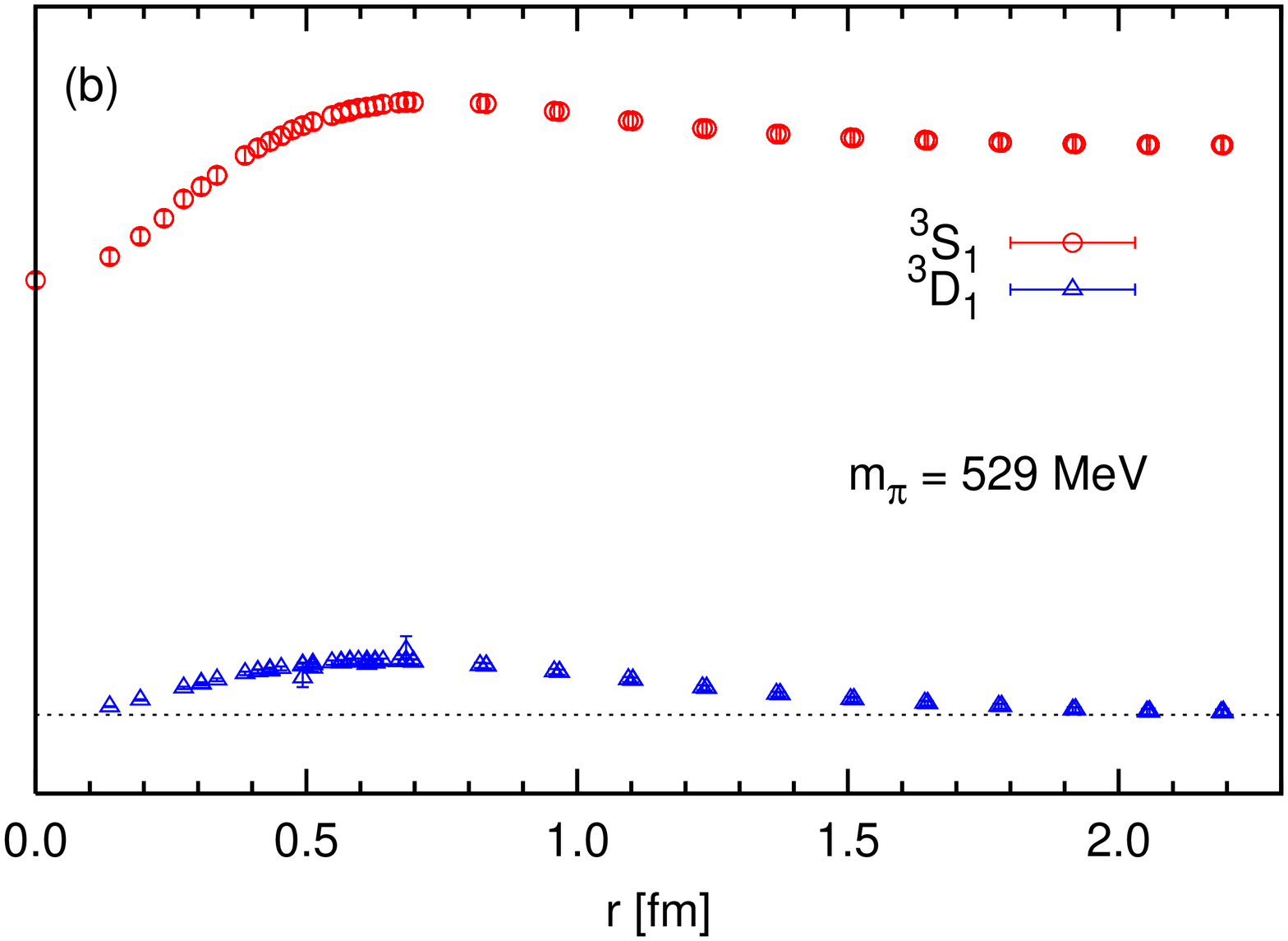}
\caption{(Left) $(\alpha,\beta)=(2,1)$ components of the orbital $A_1^+$ and non-$A_1^+$ wave functions from  $J^P=T_1^+$ (and $J_z=S_z=0$) states at $m_\pi\simeq 529$ MeV. (Right) The same wave functions but the spherical harmonics components are removed from the non-$A_1^+$ part. Taken from Ref.~\protect\cite{Aoki:2009ji}.}
\label{fig:d-wave}
\end{center}
\end{figure}

\subsection{Tensor potential}
In Fig.~\ref{fig:d-wave}(Left), we show the $A_1$ and non-$A_1$ components of the NBS wave function obtained from the $J^P=T_1^+$ (and $J_z=S_z=0$) states at $m_\pi\simeq 529$ MeV, according to eqs. (\ref{eq:projA1}) and   (\ref{eq:projE}). The $A_1$ wave function is multivalued as a function of $r$ due to its angular dependence. For example, the $(\alpha,\beta) = (2,1)$  component of the
$L=2$ part of the non-$A_1$  wave function is proportional to the spherical harmonics $Y_{20}(\theta,\phi) \propto 3\cos^2\theta -1$.  Fig.~\ref{fig:d-wave}(Right) shows the non-$A_1$ component divided by $Y_{20}(\theta,\phi)$. It is clear that the multivaluedness is mostly removed, showing that the non-$A_1$ component is dominated by the $D$ ($L=2$) state.

\begin{figure}[bt]
\begin{center}
\includegraphics[width=0.32\textwidth,angle=270]{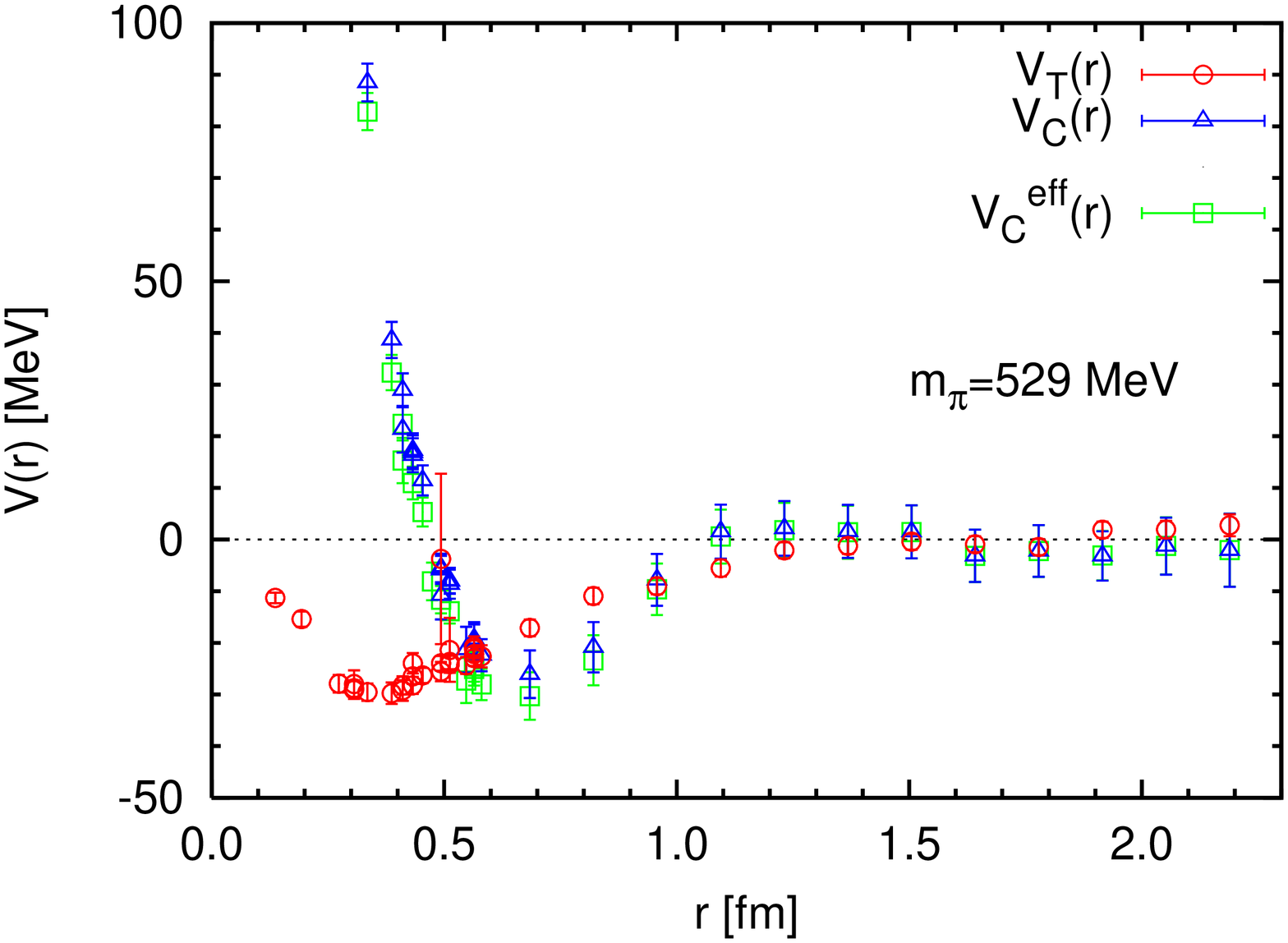}
\includegraphics[width=0.32\textwidth,angle=270]{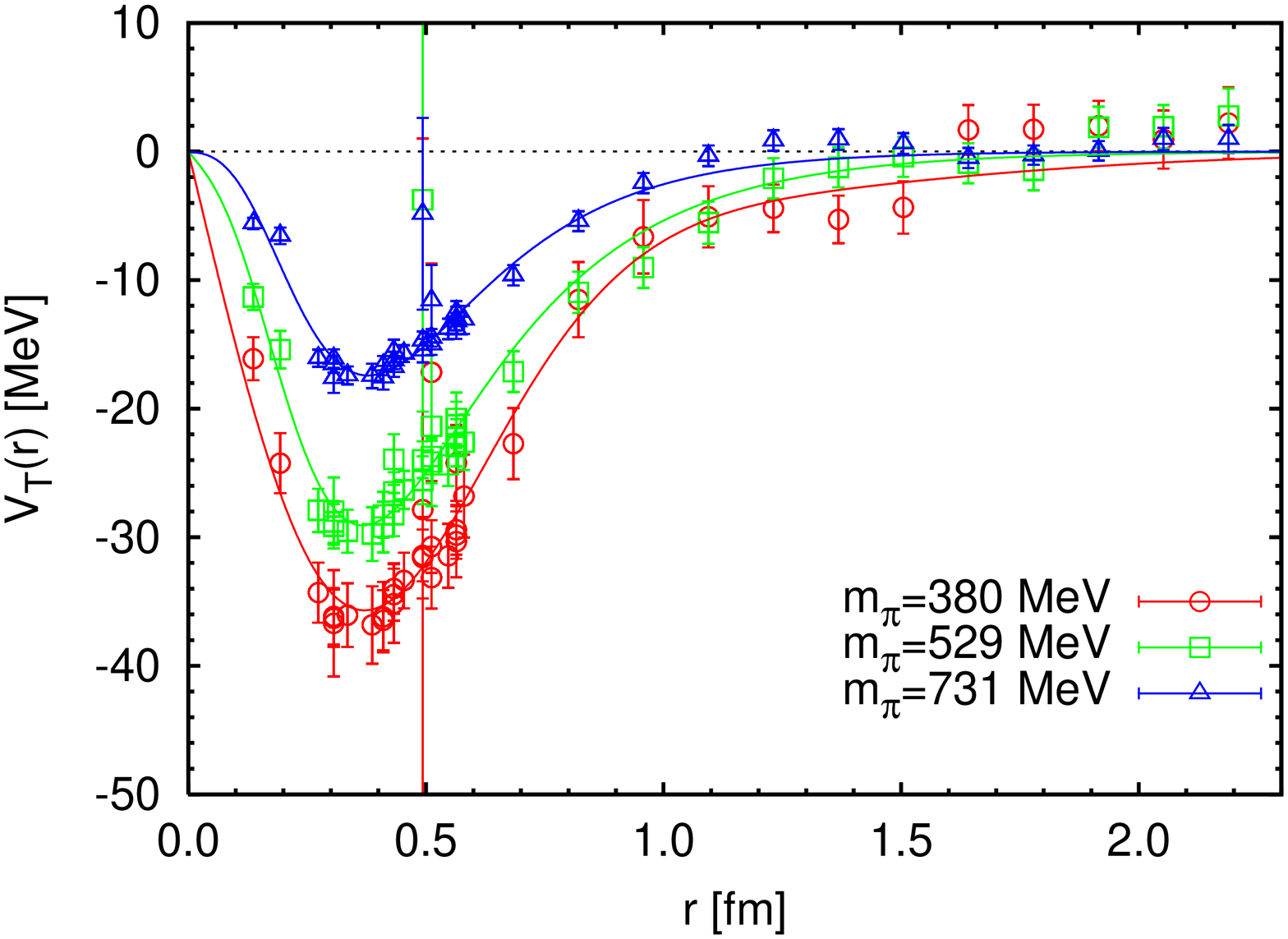}
\caption{(Left) The central potential $V_C(r)^{(1,0)}$ and  the tensor potential $V_T(r)$ obtained from  the $J^P=T_1^+$ NBS wave function, together with the effective central potential $V_C^{\rm eff} (r)^{(1,0)}$, at $m_\pi\simeq 529$ MeV. (Right) Pion mass dependence of the tensor potential. The lines are the four-parameter fit using one-pion-exchange $+$ one-rho-exchange with Gaussian form factor. Taken from Ref.~\protect\cite{Aoki:2009ji}.}
\label{fig:tensor}
\end{center}
\end{figure}
Shown in Fig.~\ref{fig:tensor} (Left) are the central potential $V_C(r)^{(1,0)}$ and tensor potential $V_T(r)$ together with the effective central potential $V_C^{\rm eff}(r)^{(1,0)}$, at the leading order of the velocity expansion as given in eqs. (\ref{eq:central}), (\ref{eq:tensor}) and  (\ref{eq:effective_central}), respectively.

Note that $V_C^{\rm eff}(r)$ contains the effect of $V_T(r)$ implicitly as higher order effects through processes such as ${}^3S_1\rightarrow {}^3D_1\rightarrow {}^3S_1$. At the physical pion mass, 
$V_C^{\rm eff}(r)$ is expected to obtain sufficient attraction from the tensor potential, which causes the appearance of a bound deuteron in the spin-triplet (and flavor-singlet) channel while an absence of the bound dineutron in the spin-singlet (and flavor-triplet) channel. The difference between $V_C(r)^{(1,0)}$ and   $V_C^{\rm eff}(r)$ in Fig.~\ref{fig:tensor} (Left) is still small in this quenched simulation due to relatively large pion mass. This is also consistent with the small scattering length 
in the previous subsection.

The tensor potential in Fig.~\ref{fig:tensor} (Left) is negative for the whole range of $r$ within statistical errors and has a minimum around 0.4 fm.  If the tensor potential receives a significant contribution from one-pion exchange as  expected from the meson theory, $V_T(r)$ is rather sensitive to the change of the pion mass.  As shown in Fig.~\ref{fig:tensor} (Right), it is indeed the case: Attraction of $V_T(r)$ is substantially enhanced as the pion mass decreases. 

The central and tensor potentials obtained from lattice QCD are given at discrete data points. For practical applications to nuclear physics, however, it is more convenient to parameterize the lattice results by known functions.  Such a fit for $V_T(r)$ is given by the form of one-pion-exchange $+$ one-rho-exchange with Gaussian form factors as
\bea
V_T(r) &=& b_1(1-e^{-b_2r^2})^2\left(1+\frac{3}{m_\rho r}+\frac{3}{(m_\rho r)^2}\right)\frac{e^{-m_\rho r}}{r} 
+
b_3(1-e^{-b_4r^2})^2\left(1+\frac{3}{m_\pi r}+\frac{3}{(m_\pi r)^2}\right)\frac{e^{-m_\pi r}}{r} ,
\nonumber\\
\eea
where $b_{1,2,3,4}$ are the fitting parameters while $m_\pi$ ($m_\rho$) is taken to be the pion mass (the rho meson mass) calculated at each pion mass.
The fit line for each pion mass is drawn in Fig.~\ref{fig:tensor} (Right).
It may be worth mentioning that the pion-nucleon coupling constant extracted from the parameter $b_3$ in the case of the lightest pion mass ($m_\pi = 380$ MeV) gives $g_{\pi N}^2/(4\pi) = 12.1
(2.7)$, which is encouragingly close to the empirical value.

\subsection{Convergence of the velocity expansion}
\begin{figure}[bt]
\begin{center}
\includegraphics[width=0.45\textwidth,angle=0]{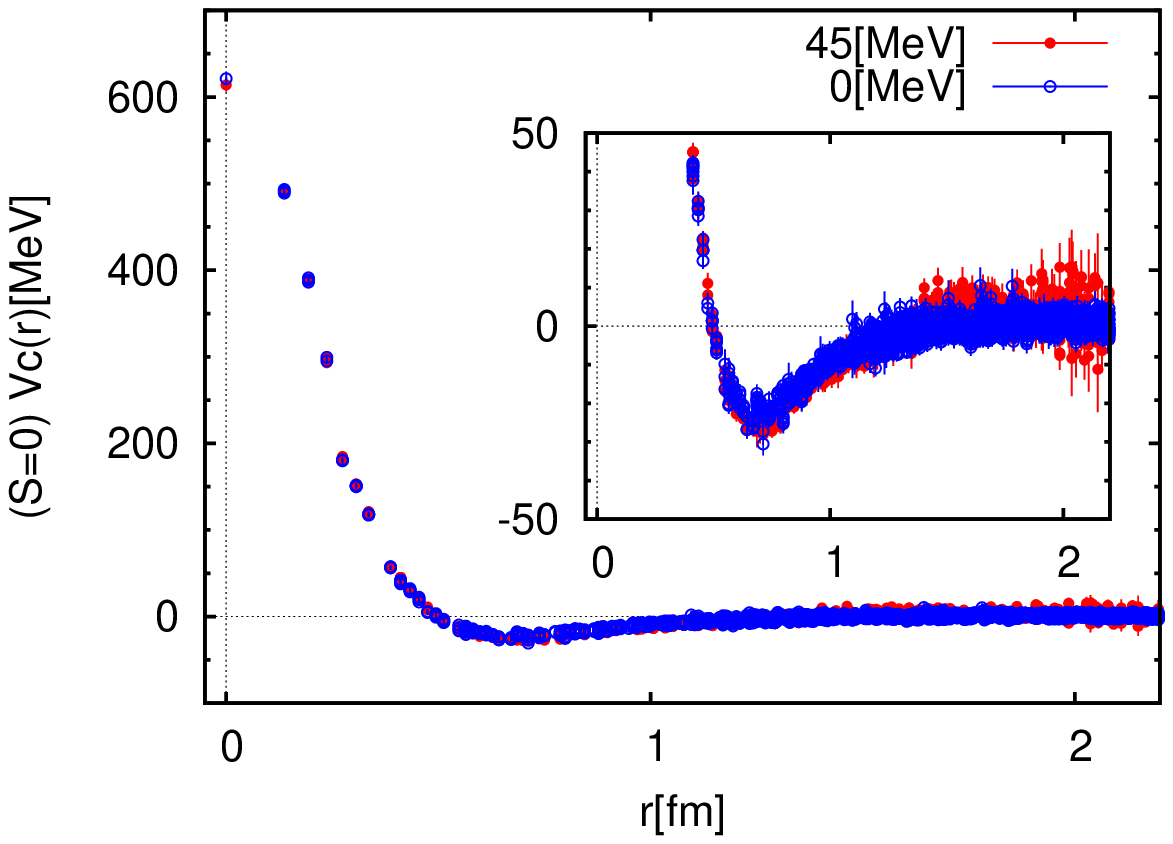}
\includegraphics[width=0.45\textwidth,angle=0]{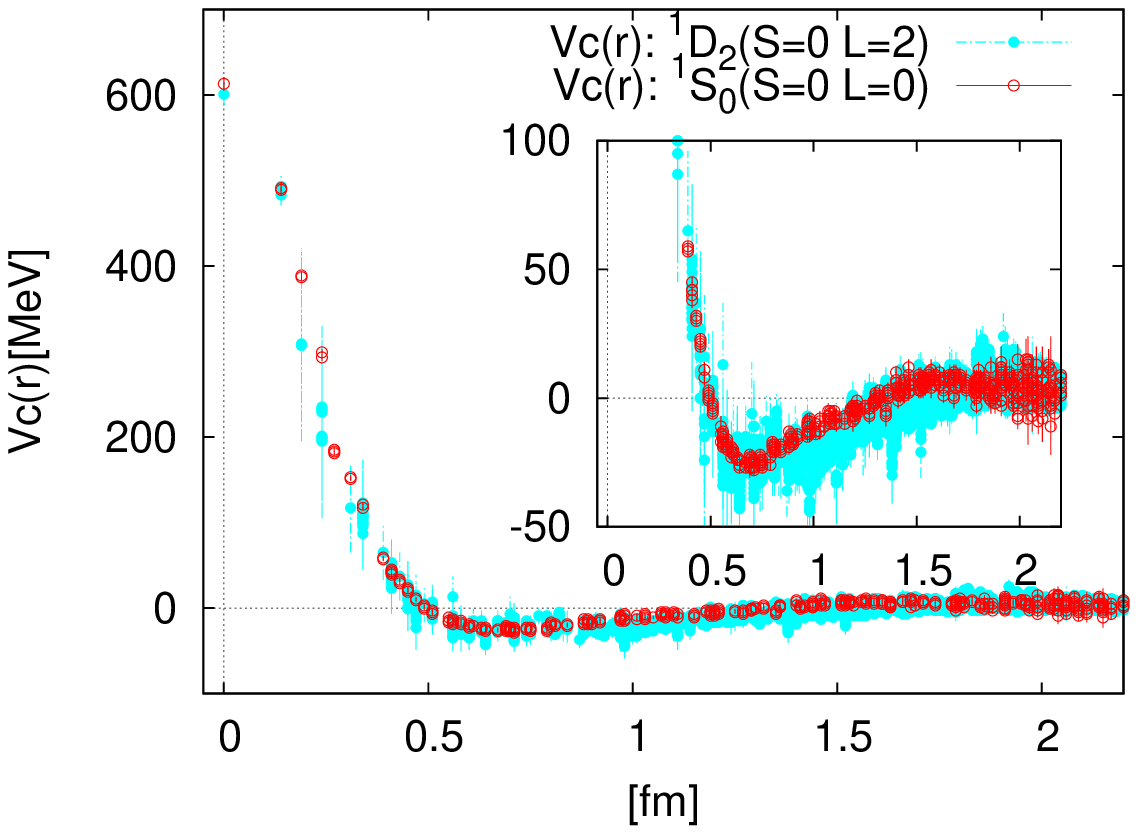}
\caption{(Left) The spin-singlet central potential $V_C(r)^{(0,1)}$  obtained from the orbital $A_1^+$ channel at $E\simeq 45$ MeV (red solid circles) and at $E\simeq 0$ MeV (blue open circles)
in quenched QCD at $m_\pi\simeq 529$ MeV.  (Right) The same potentials at $E\simeq 45$ MeV, obtained from the orbital $A_1^+$ representation (red open circles) and from the $T_2^+$ representation (cray solid circles).  Taken from Ref.~\protect\cite{Murano:2010hh}.}
\label{fig:E-depA}
\end{center}
\end{figure}
The potentials are derived so far at the leading order of the velocity expansion. It is therefore important to investigate the convergence of the  velocity expansion: How good is the leading order approximation ?  How small are higher order contributions ?  If the non-locality of the $NN$ potentials were absent,  the leading order approximation for the potentials would give exact results at all energies. The non-locality of the potentials therefore becomes manifest in the energy dependence of the potentials.

So far the LO potentials are extracted with periodic boundary conditions in the spatial directions for quark fields.  This leads to the lowest values of the effective center of mass energy $E$ almost zero. 
To study the energy dependence, the leading order local  potentials at $E\simeq 45$ MeV, realized by anti-periodic boundary conditions in the spatial directions, are calculated 
 in quenched QCD  at $m_\pi\simeq 529$ MeV and $L\simeq 4.4$ fm\cite{Aoki:2008wy,Murano:2010tc,Murano:2010hh}.
 In this case,  4 types of "momentum-wall" sources, defined by 
 \be
 q^{\rm wall}_f(t_0) \equiv \sum_{\bf x} q({\bf x},t_0) f({\bf x})
 \ee 
 are employed,
 where $f({\bf x}) =\cos( (\pm x \pm y + z)\pi/L)$. Note that $f({\bf x})=1$ corresponds to the wall source used in the periodic boundary condition.  These momentum-wall sources induce  $L = T_2^+$ as well as $L=A_1^+$ states.
\begin{figure}[bt]
\begin{center}
\includegraphics[width=0.45\textwidth,angle=0]{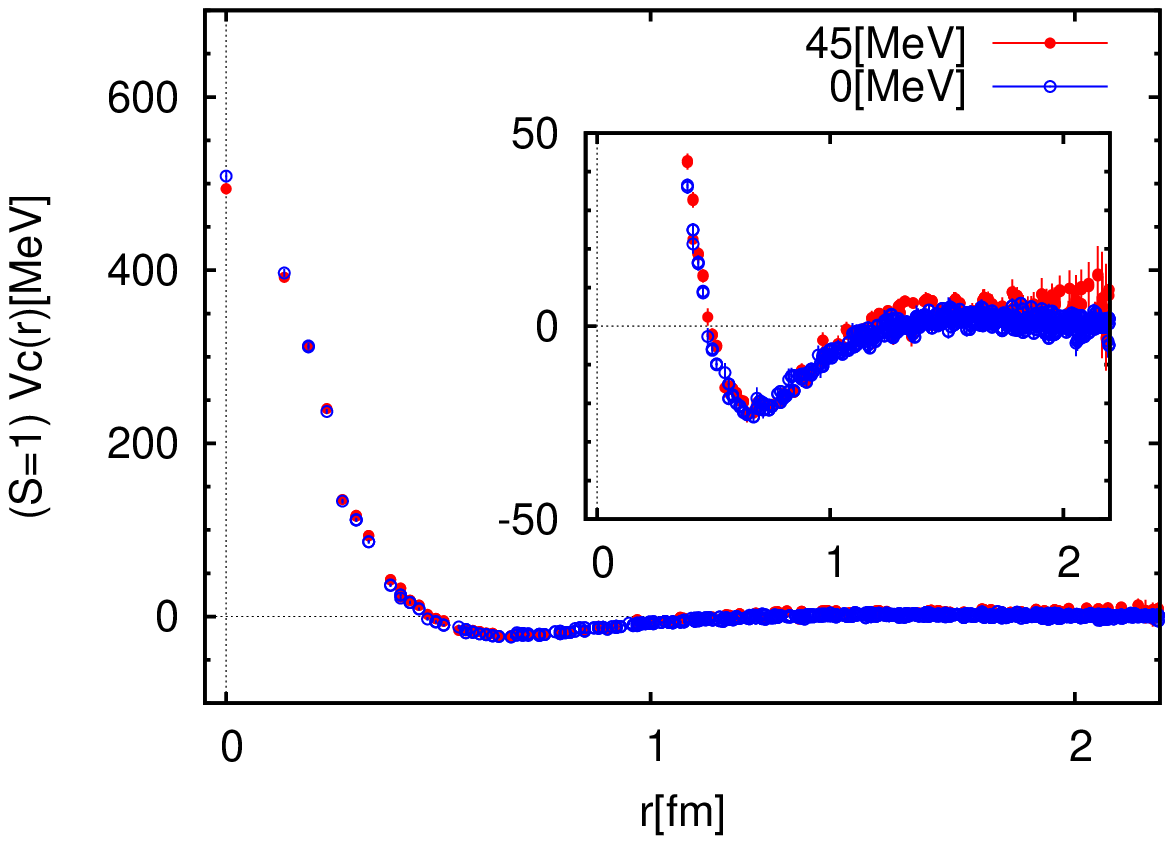}
\includegraphics[width=0.45\textwidth,angle=0]{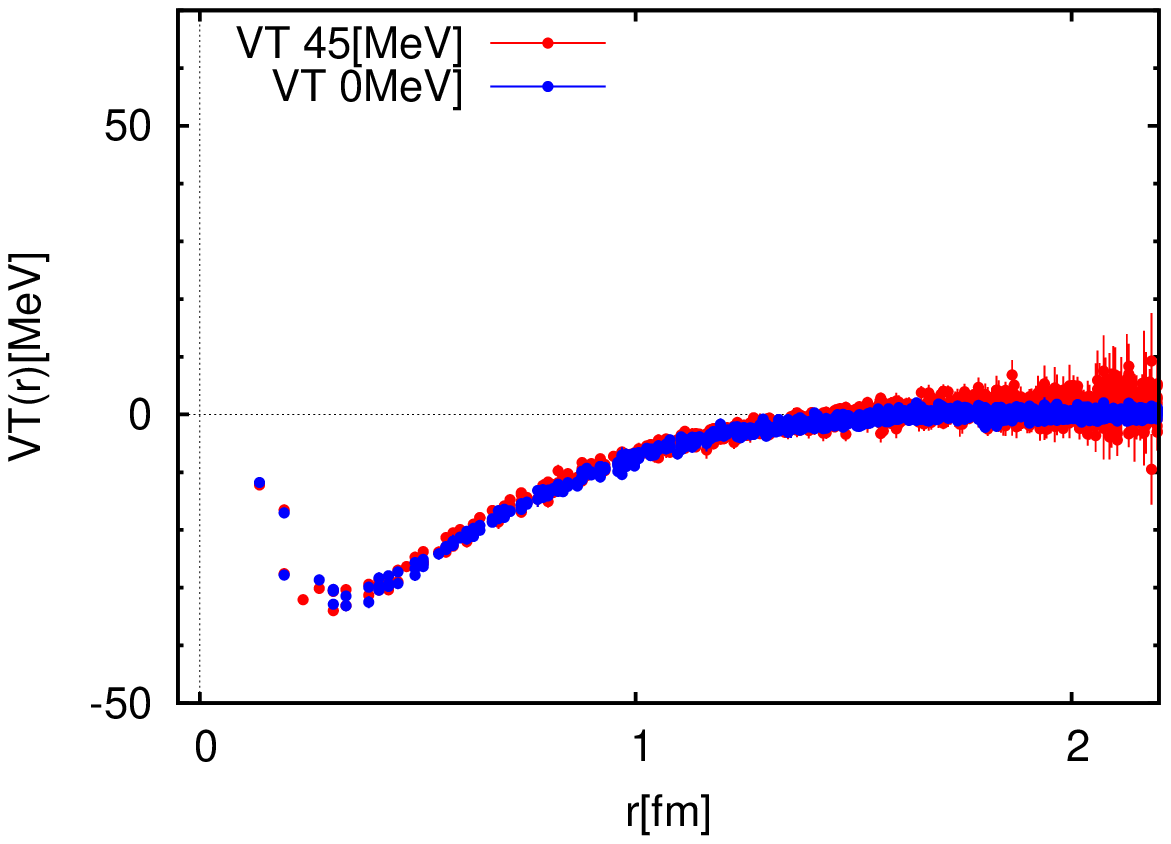}
\caption{(Left) The spin-triplet central potential $V_C(r)^{(1,0)}$ obtained from the orbital $A_1^+-T_2^+$ coupled channel in quenched QCD at $m_\pi\simeq 529$ MeV. (Right) The  tensor potential $V_T(r)$ from the orbital $A_1^+-T_2^+$ coupled channel. For these two figures, symbols are same as in Fig.~\ref{fig:E-depA}(Left). Taken from Ref.~\protect\cite{Murano:2010hh}.
}
\label{fig:E-depB}
\end{center}
\end{figure}
 
In Fig.~\ref{fig:E-depA}(Left),  the spin-singlet potential $V_C(r)^{(S,I)=(0,1)}$ obtained from the $L=A_1^+$ state  at $E\simeq 45$ MeV (red circles) is compared with that at $E\simeq 0$ MeV (blue circles),  while a comparison is made for the spin-triplet potentials in Fig.~\ref{fig:E-depB}, $V_C(r)^{(S,I)=(1,0)}$(left) and $V_T(r)$ (right).
 Good agreements between results at two energies seen in these figures indicate that higher order contributions are rather small in this energy interval.   In other words, these local potentials obtained at $E\simeq 0$ MeV can be safely used to describe the $NN$ scattering phase shift in both spin-singlet and -triplet channels  between $E=0$ MeV and $E=45$ MeV at this pion mass in quenched QCD. 
  
Non-locality of the potential may become manifest also in its angular momentum dependence, since the orbital angular momentum $L={\bf r}\times{\bf p}$ contains one derivative.
In Fig.~\ref{fig:E-depA} (Right), the spin-singlet potential $V_C(r)^{(S,I)=(0,1)}$ obtained  from the $L=T_2^+$ state, whose main component has $L=2$, is compared to the one  from the $L=A_1^+$ state, whose main component has $L=0$.
In this case local potentials are determined at the same energy, $E\simeq 45$ MeV, but different orbital angular momentum.
Although the statistical errors are rather large in the case of  $L=T_2$, a good agreement between the two is again observed, suggesting that the $L$ dependence of the potential is small for the spin-singlet state. 

By these comparisons, it is observed that both energy and orbital angular momentum dependences for local potentials are very weak within statistical errors. We therefore conclude that contributions from higher order terms in the velocity expansion are small and that the LO local potentials in the expansion obtained at $E\simeq 0$ MeV and $L= 0$ are good approximations for the non-local potentials at least up to the   energy $E\simeq 45$ MeV and orbital angular momentum $L=2$. 

Hereafter "potential" in this report means the local potential at the leading order, unless otherwise stated.

\subsection{Full QCD results}
\begin{figure}[bt]
\begin{center}
\includegraphics[width=0.32\textwidth,angle=270]{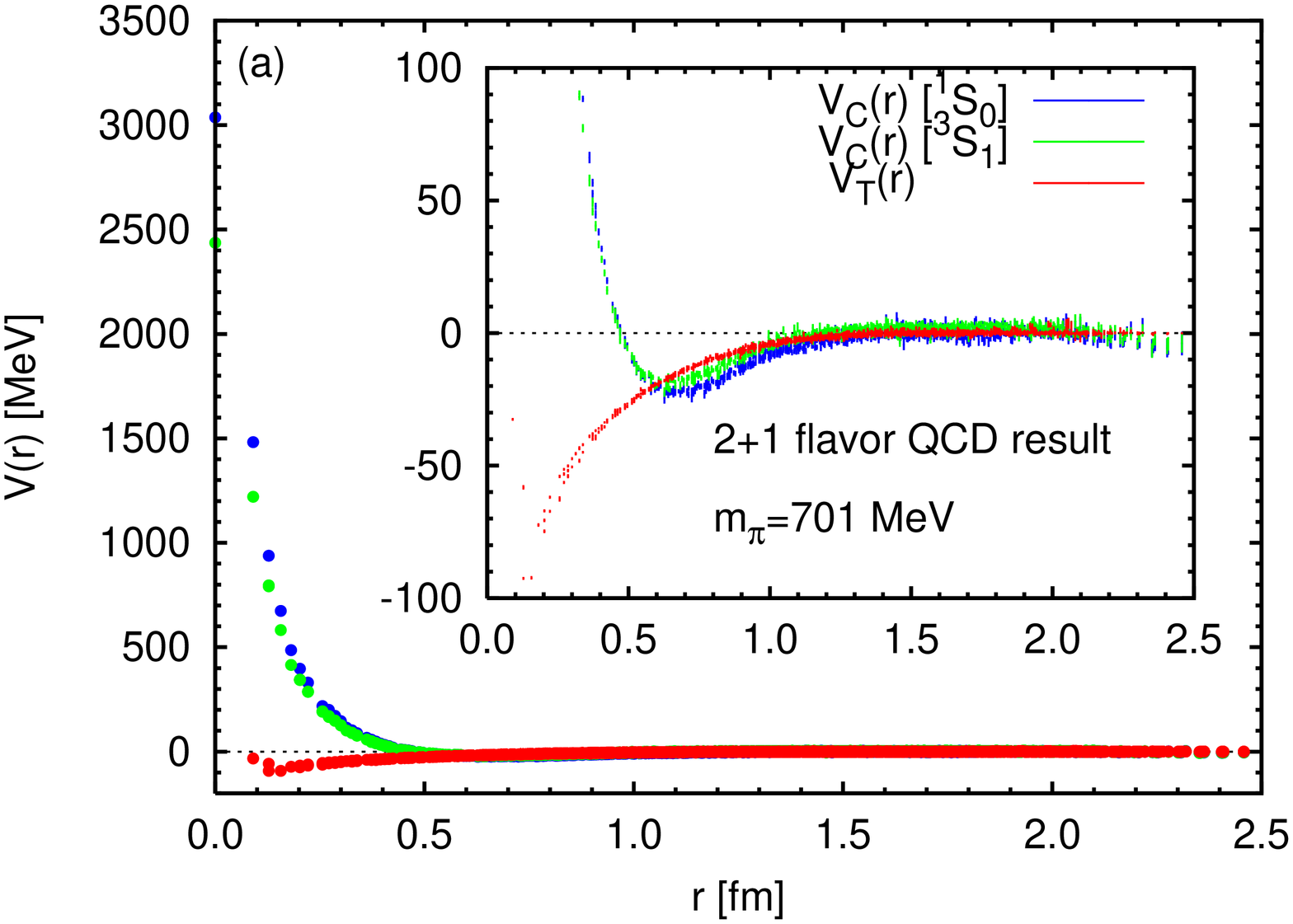}
\includegraphics[width=0.32\textwidth,angle=270]{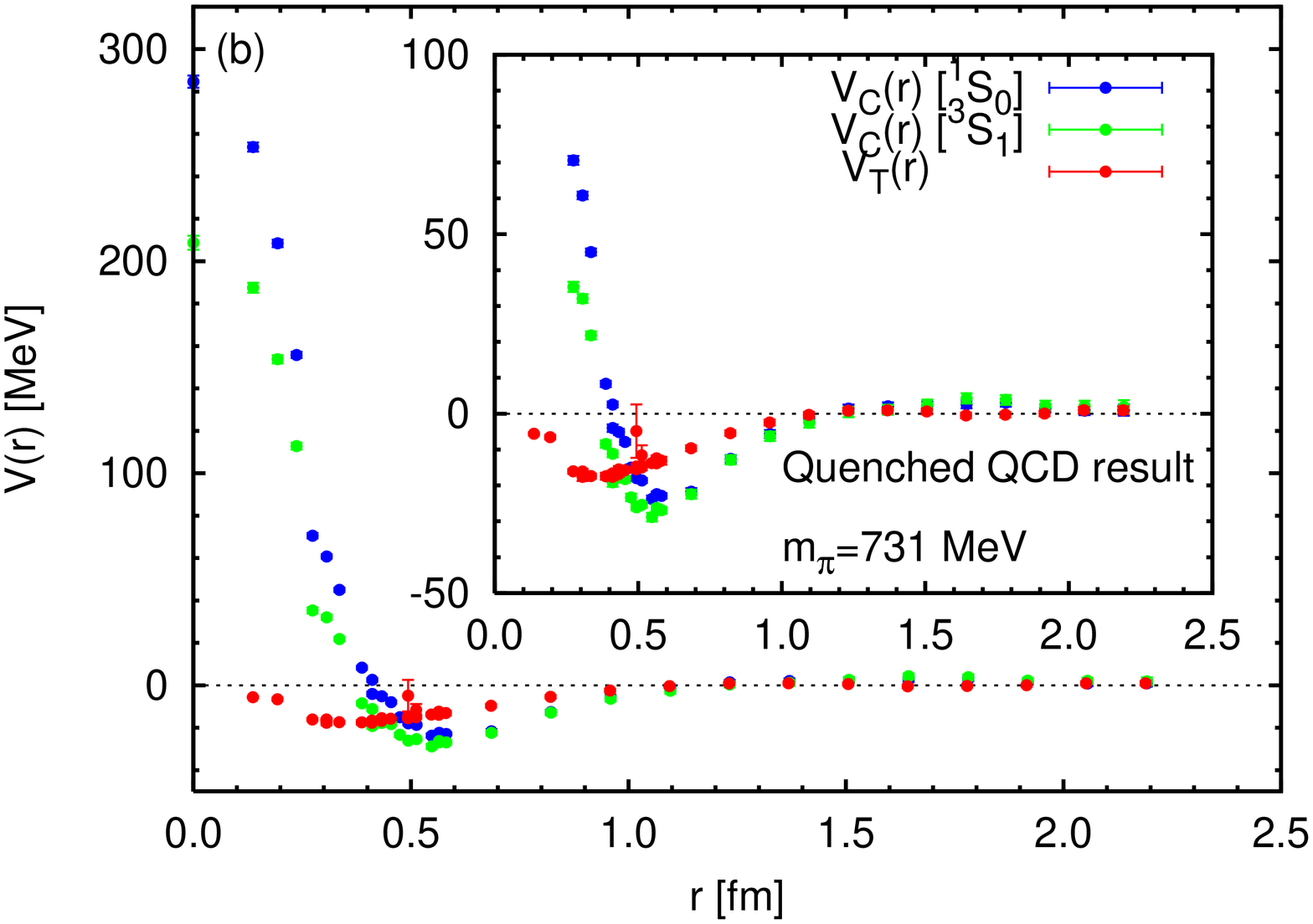}
\caption{(Left) 2+1 flavor QCD results for the central potential and tensor potentials at $m_\pi\simeq 701$ MeV. (Right) Quenched results for the same potentials at $m_\pi\simeq 731$ MeV.
Taken from Ref.~\protect\cite{Ishii:2009zr}.}
\label{fig:full}
\end{center}
\end{figure}
Needless to say, it is important to repeat calculations of $NN$ potentials in full QCD on larger volumes at lighter pion masses. The PACS-CS collaboration is performing $2+1$ flavor QD simulations, which cover the physical pion mass\cite{Aoki:2008sm,Aoki:2009ix}.
Gauge  configurations are generated with the Iwasaki gauge action and non-perturbatively $O(a)$-improved Wilson quark action on a $32^3\times 64$ lattice.
The lattice spacing  $a$ is determined from $m_\pi$, $m_K$ and $m_\Omega$ as $a\simeq 0.091$ fm,  leading to $L\simeq 2.9$ fm.
Three ensembles of gauge configurations are used 
to calculate $NN$ potentials at $(m_\pi, m_N)\simeq $(701 MeV, 1583 MeV), (570 MeV, 1412 MeV) and (411 MeV,1215 MeV )\cite{Ishii:2009zr} .
 
Fig.~\ref{fig:full}(Left) shows the $NN$ local potentials obtained from the PACS-CS configurations at $E\simeq 0$ and $m_\pi=701$ MeV, which is compared with the previous quenched results at comparable pion mass $m_\pi \simeq 731$ MeV but at $a\simeq 0.137$ fm, given in Fig.~\ref{fig:full}(Right). Both the repulsive core at short distance and the tensor potential become significantly enhanced in full QCD. The attraction at medium distances tends to be shifted to the outer region, while its magnitude remains almost unchanged. These differences may be caused by dynamical quark effects. For a definite conclusion on this point, however, a more controlled comparison at the similar lattice spacing is needed.
\begin{figure}[bt]
\begin{center}
\includegraphics[width=0.32\textwidth,angle=270]{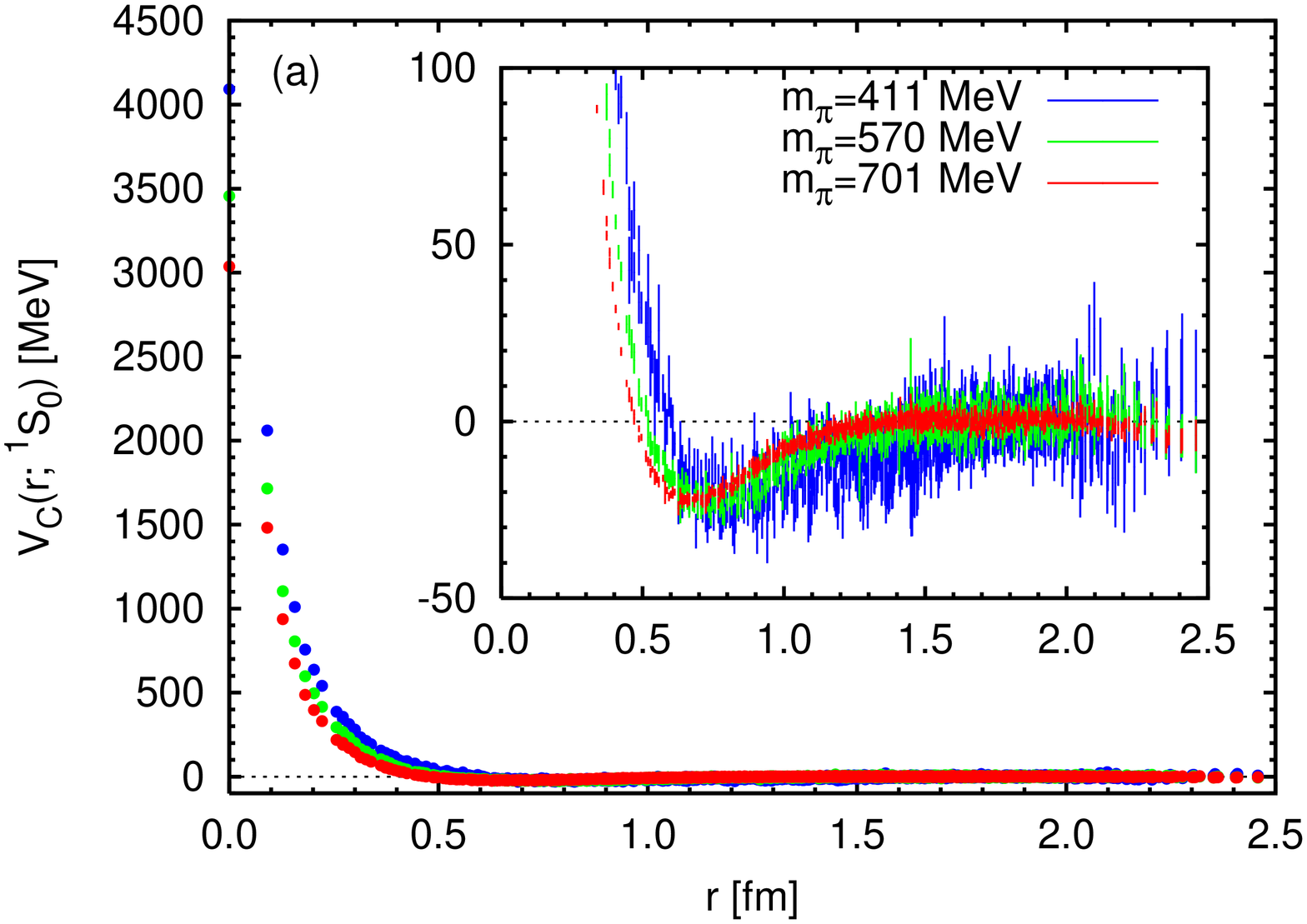}
\includegraphics[width=0.32\textwidth,angle=270]{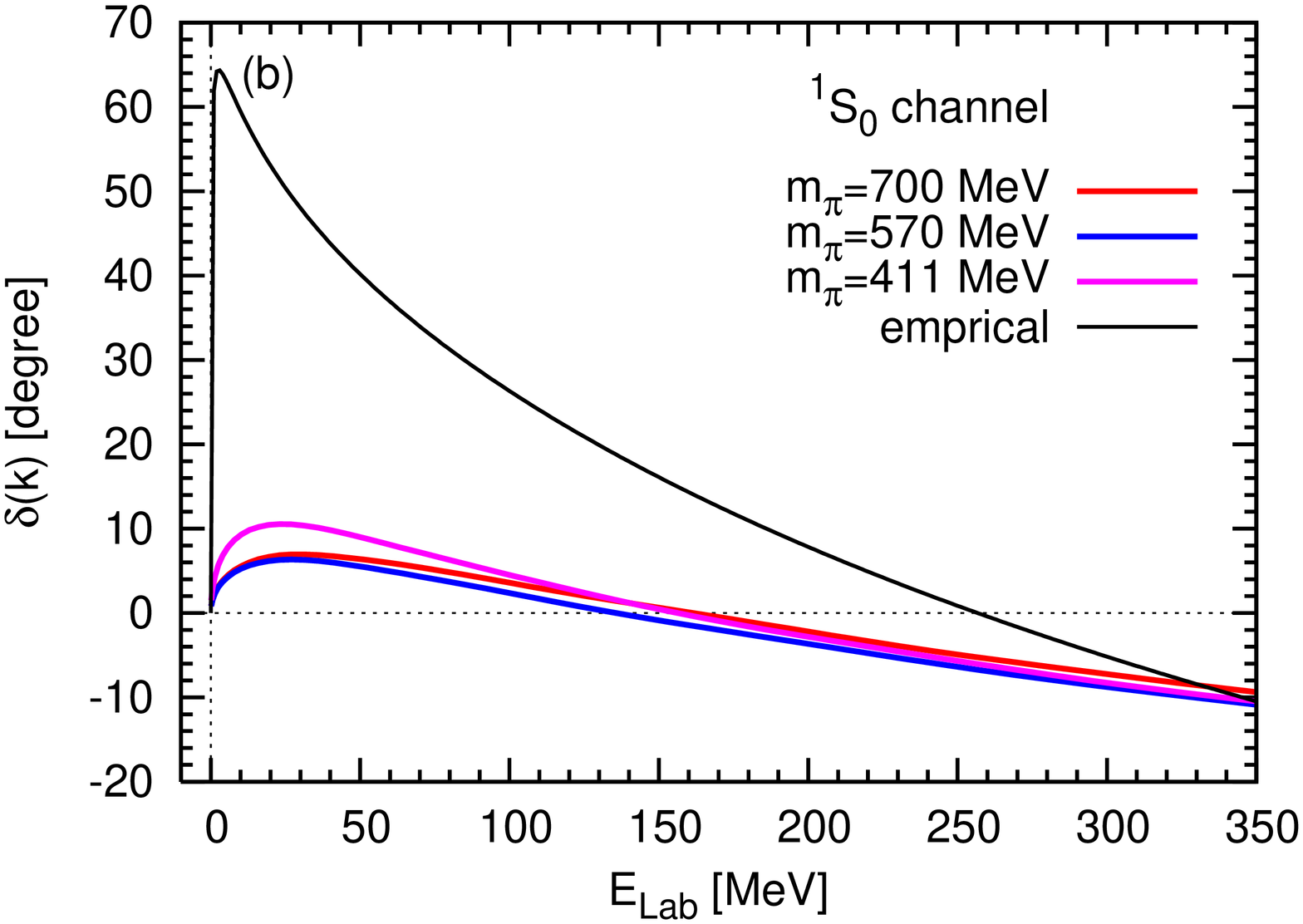}
\caption{(Left) 2+1 flavor QCD results for the spin-singlet central potential  from the orbital $A_1^+$ channel at three values of the pion mass.  (Right) Scattering phase shifts in $^1S_0$ channel from the corresponding lattice potential given in (Left), together with the empirical one.
Taken from Ref.~\protect\cite{Ishii:2009zr}. }
\label{fig:full_massA}
\end{center}
\end{figure}

\begin{figure}[bt]
\begin{center}
\includegraphics[width=0.32\textwidth,angle=270]{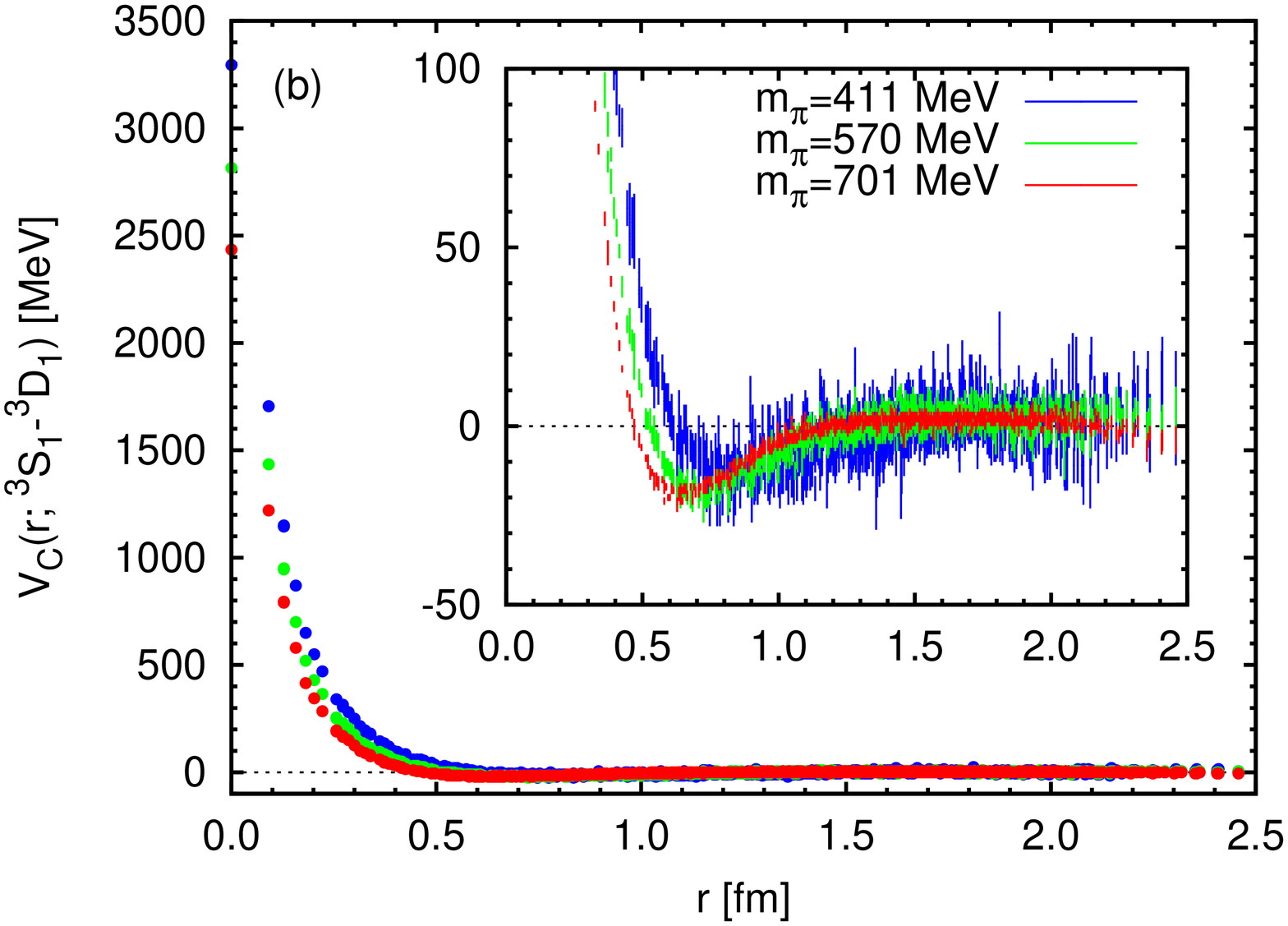}
\includegraphics[width=0.32\textwidth,angle=270]{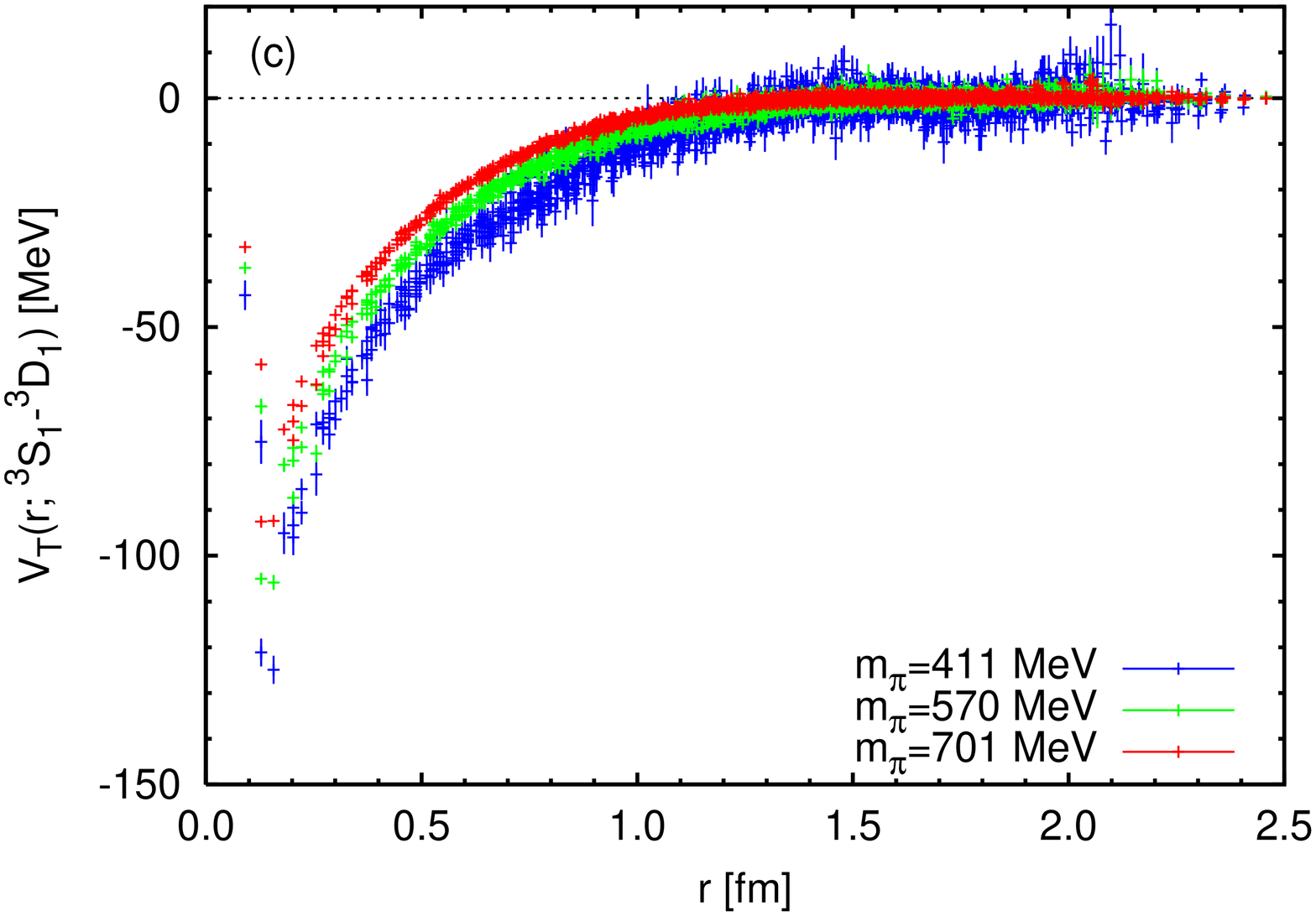}
\caption{(Left) 2+1 flavor QCD results for the spin-triplet central potential $V_C(r)^{(1,0)}$  from the orbital $A_1^+-T_2^+$ coupled channel at three values of the pion mass.  (Right) The tensor potential $V_T(r)$ at three values of the pion mass. Taken from Ref.~\protect\cite{Ishii:2009zr}.}
\label{fig:full_massB}
\end{center}
\end{figure}

In Fig.~\ref{fig:full_massA}(Left), the spin-singlet central potential $V_C(r)^{(0,1)}$ determined from the orbital $A_1$ channel is plotted at three pion masses, while the spin-triplet central potential   $V_C(r)^{(1,0)}$ and the tensor potential $V_T(r)$ from the orbital $A_1^+-T_ 2^+$ couple channel are given in Fig.~\ref{fig:full_massB}.
As in the quenched QCD,
the repulsive cores at short distance, the attractive pocket at medium distance and the strength of the tensor potential are all enhanced as pion mass decreases.

The phase shifts of the  $NN$ scattering for ${}^1S_0$ obtained from the above $V_C(r)^{(0,1)}$ are given in Fig.~\ref{fig:full_massA}(Right).
At low energy, the phase shift increases due to the attraction at medium distance, while at high energy it decreases as a consequence of the repulsive core at short distance.
The shape of the scattering phase shift as a function of energy is qualitatively similar to but   is much smaller than in magnitude the experimental one, plotted by the black solid line in Fig.~\ref{fig:full_massA}(Right).  As discussed before, the pion mass dependence is so large for the scattering phase shift that  full QCD simulations at physical pion mass are needed to reproduce the experimental behavior. 

It is noted here that ground state saturation has to be achieved  to an accuracy of around 1 MeV,  which is about 0.05 \% of the total mass of the two-nucleon system, 
in order to determine the shift of the potential ($E=k^2/m_N$) at the same accuracy. The value of $E$ has a strong influence  on the value of the scattering length.
Such  a high precision is not yet attained in full QCD calculations, since significantly larger $t$ and, accordingly, larger statistics are required. An alternative method to overcome this difficulty will be discussed in Sec.~\ref{sec:extension} .

\section{Hyperon Interactions}
\label{sec:hyperon}
Hyperon($Y$) potentials (hyperon-nucleon and hyperon-hyperon) serve as the starting point in studying
the hyper-nuclei physics. Properties of these potentials can also determine structures in the core of neutron stars. In spite of their importance, only  a limited knowledge for hyperon potentials is available so far, since experimental data such as scattering phase shifts are difficult to obtain, due to the short lifetime of hyperons. Therefore it is important to calculate hyperon potentials in lattice QCD by using the potential method.

\subsection{Quenched result for $N\Xi$ potentials}
\begin{figure}[bt]
\begin{center}
\includegraphics[width=0.45\textwidth,angle=0]{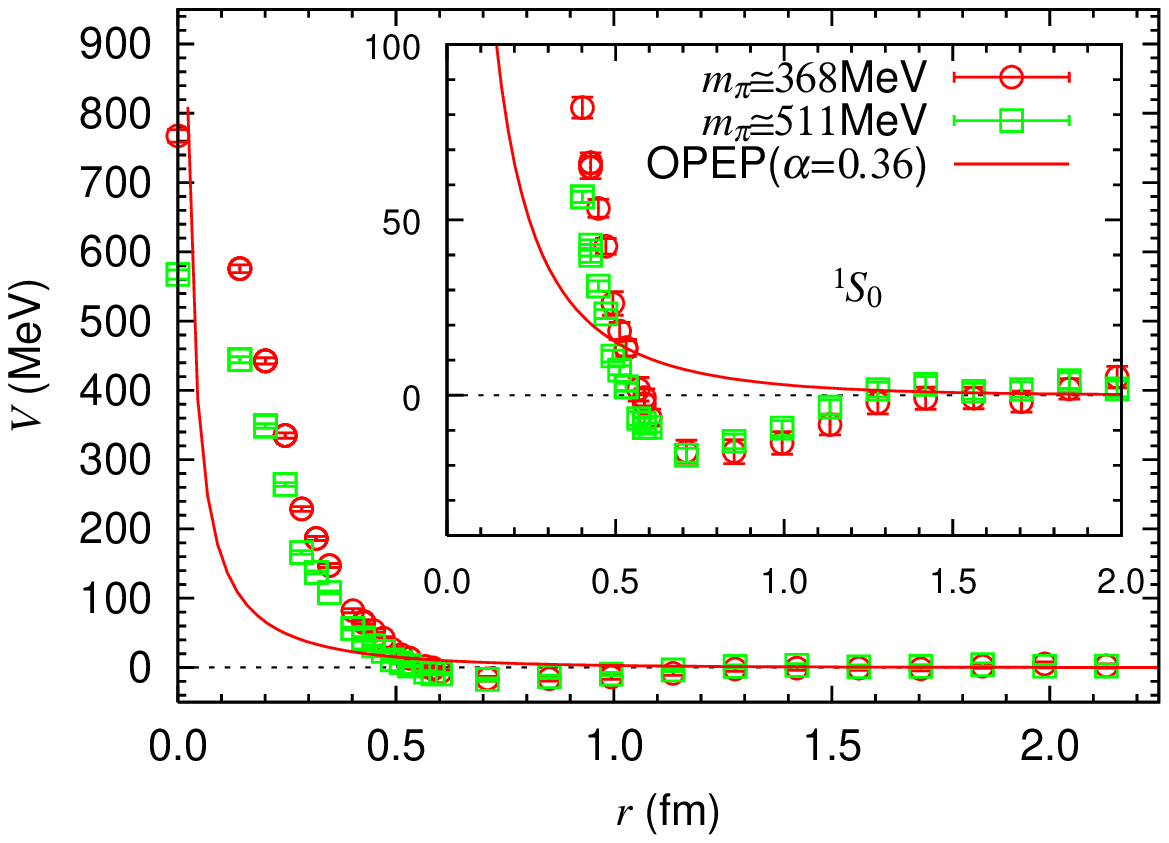}
\includegraphics[width=0.45\textwidth,angle=0]{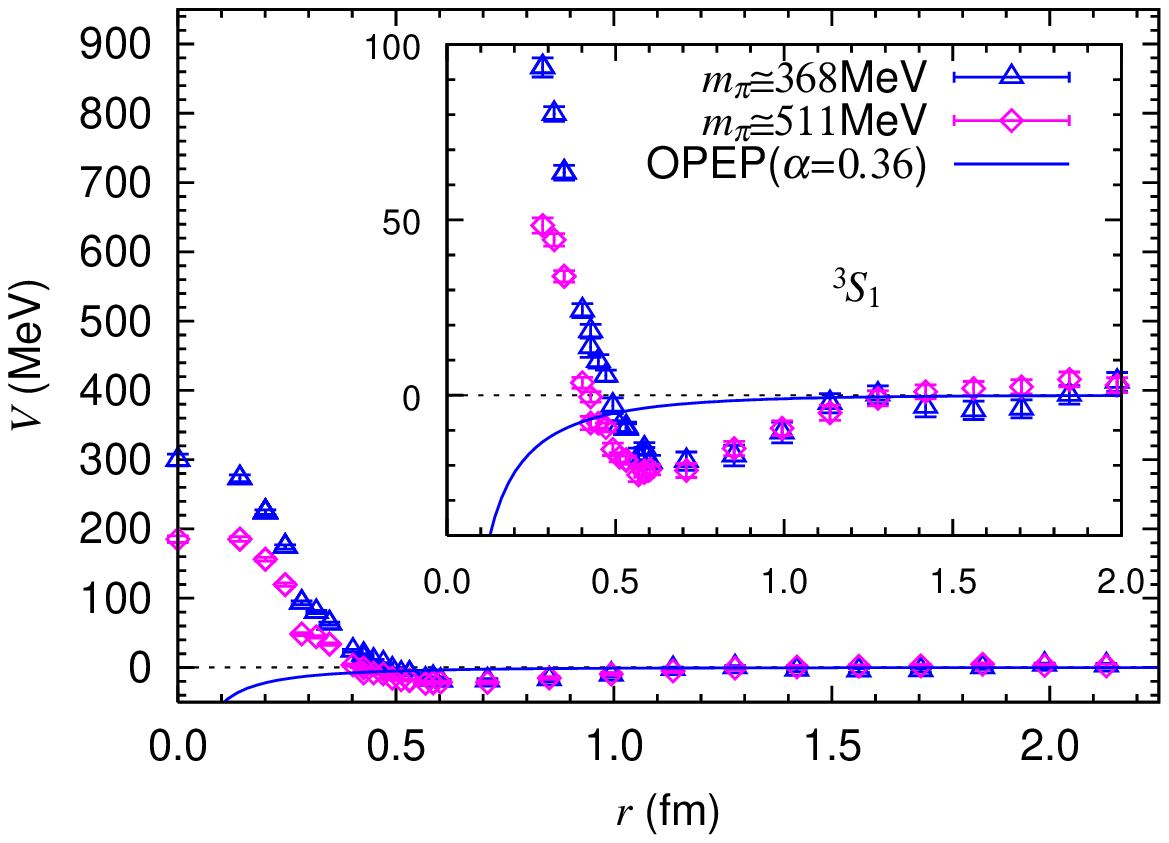}
\caption{ (Left)
 The  spin-singlet central potential for $p\Xi^0$ obtained from the orbital $A_1^+$ channel 
 at $m_\pi\simeq 368$ MeV
 (circle) and $m_\pi\simeq 511$ MeV (box). 
 The central part of the OPEP ($F/(F+D)=0.36)$ in
 Eq.~(\ref{eq:opep}) is also given by solid line. 
 (Right) The spin-triplet effective central potential from the orbital $A_1^+$ channel
 at $m_\pi\simeq 368$ MeV (triangle) and $m_\pi\simeq 511$ MeV (diamond), together with the OPEP (solid line). Taken from Ref.~\cite{Nemura:2008sp}.}
\label{fig:NXi}
\end{center}
\end{figure}
Since all octet-baryons and decuplet $\Omega$ are stable  in the strong interaction,
there are many hyperon potentials in $2+1$ flavor QCD. 
The method for the $NN$ potentials can be straightforwardly applied to the $I=1$ $N\Xi$ channel,
since $p\Xi$ is simply obtained from $ p n$ by replacing a $d$-quark in the neutron with the $s$-quark and it does not have strong decay into other channels.
Unstable channels such as  the $I=0$ $N\Xi$, which  can decay into $\Lambda\Lambda$ via the strong interaction,  will be discussed later. In addition, experimentally, not much information has been available on the $N\Xi$ interaction except for a few studies: a recent report gives the upper limit of elastic and inelastic cross sections\cite{XN_Ahn} while earlier publications suggest weak attractions of $\Xi-$ nuclear interactions\cite{Nakazawa,Fukuda,Khaustov}. The $\Xi-$nucleus interactions will be soon studied as one of the day-one experiments at J-PARC\cite{JPARC} via $(K^-,K^+)$ reaction with a nuclear target.

Ref.~\cite{Nemura:2008sp} gives the first quenched result for $I=1$ $N\Xi$ potentials.
Lattice parameters are the same as for the quenched $NN$ potential. In addition to two values of the light quark mass,  the quenched strange quark is introduced and is  fixed to one value.
The potential is calculated at $(m_\pi, m_N, m_\Xi) = (511(1){\rm MeV}, 1300(4){\rm MeV}, 1419(4){\rm MeV})$  and   $(368(1){\rm MeV}, 1167(7){\rm MeV}, 1383(6){\rm MeV})$, using the NBS wave function with the interpolation operators defined by
\be
p_\alpha (x) =\varepsilon_{abc} (u^a(x)C\gamma_5 d^b(x)) u_\alpha^c(x), \quad
\Xi_\alpha^0(x) = \varepsilon_{abc} (u^a(x)C\gamma_5 s^b(x)) s_\alpha^c(x) .
\ee 
Since both $p$ and $\Xi^0$ have $(I, I_z)=1/2,1/2$, the $p\Xi^0$ system has $I=1$ with the strangeness $S=-2$.

The left (right) of Fig.~\ref{fig:NXi} gives the (effective) central potential of the $p\Xi^0$ system obtained from the $L=A_1$ representation for the spin-singlet (triplet) at $m_\pi = 511$ MeV and 368 MeV. Potentials in the $I=1$ $N\Xi$ system for both channels show a repulsive core at $r\le 0.5$ fm surrounded by an attractive well, similar to the $NN$ systems. In contrast to the $NN$ case, however, the repulsive core of the $p\Xi^0$ potential in the spin-singlet channel is substantially stronger than in the triplet channel. The attraction in the medium to long distance region( 0.6 fm $ \le r \le 1.2$ fm ) is similar  in both channels. 
The height of the repulsive core increases as the light quark mass decreases, while a significant difference is not seen for the attraction in the medium to long distance within statistical errors.
Potentials in Fig.~\ref{fig:NXi} are weakly attractive 
on the whole in both spin channels at both pion masses, in spite of the repulsive core at short distance, though the attraction in the triplet is a little stronger than that in the singlet.

The solid lines in Fig.~\ref{fig:NXi} are the one-pion exchange potential (OPEP), given by
\be
V_C^\pi = -(1-2\alpha)\frac{g_{\pi NN}^2}{4\pi} \frac{(\vec\tau_N\cdot\vec\tau_\Xi)(\vec\sigma_N\cdot\vec\sigma_\Xi)}{3}\left(\frac{m_\pi}{2m_N}\right)^2 \frac{e^{-m_\pi r}}{r}
\label{eq:opep}
\ee
with $(m_\pi, m_N) =(368 {\rm MeV}, 1167 {\rm MeV})$,
where the pseudo-vector $\pi\Xi\Xi$ coupling $f_{\pi\Xi\Xi}$ is related to the  $\pi NN$ coupling as $f_{\pi\Xi\Xi}= -f_{\pi NN} (1-2\alpha)$ with the parameter $\alpha = F/(F+D)$, and $g_{\pi NN} = f_{\pi NN}\frac{m_\pi}{2m_N}$. The empirical vales, $\alpha \simeq 0.36$ and $g_{\pi NN}/(4\pi) \simeq 14.0$, are used for the plot.
Unlike the $NN$ potential, the OPEP in the present case has  opposite sign between the spin-singlet channel and spin-triplet channel. In addition, the absolute magnitude is smaller due to the factor $1-2\alpha$. No clear signature of the OPEP at long distance ($r\ge 1.2$ fm) is observed in Fig.~\ref{fig:NXi} within statistical errors. Furthermore, there is clear departure from the OPEP at medium distance (0.6 fm $\le r \le 1.2$ fm) in both channels. These observations may suggest an existence of state-independent attraction.

\subsection{Full and quenched QCD results for $N \Lambda$ potentials}
\begin{figure}[bt]
\begin{center}
\includegraphics[width=0.45\textwidth,angle=0]{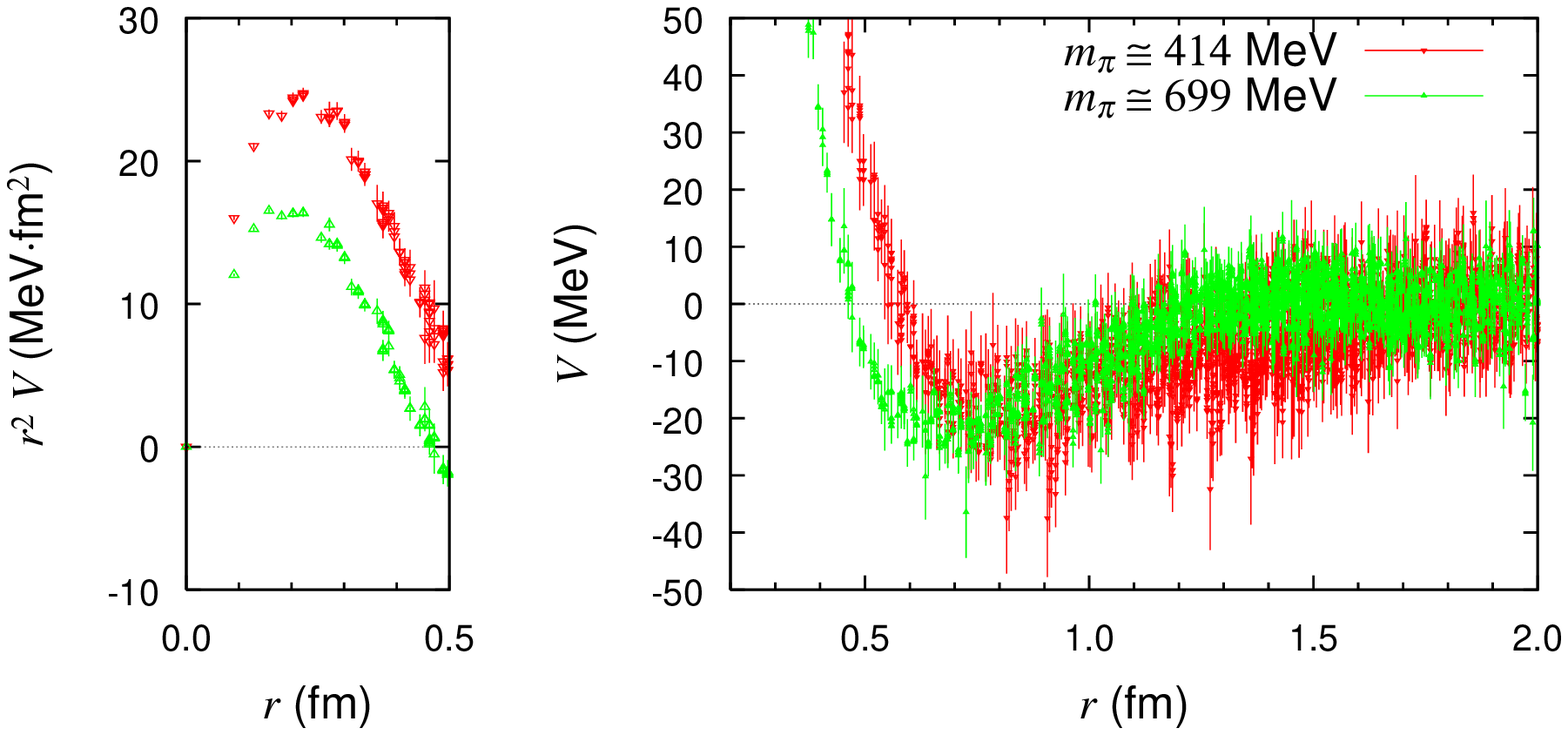}
\includegraphics[width=0.45\textwidth,angle=0]{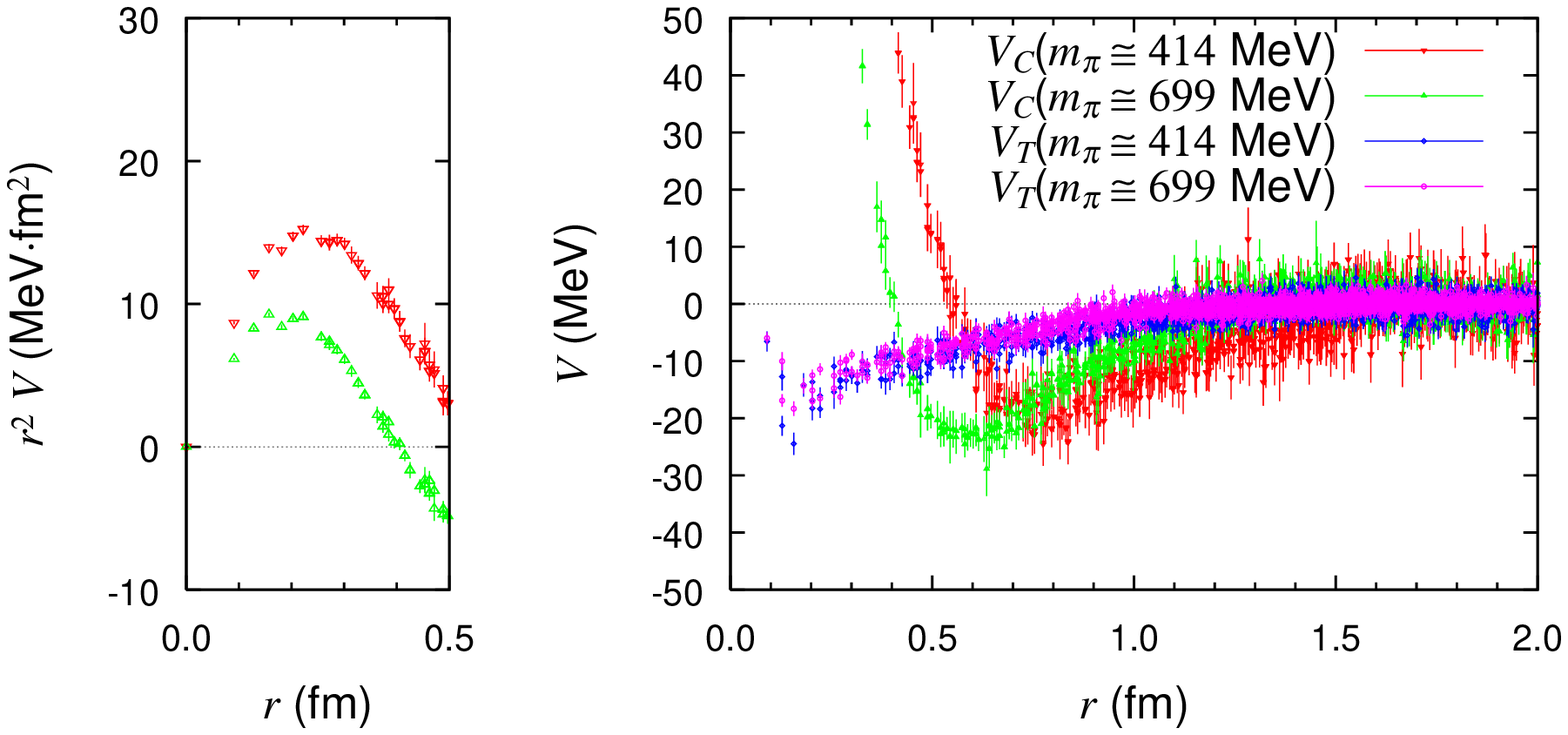}
\caption{ (Left)
 The  spin-singlet central potential for $N\Lambda$ obtained from the orbital $A_1^+$ channel 
 in 2+1 flavor QCD at $m_\pi\simeq 414$ MeV (red) and 699 MeV (green).
 (Right) The spin-triplet central potential and the tensor potential for $N\Lambda$ obtained from the orbital $A_1^+-T_2^+$ coupled channel in 2+1 flavor QCD 
 at $m_\pi\simeq 414$ MeV (red and blue) and 699 MeV (green and magenta). Taken from Ref.~\protect\cite{Nemura:2009kc}. }
\label{fig:NL_full}
\end{center}
\end{figure}

Spectroscopic studies of the $\Lambda$ and $\Sigma$ hypernuclei, carried out both experimentally and theoretically, suggest that the $\Lambda$-nucleus interaction is attractive while the $\Sigma$-nucleus interaction is repulsive. If this is the case, the $\Lambda$ particle would be the first strange baryon instead of $\Sigma^-$ to appear in the core of the neutron stars\cite{SchaffnerBielich:2008kb}.
It is therefore interesting and important to investigate the nature of the $N\Lambda$ interaction in lattice QCD, by calculating the $N\Lambda$ potential using the method of this report. 
Since the $\Lambda$ is the lightest hyperon, the $N\Lambda$ potential can be calculated as in the case of $NN$ potentials.

In Ref.~\cite{Nemura:2009kc}, the $N\Lambda$ potentials are calculated in both full and quenched QCD. The 2+1 flavor full QCD gauge configurations generated by the PACS-CS collaboration are employed for the calculations of the potentials on a $32^3\times 64$ lattice at $a= 0.091(1)$ fm, while in the quenched calculation, the potentials are obtained on a $32^3\times 48$ lattice at $a=0142(1)$ fm. Numerical values for some hadron masses for these calculations are given in Table~\ref{tab:NL_mass}, together with some lattice parameters.

Fig.~\ref{fig:NL_full}  shows the $N\Lambda$ potentials obtained from 2+1 flavor QCD calculations as a function of $r$ at $m_\pi\simeq 699$ MeV and 414 MeV. The spin-singlet central potential obtained from  the $J=A_1$ channel is plotted on the left, while the spin-triplet central potential and the tensor potential obtained from the  $J=T_1^+$ channel are given on the right. The central potential multiplied by a volume factor ($r^2 V_C(r)$) is also shown in the left panel in addition to the $V_C(r)$ itself in the right panel, in order to compare the strength of the repulsion between two quark masses.  

As can be seen in Fig.~\ref{fig:NL_full}, the attractive well of the central potentials moves to the outer region as the light quark mass decreases, while the depth of these attractive pockets does not change so much. The present results show that the tensor force is weaker than the $NN$ case in Fig.\ref{fig:full}. Moreover the quark mass dependence of the tensor force seems small. 
 Both repulsive and attractive parts of the central potentials increase in magnitude as the light quark mass decreases.

\begin{figure}[bt]
\begin{center}
\includegraphics[width=0.45\textwidth,angle=0]{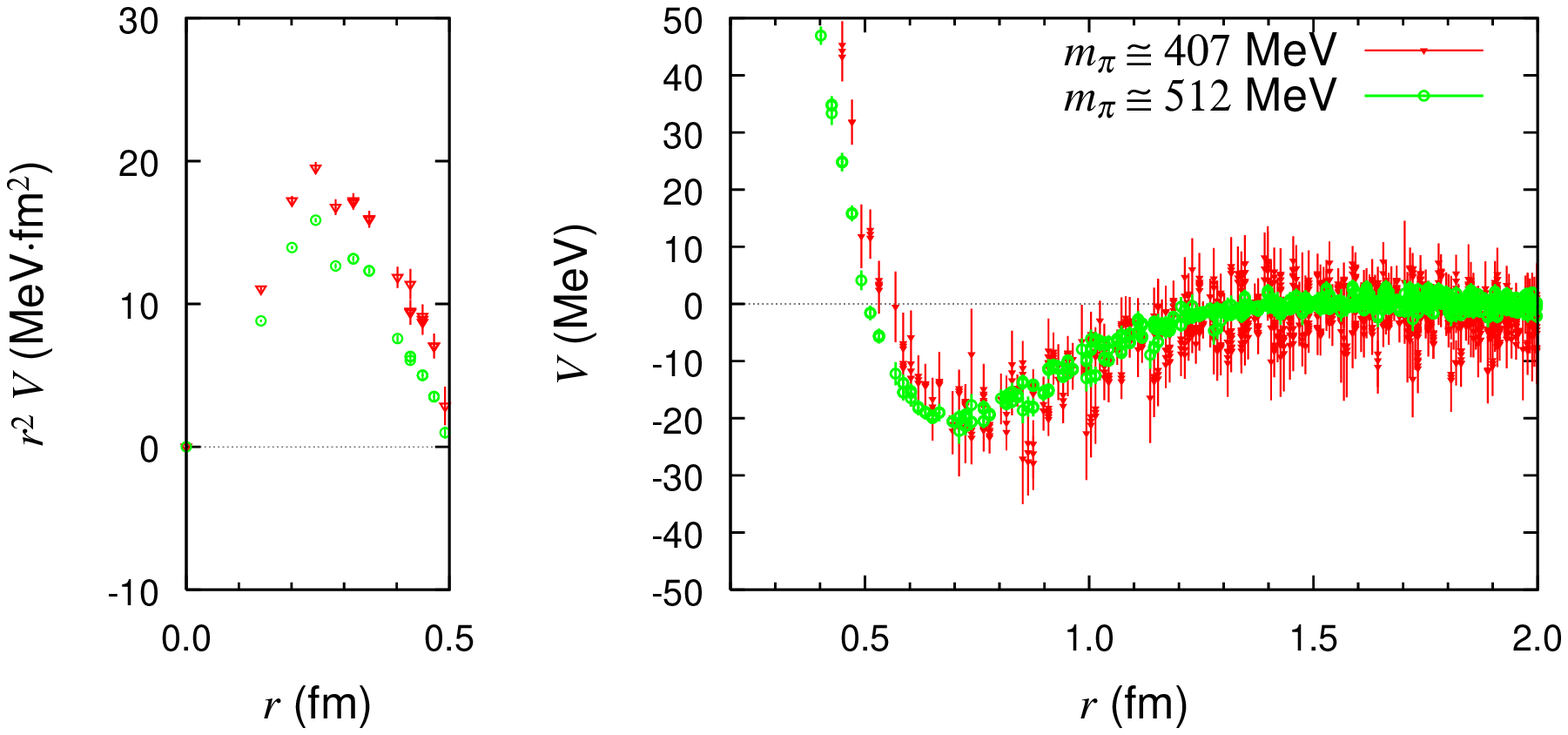}
\includegraphics[width=0.45\textwidth,angle=0]{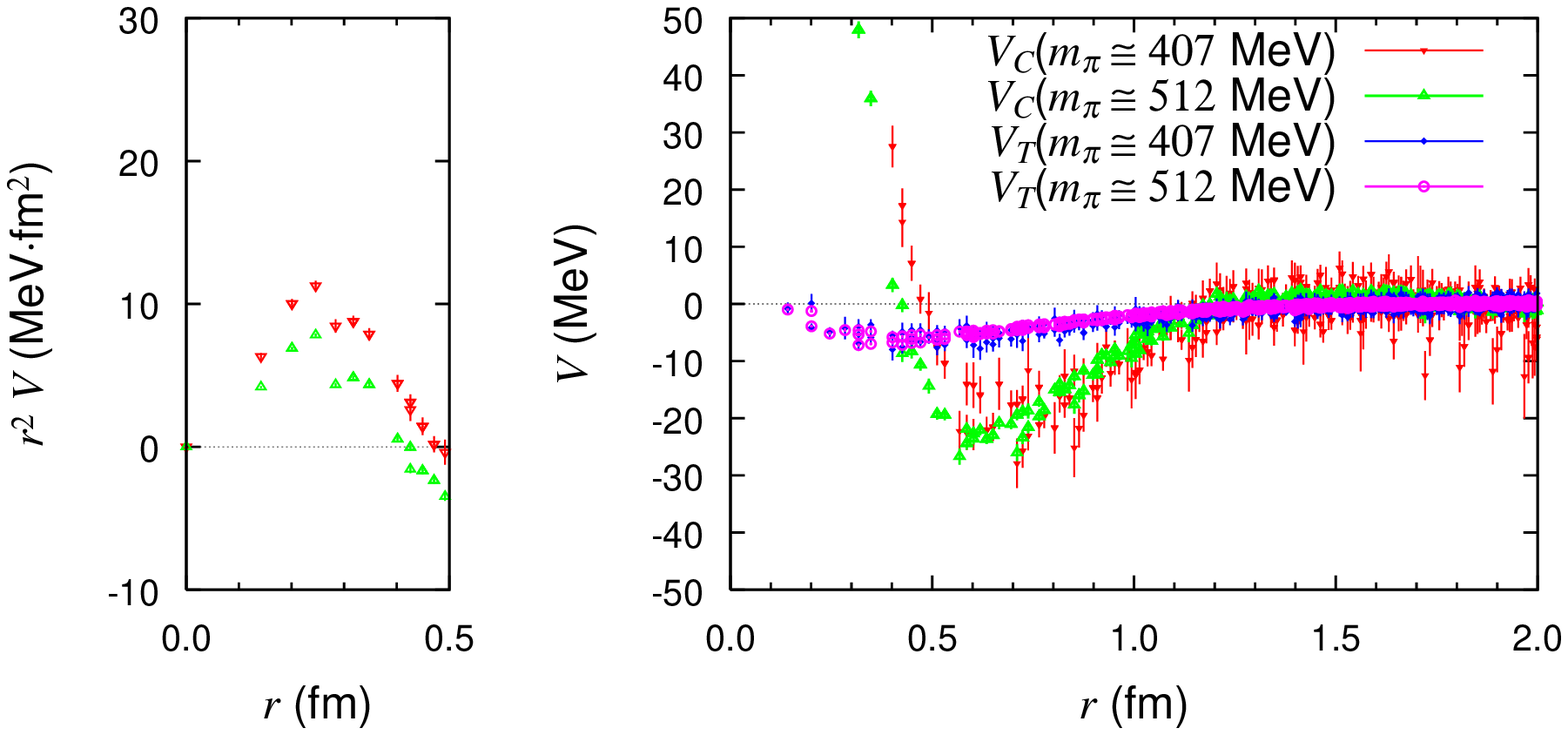}
\caption{ (Left)
 The  spin-singlet central potential for $N\Lambda$ obtained from the orbital $A_1^+$ channel 
 in quenched QCD at $m_\pi\simeq 407$ MeV (red) and 512 MeV (green).
 (Right) The spin-triplet central potential and the tensor potential for $N\Lambda$ obtained from the orbital $A_1^+-T_2^+$ coupled channel in quenched  QCD 
 at $m_\pi\simeq 407$ MeV (red and blue) and 512 MeV (green and magenta). 
 Taken from Ref.~\protect\cite{Nemura:2009kc}. }
\label{fig:NL_quench}
\end{center}
\end{figure} 
Fig.~\ref{fig:NL_quench} shows the $N\Lambda$ potentials in quenched QCD calculations at $m_\pi\simeq 512$ MeV and 407 MeV. The central potential in the spin-singlet channel from  $J=A_1$ is on the left, while the central and the tensor potential in the triplet channel from $J=T_1$ are  on the right. Qualitative features of these potentials are more or less similar to those in full QCD: the attractive pocket of the central potentials moves to the longer distance region as the quark mass decreases while the quark mass dependence of the tensor potential is small.   

\begin{table}[tb]
 \centering \leavevmode 
 \begin{tabular}{|cccccccc|}
  \hline \hline
  $m_\pi$ & $m_\rho$ & $m_K$ & $m_{K^\ast}$ & $m_N$ & $m_\Lambda$ & $m_{\Sigma}$ & $m_{\Xi}$ \\
  \hline
  \multicolumn{8}{|l|}{\underline{2+1 flavor QCD on a $32^3\times 64$ lattice at $a= 0.091(1)$ fm
  }} \\
  699.4(4) & 
  1108(3) &
  786.8(4) &
  1159(2) &
  1572(6) &
  1632(4) &
  1650(5) &
  1701(4) \\
  567.9(6) & 
  1000(4) &
  723.7(7) &
  1081(3) &
  1396(6) &
  1491(4) &
  1519(5) &
  1599(4) \\
  413.6(6) & 
  902(3) &
  636.6(4) &
  1026(3) &
  1221(7) &
  1349(4) &
  1406(8) &
  1505(4) \\
  301(3) & 
  845(10) &
  592(1) &
  980(6) &
  1079(12) &
  1248(15) &
  1308(13) &
  1432(7) \\
  \hline
  \multicolumn{8}{|l|}{ \underline{quenched QCD on a $32^3\times 48$ lattice at $a=0.142(1)$ fm
  }} \\
  511.8(5)  & 862(3)  & 605.8(5)  & 898(1) &
  1297(6) & 1344(6) & 1375(5) & 1416(3) \\
  463.6(6) & 842(4)  & 586.3(5) & 895(2) &
  1250(9)  & 1314(9) & 1351(6)  & 1404(4) \\
  407(1)   & 820(3)  & 564.9(5) & 886(3) &
  1205(13) & 1269(9) & 1326(9)  & 1383(5) \\
   \hline
  \hline
  135   &  770   & 494  & 892   &
  940   & 1116   & 1190 & 1320  \\
 \hline \hline 
  \end{tabular}
 \caption{
 Hadron masses in units of MeV in lattice QCD simulations. The last line shows experimental values. Taken from Ref.~\protect\cite{Nemura:2009kc}.
 }
 \label{tab:NL_mass}
\end{table}
      
\subsection{Flavor SU(3) limit}
In order to unravel the nature of the various channels in the hyperon interactions, it is more convenient to consider an idealized flavor SU(3) symmetric world, where $u,d$ and $s$ quarks are all degenerate with a common finite mass. In the flavor SU(3) limit, one may capture essential features of the interaction, in particular, the short range force without contamination from the quark mass difference. Calculations in the SU(3) limit allow us to extract potentials for irreducible flavor multiplets:  Potentials between asymptotic baryon states are obtained by the recombination of the multiplets with suitable Clebsh-Gordan coefficients. 

In the flavor SU(3) limit, the ground state baryon belongs to the flavor-octet with spin $1/2$,  and two baryon states with a given angular momentum are labeled by the irreducible representation of SU(3) as
\be
{\bf 8}\otimes {\bf 8} = \underbrace{{\bf 27}\oplus {\bf 8}\oplus {\bf 1}}_{\rm symmetric}\oplus 
 \underbrace{\overline{\bf 10}\oplus {\bf 10}\oplus {\bf 8}}_{\rm anti-symmetric},
\ee
where "symmetric" and "anti-symmetric" stand for the symmetry under the exchange of the flavor for two baryons. For the system with orbital S-wave, the Pauli principle imposes {\bf 27}, {\bf 8} and {\bf 1} to be spin-singlet ($^1S_0$), while  $\overline{\bf 10}$, {\bf 10}  and {\bf 8} to be spin-triplet ($^3S_1$).  

\begin{figure}[tb]
\begin{center}
 \includegraphics[width=0.45\textwidth]{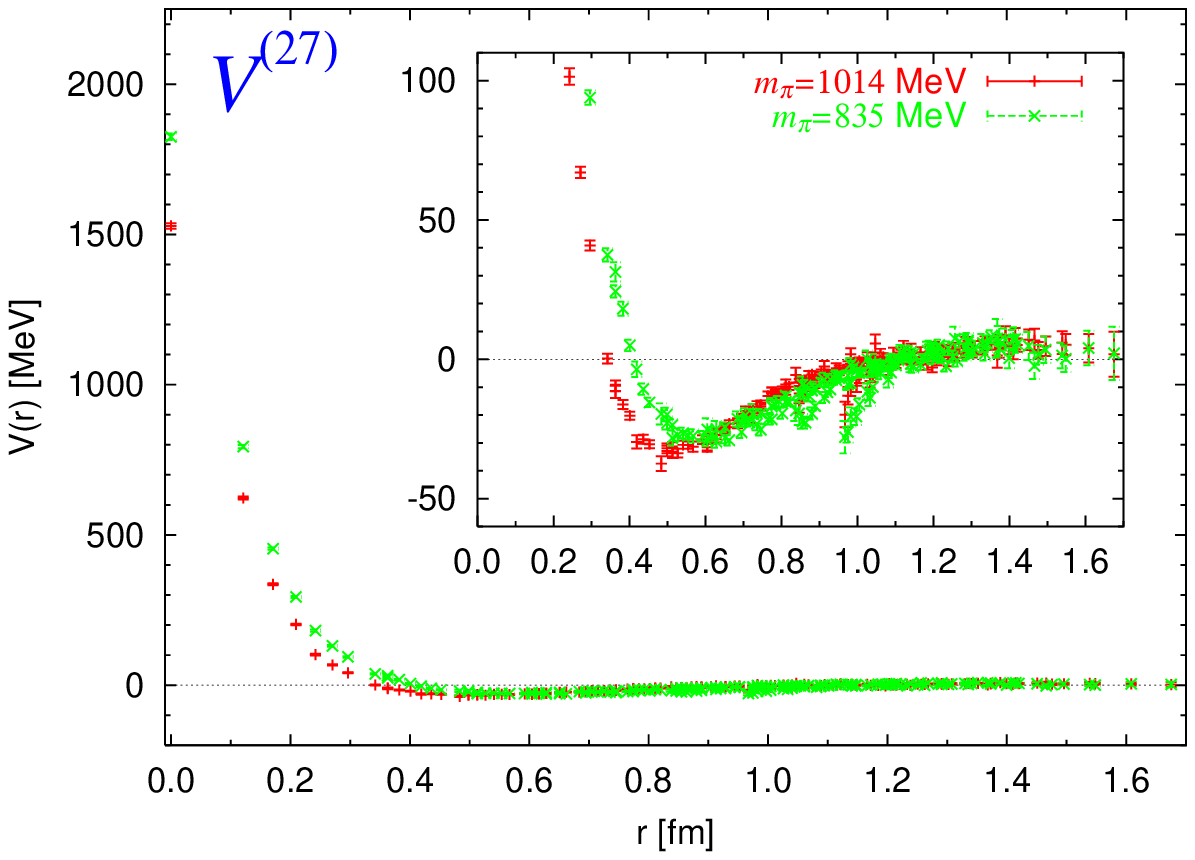}
 \includegraphics[width=0.45\textwidth]{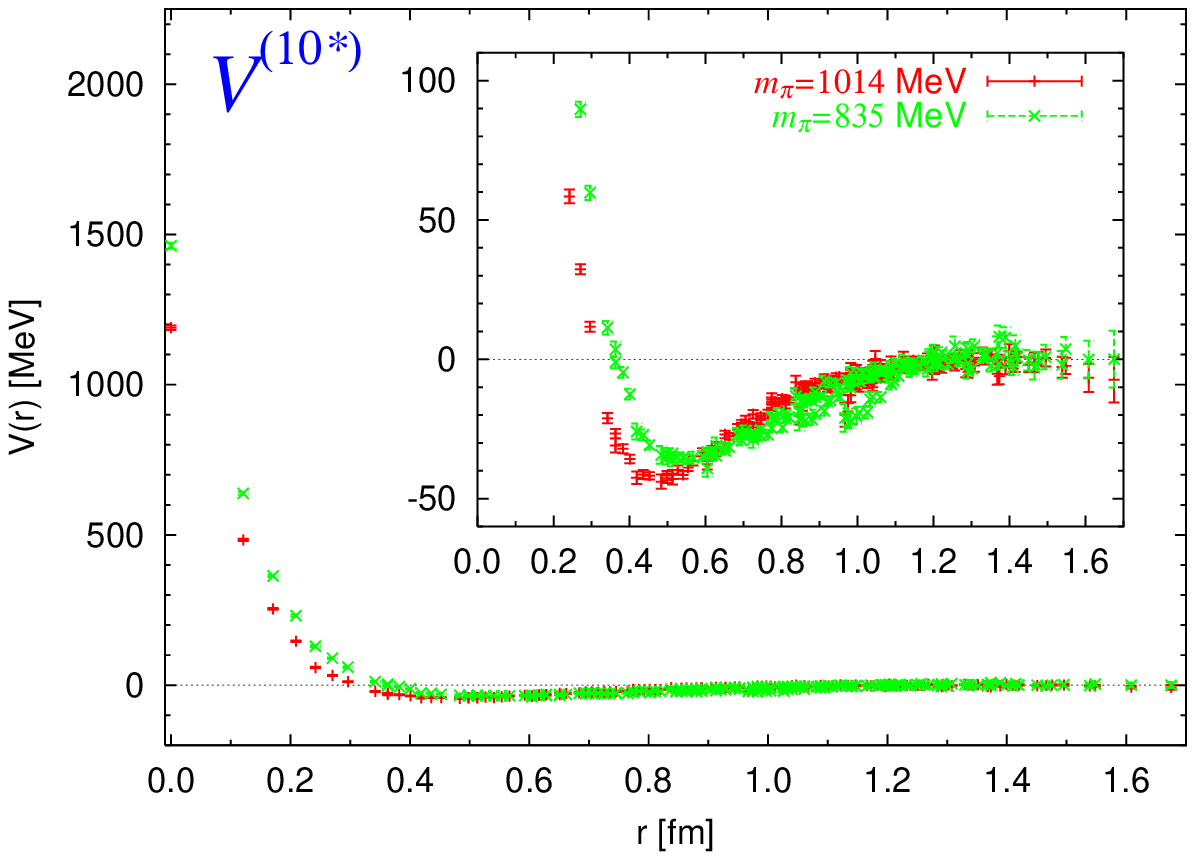}
 \end{center}
 \caption{
The BB potentials in {\bf 27} (Left) and $\overline{\bf 10}$(Right) representations from the orbital $A_1^+$ channel in the flavor SU(3) limit, 
  extracted from the lattice QCD simulation at 
  $m_{\pi}=1014$ MeV (red bars) and $m_{\pi}=835$ MeV (green crosses).
 Taken from Ref.~\protect\cite{Inoue:2010hs}.
 } 
 \label{fig:su3limitA}
\end{figure} 
\begin{figure}[tb]
\begin{center} 
 \includegraphics[width=0.45\textwidth]{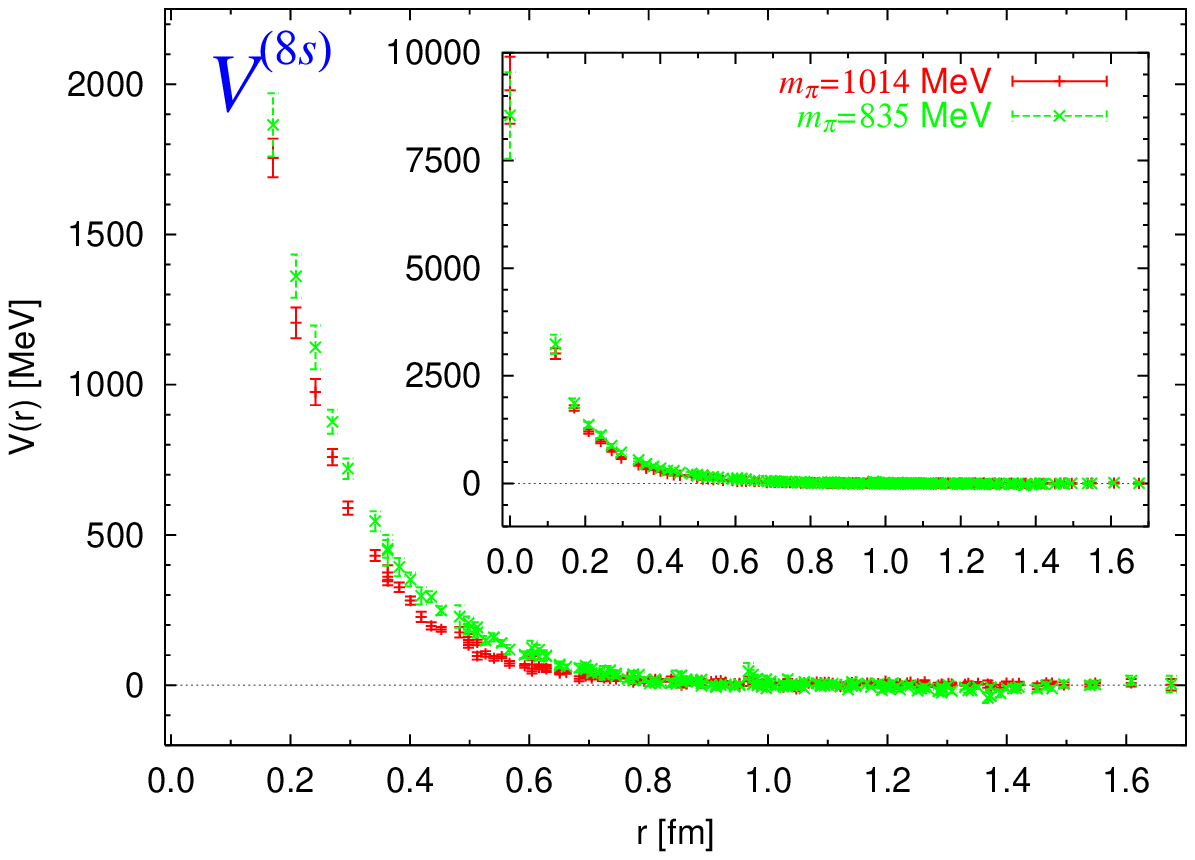}
 \includegraphics[width=0.45\textwidth]{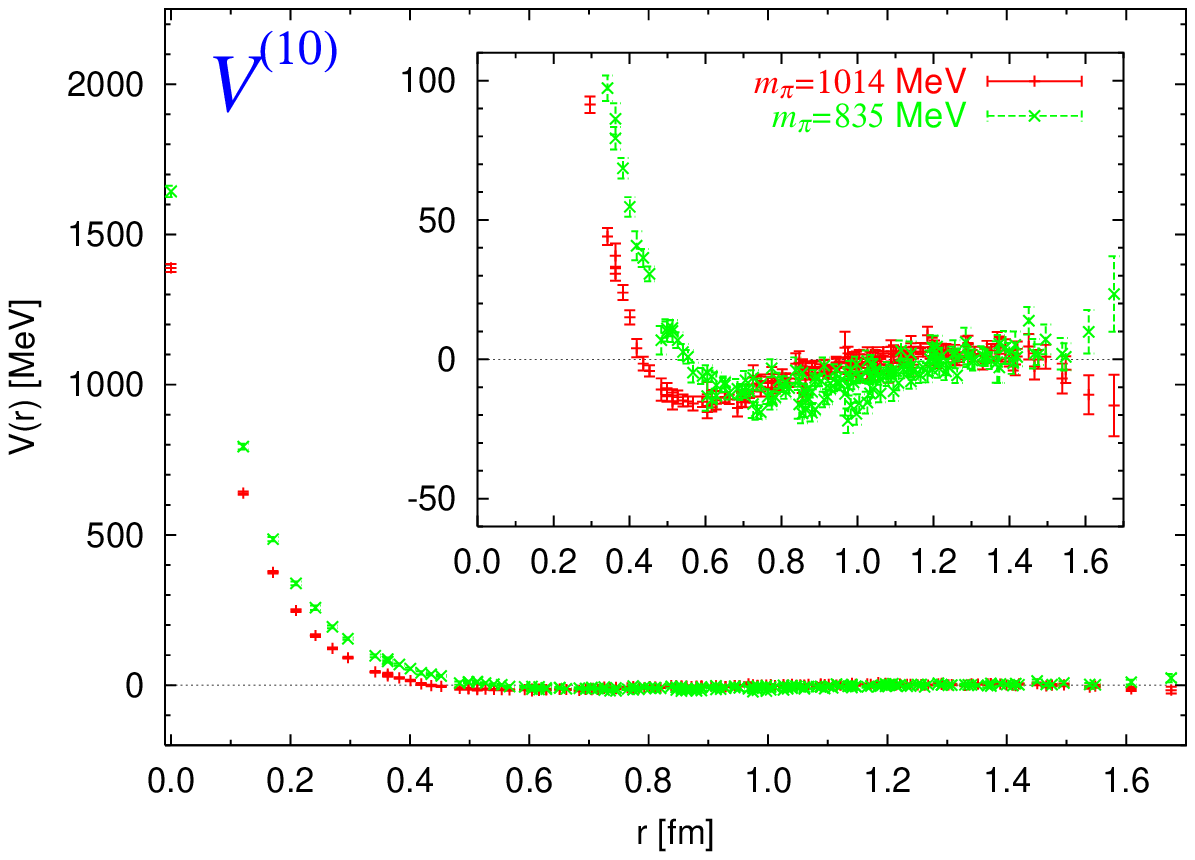}
 \includegraphics[width=0.45\textwidth]{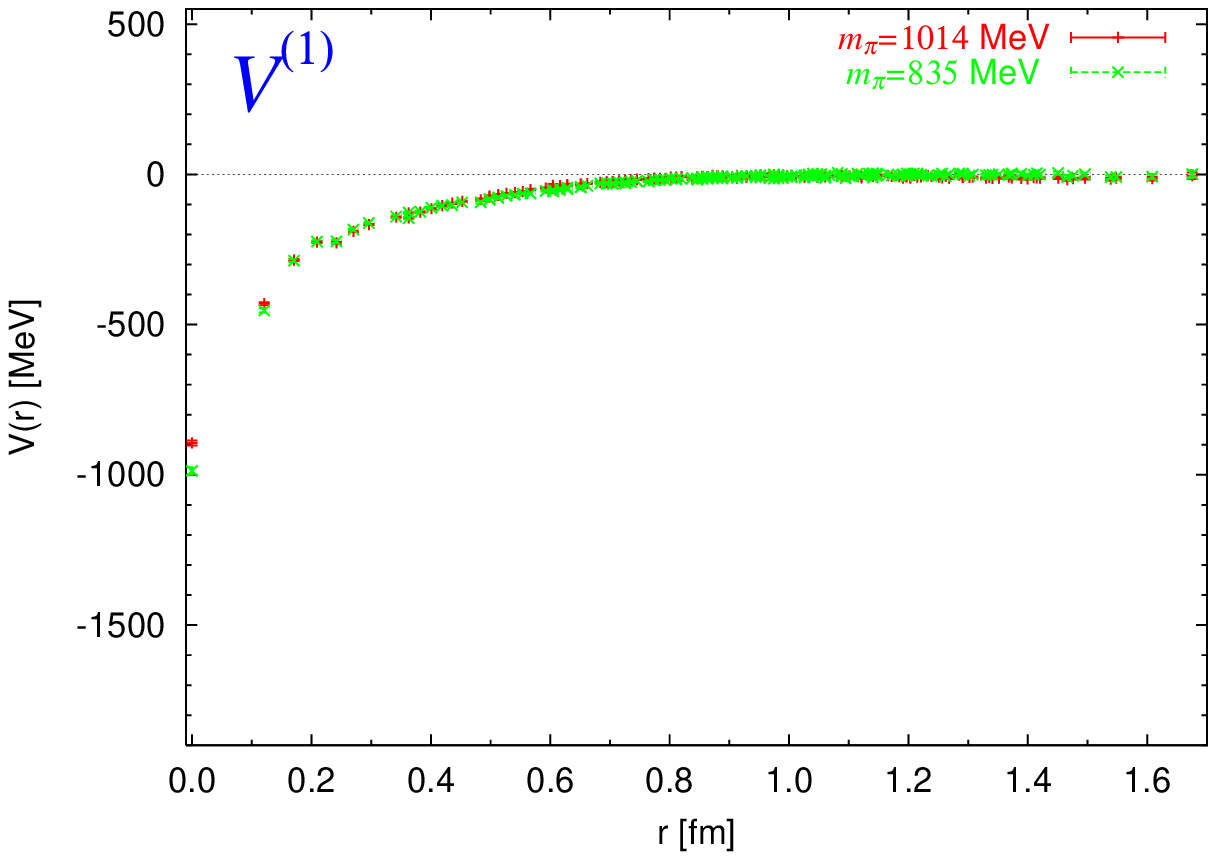}
 \includegraphics[width=0.45\textwidth]{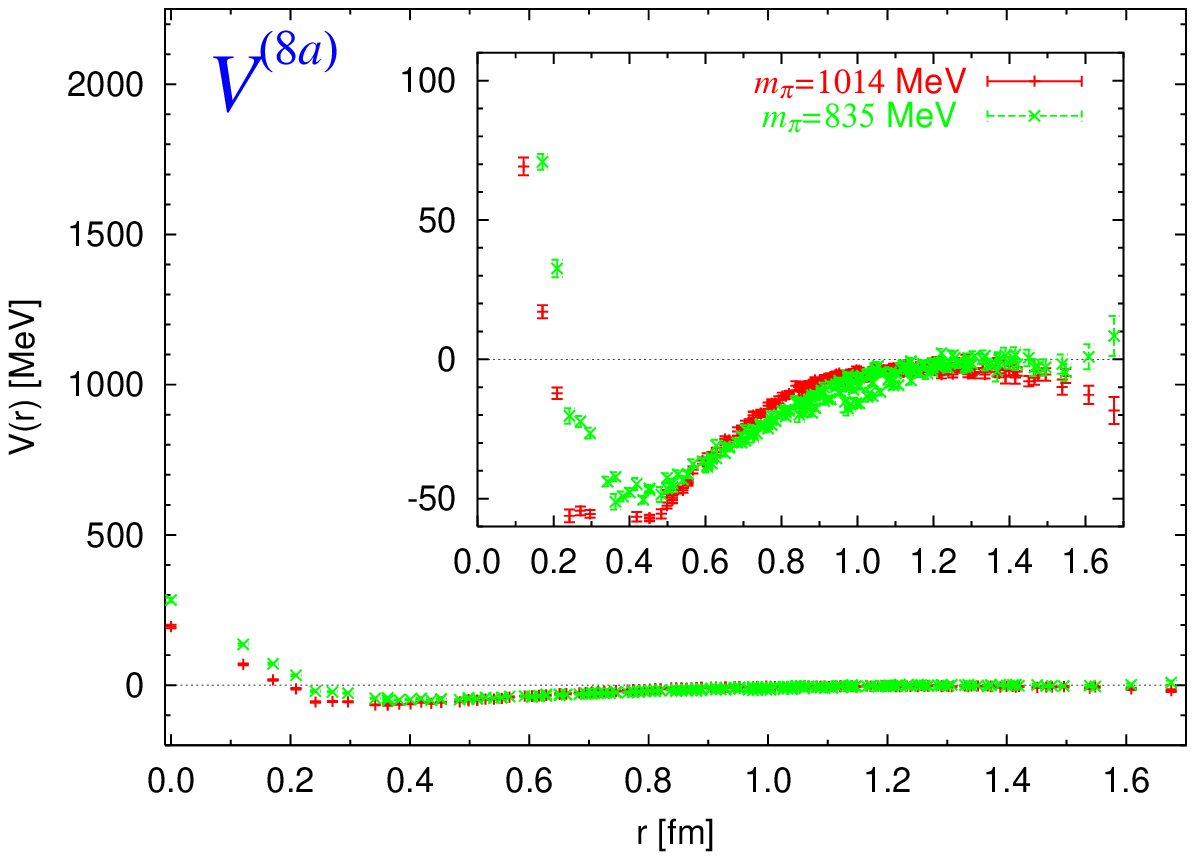}
\caption{ The BB potentials in ${\bf 8}_s$(Upper-Left), {\bf 10}(Upper-Right), {\bf 1}(Lower-Left) and ${\bf 8}_a$(Lower-Right) from the orbital $A_1^+$ channel in the flavor SU(3) limit,
  extracted from the lattice QCD simulation at   $m_{\pi}=1014$ MeV (red bars) and $m_{\pi}=835$ MeV (green crosses). 
 Taken from Ref.~\protect\cite{Inoue:2010hs}.  
 } 
\label{fig:su3limitB}
\end{center}
\end{figure} 
The NBS wave function is defined by
\be
\varphi^{W, (X)}({\bf r}) = \langle 0 \vert (BB)^{(X)}({\bf r}) \vert W, B=2, X \rangle,
\ee
from which the corresponding (effective) central potential is given as
\be
V^{(X)}(r) = \frac{1}{2\mu}\left[\frac{\nabla^2 \varphi^{W,(X)}({\bf r})}{\varphi^{W,(X)}({\bf r})}+k^2\right]
\ee
with $k^2 =W^2/4-m_B^2$ and $\mu = m_B/2$, where $m_B$ is the common octet baryon mass and 
$(BB)^{(X)}({\bf r}) =\sum_{ij} C_{ij}B_i({\bf x}+{\bf r},0) B_j({\bf x},0)$ with $X={\bf 27}, {\bf 8}_s,{\bf 1},\overline{\bf 10}, {\bf 10}, {\bf 8}_a$.
Two baryon operators $BB^{(X)}$ in the flavor basis are given in terms of the particle basis in Appendix~\ref{app:octet}. 
Potentials among octet baryons, both the diagonal part ($B_1B_2\rightarrow B_1B_2$) and the off-diagonal part ($B_1B_2\rightarrow B_3B_4$) are obtained by suitable combination of $V^{(X)}(r)$. 

Using the 3 flavor full QCD gauge configuration generated by CP-PACS/JLQCD Collaborations on a $16^3\times 32$ lattice at $a\simeq 0.12$ fm where light and strange quark masses have same values\cite{CPPACS-JLQCD},  the (effective) central potentials are calculated\cite{Inoue:2010hs} at $(m_{PS}, m_B) = (1014(1) {\rm MeV}, 2026(3) {\rm MeV})$ and $(845(1) {\rm Mev}, 1752(3) {\rm MeV})$, where $m_{PS}$ and $m_B$ denote the octet pseudo-scalar (PS) meson mass and the octet baryon mass, respectively.

Figs.~\ref{fig:su3limitA} and \ref{fig:su3limitB} give the six baryon-baryon ($BB$) potentials in the flavor basis, where red (green) data correspond to the $m_{PS} = 1014$ MeV (835 MeV). The left  panels show central potentials for the spin-singlet channel from the $J=A_1$ state, while the right panels give effective central potentials for the spin-triplet channel from  the $J=T_1$ state.
\begin{table}[tb]
   \begin{center}
   \begin{tabular}{| c|l |}
   \hline \hline
   flavor multiplet      & baryon pair (isospin)                      \\
   \hline 
   {\bf 27}       & \{NN\}(I=1), \{N$\Sigma$\}(I=3/2), \{$\Sigma\Sigma$\}(I=2),  \\
                  & \{$\Sigma\Xi$\}(I=3/2), \{$\Xi\Xi$\}(I=1) \\
   {\bf 8}$_s$      & none                             \\
   ~{\bf 1}       & none                             \\  
  \hline 
  {\bf 10}$^*$    & [NN](I=0), [$\Sigma\Xi$](I=3/2)    \\   
  {\bf 10}        & [N$\Sigma$](I=3/2), [$\Xi\Xi$](I=0)\\
  {\bf 8}$_a$     & [N$\Xi$](I=0)       \\
   \hline \hline
\end{tabular}
\caption{\label{tab:pairs}Baryon pairs in an irreducible flavor SU(3) representation,
 where $\{BB'\}$ and $[BB']$ denotes $BB' + B'B$ and $BB' - B'B$, respectively.}
 \end{center}
\end{table}
Note that some of octet-baryon pairs exclusively belong to an irreducible representation of SU(3) as shown in Table~\ref{tab:pairs}. For example, symmetric $NN$ belongs to ${\bf 27}$, which therefore can be considered as the $NN$ spin-singlet potential in the flavor SU(3) symmetric limit. Similarly $V^{(\overline{\bf 10})}$, $V^{({\bf 10})}$ and $V^{({\bf 8}_a)}$ can be considered as some $BB$ potentials of the particle basis in the SU(3) symmetric limit, while $V^{({\bf 1})}$ and $V^{({\bf 8}_s)}$ are always mixtures of different $BB$ potentials in the particle basis.

Fig.~\ref{fig:su3limitA}  shows $V^{\bf (27)}(r)$ and $V^{\bf (\overline{10})}(r)$, which correspond to  spin-singlet and spin-triplet NN potentials, respectively. Both have a repulsive core at short distance with an attractive pocket around 0.6 fm. These qualtative features are consistent with the previous results found for the NN potential in both quenched and full QCD.
The upper-right  panel of Fig.~\ref{fig:su3limitB} shows that $V^{\bf (10)}(r)$ has a stronger repulsive core and a weaker attractive pocket than $V^{\bf (27,\overline{10})}(r)$. Furthermore $V^{({\bf 8}_s)}(r)$ in the upper-left panel of Fig.~\ref{fig:su3limitB} has a very strong repulsive core among all 6 channels, while $V^{({\bf 8}_a)}(r)$ in the lower-right  panel has a very weak repulsive core. 
In contrast to all other cases,  $V^{\bf (1)}(r)$ shows attraction instead of repulsion at all distances,
as shown in the lower-left panel. 

Above features are consistent with what has been observed in a phenomenological quark model\cite{Oka:2000wj}. 
In particular, the potential in the ${\bf 8}_s$ channel in this quark model becomes strongly repulsive at short distance since the six quarks cannot occupy the same orbital state due to the Pauli exclusion for quarks. On the other hand, the potential in the {\bf 1} channel does not suffer from the quark Pauli exclusion and can become attractive due to the short-range gluon exchange. Such agreements between the lattice data and the phenomenological model suggest that the quark Pauli exclusion plays an essential role for the repulsive core in BB systems.

The BB potentials in the baryon basis can be obtained from those in the SU(3) basis by a unitary rotation as
\be
V_{ij}(r) = \sum_X U_{i X} V^{(X)}(r)U_{X j}^\dagger
\label{eq:rotation}
\ee
where $U$ is a unitary matrix which rotates the flavor basis $\ketv{X}>$ 
to the baryon basis $\ketv i>$, {\it i.e.} $\ketv i> = U_{i X} \ketv{X}>$.
The explicit forms of the unitary matrix $U$ in terms of the 
CG coefficients are found in Appendix \ref{app:octet}.

\begin{figure}[tb]
\begin{center}
 \includegraphics[width=0.45\textwidth]{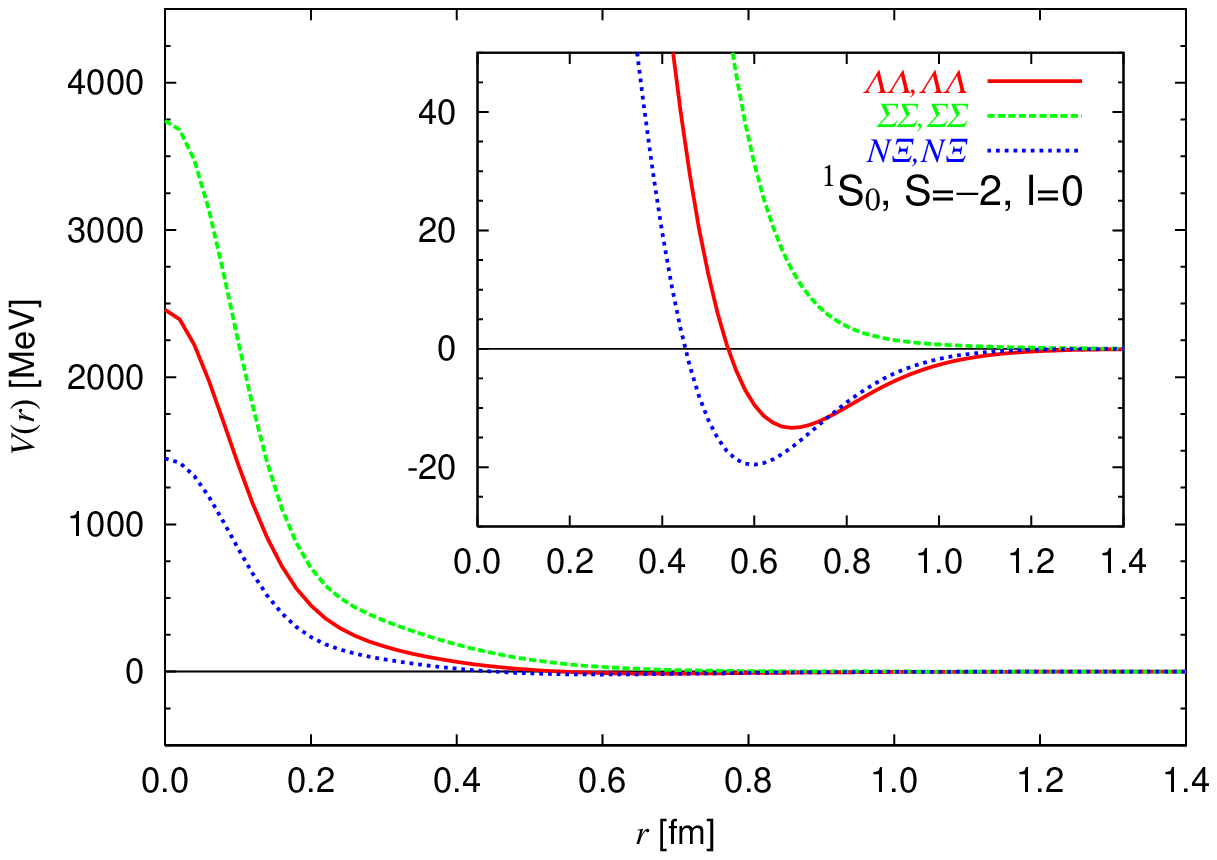}\hfill
 \includegraphics[width=0.45\textwidth]{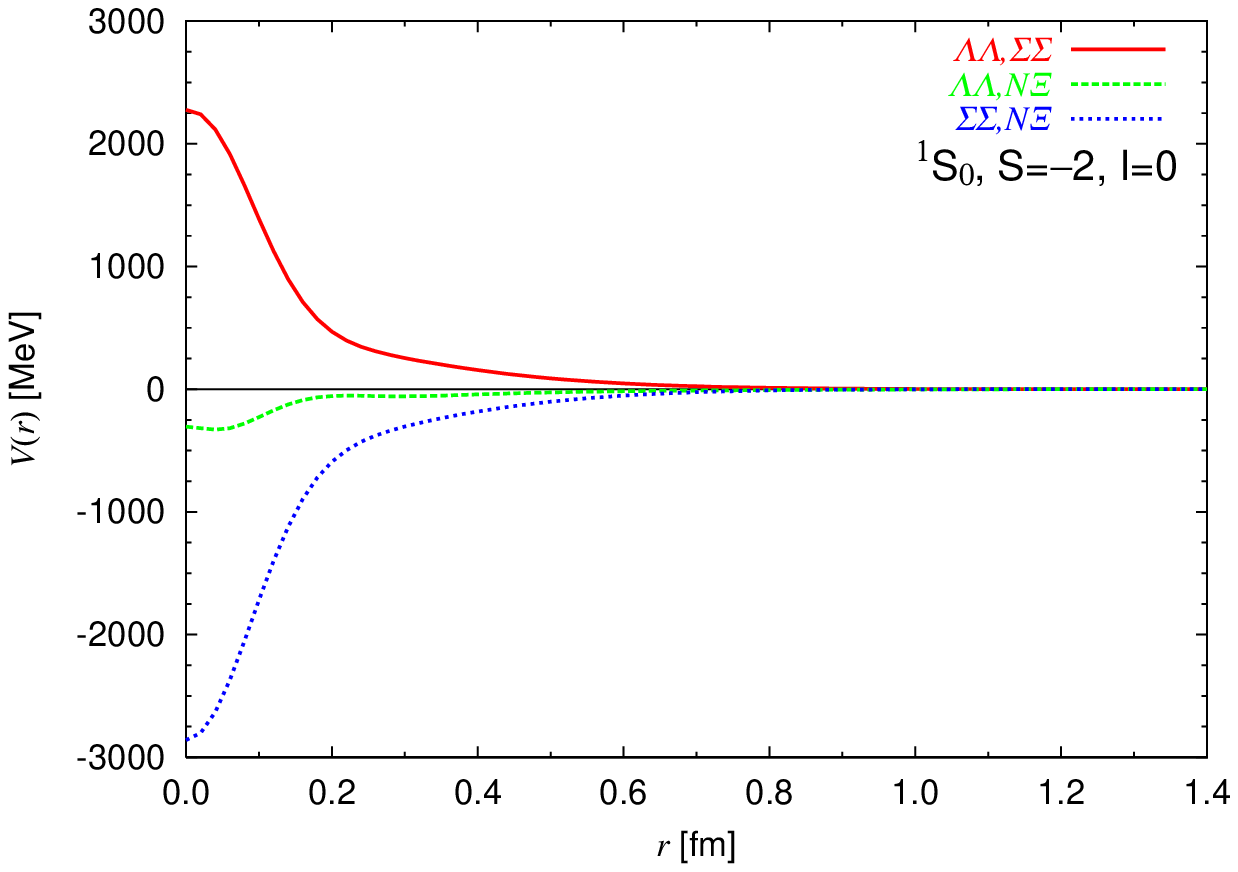}
 \caption{\label{fig:pot_lamlam}
  BB potentials in baryon basis for the S=$-$2, I=0, $^1S_0$ sector. 
  Three diagonal(off-diagonal) potentials are shown in left(right) panel.
  Taken from Ref.~\protect\cite{Inoue:2010hs}.  
 }
 \end{center}
\end{figure}

In Fig.~\ref{fig:pot_lamlam}, as characteristic examples,
let us show the spin-singlet potentials for S=$-$2, I=0 channel determined from the orbital $A_1^+$ representation   at $m_\pi = 835$ MeV. 
To obtain $V_{ij}(r)$,  the potentials in the SU(3) basis are fitted 
by the following form with five parameters $b_{1,2,3,4,5} $,
\begin{equation}
V(r) = b_1 e^{-b_2\,r^2} + b_3(1 - e^{-b_4\,r^2})\left( \frac{e^{-b_5\,r}}{r} \right)^2 .
\end{equation}
Then the right hand side of  Eq. (\ref{eq:rotation}) is used 
to obtain the potentials in the baryon basis.
The left panel of  Fig.~\ref{fig:pot_lamlam} shows the diagonal part of the potentials.
The strong repulsion in the ${\bf 8}_s$  channel is reflected most in the $\Sigma\Sigma$(I=0) potential
due to its largest CG coefficient among three channels.
The strong attraction in the ${\bf 1}$ channel is reflected most in the $N\Xi$(I=0) potential
due to its largest CG coefficient. 
Nevertheless, all three diagonal potentials have a repulsive core originating from the ${\bf 8}_s$ component.
 The right panel of  Fig.~\ref{fig:pot_lamlam} shows the off-diagonal parts of the potentials
 which are comparable in magnitude to the diagonal ones.
 Since the off-diagonal parts are not negligible in the baryon basis,
 fully coupled channel analysis is necessary to study observables.
 A similar situation  holds even in (2+1)-flavors where the strange quark is heavier than up and down quarks: The SU(3) basis with approximately
  diagonal potentials is useful for obtaining essential features of the BB interactions, while
   the baryon basis with substantial magnitude of the  off-diagonal potentials is
  necessary for practical applications.

Other potentials in the baryon basis are given in Ref.~\cite{Inoue:2010hs}. 
Since the ${\bf 8}_s$ state does not couple to the spin-triplet channel,
the repulsive cores in the spin-triplet channel 
are relatively small.  
The off-diagonal potentials are not generally small: 
For example,
the  $N\Lambda$-$N\Sigma$ potential in the spin-triplet channel 
is comparable in magnitude at short distances with the 
diagonal $N\Lambda$-$N\Lambda$ and $N\Sigma$-$N\Sigma$ potentials.
Although all quark masses of 3 flavors are degenerate and rather heavy in these simulations,
the coupled channel potentials in the baryon basis may give 
useful hints for the behavior of hyperons ($\Lambda$, $\Sigma$ and $\Xi$)
in hyper-nuclei and in neutron stars \cite{Hashimoto:2006aw,SchaffnerBielich:2010am}.

The flavor singlet channel has attraction for all distances, which might produce the bound state, the $H$-dibaryon, in this channel.  The present data, however, are not sufficient to make a definite
conclusion on the $H$-dibaryon, since  a single lattice with small extension $L\simeq 2$ fm is employed. In order to investigate whether the $H$-dibaryon exists  or not in the flavor SU(3)  limit, data on several different  volumes are needed. Such a study on the $H$-dibaryon will be discussed in Sec.~\ref{sec:extension}.

In order to extend the study in the flavor SU(3) limit to the real world where the strange quark is much heavier than light quarks, the potential method used so far has to be extended to more general cases, which will also be considered in Sec.~\ref{sec:extension}. 

\section{Origin of repulsive core}
\label{sec:OPE}
As seen in the previous sections, lattice QCD calculations show that the $NN$ potential defined through the NBS wave function has not only the attraction at medium to long distance
that has long been well understood in terms of pion and other heavier meson exchanges, but also a characteristic repulsive core at short distance, whose origin is still theoretically unclear.
Furthermore, the $BB$ potentials in the flavor SU(3) limit show several different behaviors at short distance: some have a stronger/weaker repulsive core than $NN$ while the singlet has an attractive core. In this section,      
recent attempts \cite{Aoki:2010kx, Aoki:2009pi,Aoki:2010uz} to theoretically understand  the short distance behavior of the potential in terms of  the operator product expansion (OPE)  is explained. 

\subsection{OPE and repulsive core}
Let us first explain the  basic idea. We consider the equal time NBS wave function defined  by
\begin{eqnarray}
\varphi^E_{AB}(\br) &=& \langle 0\vert T\{ O_A(\br/2,0) O_B(-\br/2,0)\} \vert E\rangle,
\end{eqnarray}
where $\vert E \rangle $ is some eigenstate of a certain system with total energy $E$, and $O_A$, $O_B$ are some operators of this system. (We suppress other quantum numbers of the state $\vert E\rangle$ for simplicity.)  The OPE reads
\begin{eqnarray}
O_A(\br/2,0) O_B(-\br/2,0) \simeq \sum_C D_{AB}^C(\br) O_C({\bf 0},0),
\end{eqnarray}
Suppose that the coefficient function of the OPE behaves 
in the small $r(=\vert \br\vert)$ limit as
\begin{eqnarray}
D_{AB}^C(\br ) &\simeq & r^{\alpha_C}( - \log r)^{\beta_C} f_C(\theta,\phi),
\end{eqnarray}
where $\theta,\phi$ are the angles of $\br$, the NBS wave function becomes
\begin{eqnarray}
\varphi_{AB}^E(\br) &\simeq& \sum_C r^{\alpha_C} (-\log r)^{\beta_C} f_C(\theta,\phi) D_C(E), 
\end{eqnarray}
where
\begin{eqnarray}
D_C(E) &=& \langle 0 \vert O_C({\bf 0},0)\vert E\rangle.
\end{eqnarray}
The potential at short distances can be calculated from this expression. 
For example, in the case of the Ising field theory in two dimensions, the OPE for the spin field $\sigma$ is given by
\begin{eqnarray}
\sigma(x,0) \sigma(0,0) \simeq G(r) {\bf 1} + c\, r^{3/4} O_1(0) + \cdots, \quad r=\vert x\vert,
\end{eqnarray}
where $O_1(x)$ ($= :\bar\psi\psi(x):$ in terms of free fermion fields) is an operator of dimension one.  This leads to
\begin{eqnarray}
\varphi (r,E) \simeq  r^{3/4} D(E) + O(r^{7/4}),\quad D(E)= c \langle 0 \vert   O_1(0) \vert E \rangle ,
\end{eqnarray}
where $ \vert E \rangle $ is a two-particle state with energy $E=2\sqrt{k^2+ m^2}$. 
From this expression the potential becomes 
\begin{eqnarray}
V(r) &=& \frac{\varphi^{\prime\prime}(r,E) + k^2\varphi(r,E)}{m  \varphi(r,E) }
\simeq -\frac{3}{16}\frac{1}{m r^2}
\end{eqnarray}
in the $r\rightarrow 0$ limit. The OPE predicts not only the $r^{-2}$ behavior of the potential at short distance but also its coefficient $-3/16$. Furthermore the potential at short distance does not depend on the energy of the state in this example\cite{Aoki:2008wy,Aoki:2008yw}. 

In QCD the dominant terms at short distance have  $\alpha_C=0$. Among these terms, we assume that $C$ has the largest contribution such that
$\beta_C > \beta_{C^\prime}$ for $\forall C^\prime\not= C$. Since such dominant operators with $\alpha_C=0$ mainly couple to the zero angular momentum ($L=0$) state, 
let us consider the NBS wave function with $L=0$.
Applying $\nabla^2$ to this wave function,
we obtain the following classification of the short distance behavior of the potential.
\begin{enumerate}
\item[(1)] $\beta_C\not=0$: The potential at short distance is energy independent and becomes
\begin{eqnarray}
V(r) \simeq -\frac{\beta_C}{m r^2(-\log r)} \,,
\end{eqnarray}
which is attractive for $\beta_C > 0$ and repulsive for $\beta_C<0$.
\item[(2)] $\beta_C=0$: In this case the potential becomes
\begin{eqnarray}
V(r) \simeq \frac{ D_{C^\prime}(E)}{D_C(E)}\frac{-\beta_{C^\prime}}{m r^2}(-\log r)^{\beta_{C^\prime}-1} \,,
\label{eq:pot_ope}
\end{eqnarray}
where $\beta_{C^\prime} < 0$ is the second largest exponent. The sign of the potential at short distance depends on the sign of $ D_{C^\prime}(E)/D_{C}(E)$.
\end{enumerate}

On the lattice, we do not expect divergence at $r=0$ due to lattice artifacts at short distance. The above classification holds at $a \ll r \ll 1/\Lambda_{\rm QCD}$, while the potential becomes finite even at $r=0$ on the lattice.

Since QCD is an asymptotic free theory, the 1-loop calculation for anomalous dimensions becomes
exact at short distance. 
The OPE in QCD is written as
\begin{eqnarray}
O_A(y/2) O_B(-y/2)&=&\sum_C D_{AB}^C(r,g,m,\mu)O_C(0)
\end{eqnarray}
where $g$ ($m$) is the renormalized coupling constant (quark mass) at scale $\mu$.
In the limit that $r=\vert y\vert  = e^{-t} R \rightarrow 0 $ ( $t\rightarrow\infty$ with fixed $R$),
the renormalization group analysis leads to
\begin{eqnarray}
\lim_{r\rightarrow 0} D_{AB}^C(r,g,m,\mu) = (-2\beta^{(1)} g^2\log r)^{\gamma_{AB}^{C,(1)}/(2\beta^{(1)})} D_{AB}^C(R, 0,0,\mu),
\end{eqnarray}
where 
$
\beta^{(1)} = \displaystyle \frac{1}{16\pi^2}\left(11-\frac{2N_f}{3}\right)
$
 is the QCD beta-function at 1-loop, and 
\begin{eqnarray}
\gamma_{AB}^{C,(1)} = \gamma_C^{(1)}-\gamma_A^{(1)} - \gamma_B^{(1)} \equiv \frac{1}{48\pi^2}\gamma .
\end{eqnarray}
Here $\gamma_X^{(1)}$ is the 1-loop anomalous dimension of the operator $O_X$. 
An appearance of $D_{AB}^C(R,0,0,\mu)$ on the right-hand side tells us that
it is enough to know the OPE only at tree level. 
From the above expression, $\beta_C$  is given by
$
\beta_C = \displaystyle \frac{\gamma_{AB}^{C,(1)}}{2\beta^{(1)}}.
$

\subsection{Two flavor case}
We first consider the OPE for  $N_f=2$ QCD\cite{Aoki:2010kx}. The 1-loop calculations show that the largest value of $\beta_C$ is always zero for both spin-singlet and spin-triplet channels for $N_f=2$ QCD and that   
the second largest value of  $\beta_{C^\prime}$ is given by 
\be
\beta_{C^\prime}^{S=0} = -\frac{6}{33-2N_f}, \quad \beta_{C^\prime}^{S=1} = -\frac{2}{33-2N_f},
\ee
where $S=0,1$ denotes the total spin.
This corresponds to the case (2) in the previous subsection. Therefore
the OPE and renormalization group analysis in QCD predicts the universal functional form of the $NN$ central potential at short distance as
\be
V_C^S(r) \simeq \frac{D_{C^\prime}(E)}{D_C(E)} \frac{-\beta_{C^\prime}^S (-\log r)^{\beta_{C^\prime}^S -1}}{m_N r^2}, \qquad r\rightarrow 0,
\ee 
which is a little weaker than a $1/r^2$ singularity, while for the tensor potential we have
\be
V_T(r)\simeq 0
\ee 
at the 1-loop order.

The OPE, however, can not tell whether the potential at short distance is repulsive or attractive, which is determined by the sign of the coefficient. If $D_X(E)$ and $D_Y(E)$ are evaluated by the non-relativistic quark model at the leading order, we obtain
\be
\frac{D_{C^\prime}(E)}{D_C(E)}(S=0) \simeq \frac{D_{C^\prime}(E)}{D_C(E)}(S=1)\simeq 2 .
\ee
For both cases, the ratio has positive sign, which gives repulsion at short distance, the repulsive core.

\subsection{Extension to three flavors}
The above calculation has been extended to $N_f=3$ QCD\cite{Aoki:2010uz}.
In the 3-flavor case, some channel may become attractive at short distance since the Pauli exclusion principle is less significant than in the 2-flavor case.  Indeed the lattice QCD calculations in the flavor SU(3) limit shows the attractive potential for the singlet channel, as seen in the previous section.

The largest  value of $\beta_C$ is given for each $X$ representation of the flavor SU(3) in unit of $1/(33-2N_f)$  in the table~\ref{tab:ope}, where we define
\be
\beta_C =\frac{\gamma^{(X)}}{33- 2 N_f} .
\ee
\begin{table}[t]
    \begin{center}
   \begin{tabular}{| c | llllll  |}
   \hline \hline 
   $X$    & ${\bf 27}$ & ${\bf 8}_s$ & ${\bf 1}$ & $\overline{\bf 10}$ & ${\bf 10}$ & ${\bf 8}_a$    \\
   \hline 
  $\gamma^{(X)}$       & 0 &  6 & 12 & 0 & 0 & 4 \\
Non-relativistic  op. & yes & no & yes & yes & yes & yes \\ 
   \hline \hline
    \end{tabular}
   \caption{\label{tab:ope} The largest value of $\beta_C$ in unite of $1/(33-2N_f)$ of 3-flavor QCD for each representation. The last line indicates that the operator corresponding to the largest value of $\beta_C$ exists or not in the non-relativistic limit.}
 \end{center}
\end{table}

From the table, we observe that the largest value of $\beta_C$ is zero in the ${\bf 27}$, $\overline{\bf 10}$ and ${\bf 10}$ channels. This is consistent with the nucleon case in the previous subsection, which belong to ${\bf 27}$ (spin-singlet) and $\overline{\bf 10}$(spin-triplet). These three channels correspond to the case (2), so that the potentials are given by eq. (\ref{eq:pot_ope}).
On the other hand, the largest value of $\beta_C$ becomes positive in the ${\bf 8}_s$, $\overline{\bf 8}_a$ and ${\bf 1}$ channels, which correspond to the case (1). Therefore the (effective) central potential becomes attractive at short distance as
\be
V_C^{(X)}(r) \simeq -\frac{\gamma^{(X)}}{(33-2N_f)}\frac{1}{ m_B r^2(-\log r)} ,
\ee
where $m_B$ is the octet baryon mass.

The attractive core of the potential in the flavor singlet channel agrees with the behavior  of the potential found for the numerical simulation of lattice QCD in the previous section,
while for other two channels, ${\bf 8}_s$ and ${\bf 8}_a$, the prediction by the OPE disagrees with
the lattice QCD results:  The potential in the ${\bf 8}_s$ channel is most repulsive among 6 channels and the potential in the  ${\bf 8}_a$ channel still has a repulsive core, which is however weaker than others. 
The disagreement between the OPE and the lattice QCD result for the ${\bf 8}_s$ channel may be understood by the fact that no local 6 quark operator exists for this channel in the non-relativistic limit, as shown in the table~\ref{tab:ope}: the ${\bf 8}_s$ operator with the largest positive $\beta_C$ has a very small coefficient at low energy, so that the other operators with zero or negative $\beta_C$ may still dominate at distance scales comparable to  the lattice spacing $a= 0.1-0.2$ fm.   For the ${\bf 8}_a$ case, the weakest repulsive core in the lattice QCD simulation suggests that 
the attraction from the leading operator with the positive $\beta_C$ may be cancelled by other contributions from the sub-leading operators with zero or negative $\beta_C$ at the distance scale comparable to the lattice spacing of the simulations.  
It is therefore important to confirm the prediction from the OPE, by investigating the behavior of the repulsive core for each channel in the flavor SU(3) limit at finer lattice spacings and hopefully in the continuum limit.

\section{Extensions}
\label{sec:extension}
In this section, recent extensions of the potential method are considered.

\subsection{Inelastic scattering}
The potential method discussed so far is shown to be quite successful in order to describe elastic hadron interactions. Hadron interactions in general, however, lead to inelastic scatterings as the total energy of the system increases.  In order to extract hadron interactions which describe such inelastic scatterings from lattice QCD,  an extension of the potential method  is considered in this subsection. 

Let us first discuss the case of $A+B\rightarrow C+D$ scattering where $A,B,C,D$ represent some 1-particle states. This is a simplified version of the octet baryon scattering in the strangeness $S=-2$ and isospin $I=0$ channel, where $\Lambda\Lambda$, $N\Xi$ and $\Sigma\Sigma$ appear as asymptotic states of the strong interaction if the total energy is larger than $2m_\Sigma$.

We here assume $m_A + m_B < m_C + m_D < W $, where $W=E_k^A + E_k^B$ is the total energy of the system, and $E_k^X =\sqrt{m_X^2 +\bk^2}$. In this situation, the QCD eigen-state with the quantum numbers of the $AB$ state  and center of mass energy $W$  is expressed in general as
\begin{eqnarray}
\vert W \rangle &=& c_{AB} \vert AB,W\rangle + c_{CD} \vert CD, W\rangle +\cdots\\
\vert AB, W\rangle &=&  \vert A, \bk \rangle_{\rm in}\otimes \vert B, -\bk \rangle_{\rm in}, \quad
\vert CD, W\rangle =  \vert C, \bq \rangle_{\rm in}\otimes \vert D, -\bq \rangle_{\rm in},
\end{eqnarray}
where $W=E_k^A+E_k^B = E_q^C+E_q^D$.
We define the following NBS wave functions,
\begin{eqnarray}
\varphi_{AB}(\br,\bk )e^{-Wt} &=& \langle 0 \vert T\{ \varphi_A(\bx+\br,t) \varphi_B(\bx,t) \}\vert W \rangle, \\
\varphi_{CD}(\br,\bq )e^{-Wt} &=&  \langle 0 \vert T\{ \varphi_C(\bx+\br,t) \varphi_D(\bx,t) \}\vert W \rangle .
\end{eqnarray}
Using the partial wave decomposition such that\footnote{Here we ignore spins for simplicity.}
\bea
\varphi_{XY}(\br,\bk ) &=& 4\pi\sum_{l,m} i^l \varphi^{\ell}_{XY}(r,k)Y_{lm}(\Omega_{\brs}) \overline{Y_{lm}(\Omega_{\bks})} ,
\eea
the NBS wave function of the 2-channel system behaves for large $r$ as
\begin{eqnarray}
\left(
\begin{array}{l}
\varphi_{AB}^{\ell}(r,k) \\
\varphi_{CD}^{\ell}(r,q) \\
\end{array}
\right)
&\simeq&
\left(
\begin{array}{ll}
 j_l(k r) & 0 \\
 0 &  j_l(q r) \\
 \end{array}
 \right)
\left(
\begin{array}{l}
c_{AB} \\
c_{CD}\\
\end{array}
\right) 
+
\left(
\begin{array}{ll}
 n_l(kr)+i j_l(k r) & 0 \\
 0 & n_l(qr)+i  j_l(q r) \\
 \end{array}
 \right)
\nonumber \\
\nonumber \\
&\times&
O(W)
\left(
\begin{array}{cc}
e^{i\delta_l^1(W)}\sin \delta_l^1(W) & 0 \\
0 & e^{i\delta_l^2(W)}\sin \delta_l^2(W) \\
\end{array}
\right) 
O^{-1}(W)
\left(
\begin{array}{l}
c_{AB} \\
c_{CD}\\
\end{array}
\right) , \\
O(W) &=& \left(
\begin{array}{cc}
\cos\theta(W) & -\sin\theta(W) \\
\sin\theta(W) & \cos\theta(W) \\
\end{array}
\right) , 
\end{eqnarray}
where $\delta_l^i(W)$ is the scattering phase shift , whereas $\theta(W)$ is the mixing angle. 
This expression shows that the NBS wave functions for large $r$ agree with scattering waves described by two scattering phases $\delta_l^i(W)$ ($i=1,2$) and one mixing angle $\theta(W)$.
Because of this property, these wave functions satisfy
\be
(\nabla^2 + \bk^2 )\varphi_{AB}(\br,\bk) =0, \quad
(\nabla^2 + \bq^2 )\varphi_{CD}(\br,\bq) =0
\ee
for $r\rightarrow \infty$.

Let us now consider QCD in the finite volume $V$.
In the finite volume, $ \vert AB, W \rangle $ and $ \vert CD, W \rangle $ are no longer eigen-states of the hamiltonian. 
True eigenvalues are shifted from $W$ to $W_i = W + O(V^{-1})$ ($i=1,2$).
By diagonalization method in lattice QCD simulations, 
it is relatively easy to determine $W_1$ and $W_2$. With these values L\"uscher's finite volume formula gives two conditions, which,
however, are insufficient to determine three observables,
$\delta_l^1$, $\delta_l^2$ and $\theta$. (See \cite{Liu:2005kr, Lage:2009zv,Bernard:2010fp} for recent proposals to overcome this difficulty.)
An alternative approach to extract three observables, $\delta_l^1$, $\delta_l^2$ and $\theta$, has been proposed in lattice QCD through the above NBS wave functions\cite{Ishii:2011,HALQCD:2011}.
We consider the NBS wave functions at two different values of energy, $W_1$ and $W_2$,
in the finite volume:
\begin{eqnarray}
\varphi_{AB}(\br,\bk_i )e^{-W_i t} &=& \langle 0 \vert T\{ \varphi_A(\bx+\br,t) \varphi_B(\bx,t) \}\vert W_i \rangle \\
\varphi_{CD}(\br,\bq_i )e^{-W_i t} &=& \langle 0 \vert T\{ \varphi_C(\bx+\br,t) \varphi_D(\bx,t) \}\vert W_i \rangle, \quad i=1,2 .
\end{eqnarray}
We then define the coupled channel non-local potentials from the coupled channel Schr\"odinger equation as
\begin{eqnarray}
\left[\frac{k_i^2}{2\mu_{AB}} - H_0\right] \varphi_{AB}(\bx,\bk_i) &=&
\int d^3 y\ U_{AB,AB}(\bx;\by)\ \varphi_{AB}(\by,\bk_i)+ \int d^3 y\ U_{AB,CD}(\bx;\by)\ \varphi_{CD}(\by,\bq_i)\nonumber \\
\\
\left[\frac{q_i^2}{2\mu_{CD}} - H_0\right] \varphi_{CD}(\bx,\bk_i) &=&
\int d^3 y\ U_{CD,AB}(\bx;\by)\ \varphi_{AB}(\by,\bk_i)+ \int d^3 y\ U_{CD,CD}(\bx;\by)\ \varphi_{CD}(\by,\bq_i) \nonumber \\
\end{eqnarray}
for $i=1,2$.
As before the velocity expansion is introduced as
\begin{eqnarray}
U_{XY,VZ} (\bx;\by) &=& V_{XY,VZ}(\bx,\nabla)\delta^3(\bx-\by)
= \left[V_{XY,VZ}(\bx) + O(\nabla)\right]\delta^3(\bx-\by)
\end{eqnarray}
and at the leading order of the expansion, we have
\begin{eqnarray}
K_{AB}(\bx,\bk_i)\equiv \left[\frac{k_i^2}{2\mu_{AB}} - H_0\right] \varphi_{AB}(\bx,\bk_i) &=&
 V_{AB,AB}(\bx)\ \varphi_{AB}(\bx,\bk_i)+  V_{AB,CD}(\bx)\ \varphi_{CD}(\bx,\bq_i)\nonumber \\
\\
K_{CD}(\bx,\bq_i)\equiv \left[\frac{q_i^2}{2\mu_{CD}} - H_0\right] \varphi_{CD}(\bx,\bk_i) &=&
V_{CD,AB}(\bx)\ \varphi_{AB}(\bx,\bk_i)+ V_{CD,CD}(\bx)\ \varphi_{CD}(\bx,\bq_i). \nonumber \\
\end{eqnarray}
These equations for $i=1,2$  can be solved as
\begin{eqnarray}
\left(
\begin{array}{ll}
V_{AB,AB}(\bx) &  V_{AB,CD}(\bx) \\
V_{CD,AB}(\bx) & V_{CD,CD}(\bx)\\
\end{array}
\right) &=&
\left(
\begin{array}{ll}
K_{AB}(\bx,\bk_1) & K_{AB}(\bx,\bk_2)\\
K_{CD}(\bx,\bq_1) & K_{CD}(\bx,\bq_2)\\
\end{array}
\right) \nonumber \\
&\times &
\left(
\begin{array}{ll}
\varphi_{AB}(\bx,\bk_1) & \varphi_{AB}(\bx,\bk_2) \\
\varphi_{CD}(\bx,\bq_1) & \varphi_{CD}(\bx,\bq_2) \\
\end{array}
\right)^{-1}. 
\end{eqnarray}

Once we obtain the coupled channel local potentials $V_{XY, VZ}(\bx)$, we solve the coupled channel Schr\"odinger equation in {\it infinite} volume with some appropriate boundary condition such that the incoming wave has a definite $\ell$ and consists of the $AB$ state only , in order to extract
three observables for each $\ell$ ($\delta_l^1(W)$, $\delta_l^2(W)$ and $\theta(W)$)  
at all values of $W$.
Of course, since $V_{XY,VZ}$ is the leading order approximation in the velocity expansion of $U_{XY,VZ}(\bx;\by)$,  results for three observables $\delta_l^1(W)$, $\delta_l^2(W)$ and $\theta(W)$  at $W\not=W_1, W_2$ are also approximate ones and might be different from the exact values.  By performing an  additional extraction of $V_{XY, VZ}(\bx)$ at $(W_3,W_4)\not=( W_1,W_2)$, we can test  how good the leading order approximation is.

The method considered above can be generalized to inelastic scattering where a number of particles is not conserved. For illustration, let us consider the scattering $A+B\rightarrow A+B$ and $A+B\rightarrow A+B+C$ where the total energy $W$ satisfies 
$ m_A+m_B+m_C < W < m_A+m_B+2m_C$.

 The following NBS wave functions at the center of mass system are used:
\begin{eqnarray}
\varphi_{AB}^W(\bx)e^{-Wt} &=& \langle 0 \vert T\{\varphi_A(\br+\bx,t)\varphi_B(\br,t)\}\vert W \rangle \\
\varphi_{ABC}^W(\bx,\by)e^{-Wt} &=& \langle 0 \vert T\{\varphi_A\left(\br+\bx+\frac{\by\,\mu_{BC}}{m_C},t\right)\varphi_B(\br+\by,t)\varphi_C(\br,t)\}\vert W \rangle ,
\end{eqnarray}
where
\begin{eqnarray}
\vert W \rangle &=& c_1\ \vert \bk \rangle_{\rm in}\otimes \vert -\bk \rangle_{\rm in}  +
c_2\ \vert \bq_x \rangle_{\rm in}\otimes \left\vert \bq_y-\frac{\bq_x\mu_{BC}}{m_C} \right\rangle_{\rm in}\otimes \left\vert -\bq_y- \frac{\bq_x\mu_{BC}}{m_B}\right\rangle_{\rm in}
\end{eqnarray}
with
\begin{eqnarray}
W &=& \sqrt{\bk^2+m_A^2} +\sqrt{\bk^2+m_B^2} \nonumber \\
&=& \sqrt{\bq_x^2+m_A^2}+\sqrt{(\bq_y-\frac{\bq_x\mu_{BC}}{m_C})^2+m_B^2} +\sqrt{(\bq_y+\frac{\bq_x\mu_{BC}}{m_B})^2+m_C^2} 
\end{eqnarray}
and $1/\mu_{AB} =1/m_B + 1/m_C$. Here  $\by = \br_B-\br_C$ is a relative coordinate between 
$B$ and $C$ with the reduced mass $\mu_{BC}$, while $\bx =\br_A-{\bf R}_{BC}$ is the one between $A$ and the center of mass of $B$ and $C$ with ${\bf R}_{BC} = (m_B\br_B + m_C \br_C)/(m_B+ m_C)$.  

We define the non-local potential from the coupled channel equations as
\begin{eqnarray}
K_{AB}^W(\bx)&\equiv& \left[\frac{\bk^2}{2\mu_{AB}}-H_0^{AB}\right] \varphi_{AB}^W(\bx)= 
\int d^3\,z\ U_{AB,AB}(\bx;\bz)\, \varphi_{AB}^W(\bz) \nonumber \\
&+& \int d^3\,z\ d^3\,w\ U_{AB,ABC}(\bx;\bz,\bw)\, \varphi_{ABC}^W(\bz,\bw)\\
K_{ABC}^W(\bx,\by)&\equiv&\left[\frac{\bq_x^2}{2\mu_{A,BC}} + \frac{\bq_y^2}{2\mu_{BC}} -H_0^{A,BC} - H_0^{BC}\right]
\varphi_{ABC}^W(\bx,\by)= 
\int d^3\,z\ U_{ABC,AB}(\bx,\by;\bz)\nonumber \\
&\times& \varphi_{AB}^W(\bz)\nonumber 
+ \int d^3\,z\ d^3\,w\ U_{ABC,ABC}(\bx,\by;\bz,\bw)\, \varphi_{ABC}^W(\bz,\bw)
\end{eqnarray}
where
\begin{eqnarray}
H_0^{AB}&=&-\frac{\nabla_{\bxs}^2}{2\mu_{AB}}, \quad
H_0^{A,BC}=-\frac{\nabla_{\bxs}^2}{2\mu_{A,BC}}, \quad
H_0^{BC}=-\frac{\nabla_{\bys}^2}{2\mu_{BC}}
\end{eqnarray}
with another reduced mass defined by $1/\mu_{A,BC}=1/m_A+1/(m_B+m_C)$.

We consider the following velocity expansions
\begin{eqnarray}
U_{AB,AB} (\bx; \bz) &=& \left[V_{AB,AB}(\bx) + O(\nabla^x)\right]\delta^3(\bx-\bz) \\
U_{AB,ABC} (\bx; \bz,\bw) &=& \left[V_{AB,ABC}(\bx,\bw) + O(\nabla^x)\right]\delta^3(\bx-\bz)  \\
U_{ABC,AB} (\bx,\by; \bz) &=& \left[V_{AB,ABC}(\bx,\by) + O(\nabla^x)\right]\delta^3(\bx-\bz)\\
U_{ABC,ABC} (\bx,\by; \bz,\bw) &=& \left[V_{ABC,ABC}(\bx,\by) + O(\nabla^x,\nabla^y)\right]\delta^3(\bx-\bz) \delta^3(\by-\bw) ,
\end{eqnarray}
where the hermiticity of the non-local potentials gives $V_{AB,ABC}(\bx,\by) = V_{ABC,AB}(\bx,\by)$.

At the leading order of the velocity expansions, the coupled channel equations become
\begin{eqnarray}
K_{AB}^W(\bx) &=& V_{AB,AB}(\bx) \varphi_{AB}^W(\bx) + \int d^3\,w\ V_{AB,ABC}(\bx,\bw)\varphi_{ABC}^W(\bx,\bw ) \\
K_{ABC}^W(\bx,\by) &=& V_{ABC,AB}(\bx,\by)\varphi_{AB}^W(\bx) + V_{ABC,ABC}(\bx,\by)\varphi_{ABC}^W(\bx,\by) .
\end{eqnarray}
By considering two values of energy such that $W=W_1, W_2$, 
we can determine $V_{ABC,AB}$ and $V_{ABC,ABC}$ from the second equation as
\begin{eqnarray}
\left(
\begin{array}{ll}
V_{ABC,AB}(\bx,\by) & V_{ABC,ABC}(\bx,\by) \\
\end{array}
\right) &=& 
\left(
\begin{array}{ll}
K_{ABC}^{W_1}(\bx,\by) & K_{ABC}^{W_2}(\bx,\by) \\
\end{array}
\right) \nonumber \\
&\times&
\left(
\begin{array}{ll}
\Psi_{AB}^{W_1}(\bx) & \Psi_{AB}^{W_2}(\bx) \\
\Psi_{ABC}^{W_1}(\bx,\by) & \Psi_{ABC}^{W_2}(\bx,\by) \\
\end{array}
\right)^{-1} .
\end{eqnarray}
Using the hermiticity relation $V_{AB,ABC}(\bx,\by) = V_{ABC,AB}(\bx,\by)$, we can extract $V_{AB,AB}$ from the first equation as
\begin{eqnarray}
V_{AB,AB}(\bx) &=&\frac{1}{\Psi_{AB}^W(\bx)}\left[
K_{AB}^{W}(\bx) -\int d^3\,w\ V_{ABC,AB}(\bx,\bw)\Psi_{ABC}^W(\bx,\bw ) 
\right]
\end{eqnarray}
for $W=W_1, W_2$. A difference of $V_{AB,AB}(\bx)$ between two estimates at $W_1$ and $W_2$ gives an estimate for higher order contributions in the velocity expansions. 

Once we obtain $V_{AB,AB}$, $V_{AB,ABC}=V_{ABC,AB}$ and $V_{ABC,ABC}$, we can solve the coupled channel Schr\"odinger equations in the {\it infinite} volume, in order to extract physical observables. 
 As $W$ increases and becomes larger than $m_A+m_B + n m_C$, the inelastic scattering  $A+B\rightarrow A+B+n C $ becomes possible. As in the case of $A+B\rightarrow A+B+C$ in the above,  we can define the coupled channel potentials including this channel, though calculations of the NBS wave functions for multi-hadron operators become more and more difficult in practice.  
      
\subsection{Coupled channels with $S=-2$ and $I=0$}
\label{sec:coupled_channel}
\begin{table}[tb]
\begin{center}
  \begin{tabular}{|c|c|cccccc|}
  \hline \hline
   & $N_{conf}$ & $m_\pi$ & $m_K$ & $m_N$ & $m_\Lambda$ & $m_\Sigma$ & $m_\Xi$ \\
  \hline
  Set 1 & $700$ & $875(1)$ & $916(1)$ & $1806(3)$ & $1835(3)$ & $1841(3)$ & $1867(2)$ \\
  Set 2 & $800$ & $749(1)$ & $828(1)$ & $1616(3)$ & $1671(2)$ & $1685(2)$ & $1734(2)$ \\
  Set 3 & $800$ & $661(1)$ & $768(1)$ & $1482(3)$ & $1557(3)$ & $1576(3)$ & $1640(3)$ \\
  \hline \hline
   \end{tabular}
    \caption{Hadron masses in units of [MeV]  and number of configurations for each set.}
\label{tab:hadron_2+1}
 \end{center}
\end{table}
As an application of the method in the previous subsection, let us consider  $BB$ potentials for the $S=-2$ and $I=0$ channel, which consist of the $\Lambda\Lambda$, $N\Xi$ and $\Sigma\Sigma$ components in terms of low-lying octet baryons. Mass differences of these components are quite small such that $2m_\Lambda = 2232$ Mev, $m_N+m_\Sigma = 2257$ MeV and $2m_\Sigma = 2386$ MeV. Using diagonalized source operators, the NBS wave functions  at three different values of energy,
\be
\varphi_{AB}^{W_i}(\br,\bk_{AB}^i)e^{-W_i t} = \langle 0 \vert T\{ \varphi_A(\br+\bx,t)\varphi_B(\br,t)\}\vert W_i \rangle 
\ee
for $i=0,1,2$, are extracted in lattice QCD simulations,  
 where $ AB = \Lambda\Lambda$, $N\Xi$ and $\Sigma\Sigma$, and $\bk_{AB}^i$ satisfies
 \be
 W_i =\sqrt{(\bk_{AB}^i)^2+ m_A^2} + \sqrt{(\bk_{AB}^i)^2+ m_B^2} .
 \ee 
 Using the notation
 \be
K_{AB}^i ( \br) = \frac{1}{2\mu_{AB}}\left(\nabla^2+(\bk_{AB}^i)^2\right)\varphi_{AB}^{W_i}(\br,\bk_{AB}^i)
 \ee
 where $1/\mu_{AB}=1/m_A + 1/m_B$,
the coupled channel $3\times 3$ potential matrix is given by
\be
V_{AB,CD}(\br) = \sum_i K_{AB}^i (\br) \left[\varphi_{CD}^{W_i}(\br,\bk_{CD}^i)\right]^{-1} .
\ee
Here the last factor is the inverse of  the $3\times 3$ matrix $\varphi_{CD}^{W_i}(\br,\bk_{CD}^i)$
with indices $i$ and $CD$.

\begin{figure}[tb]
\begin{center}
 \includegraphics[width=\textwidth]{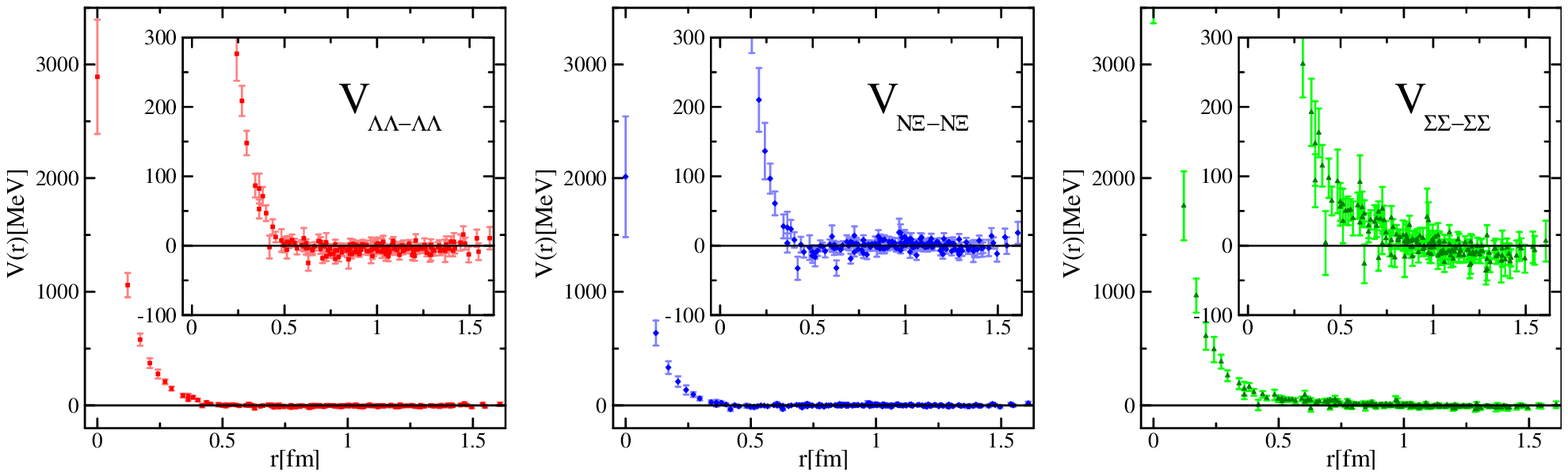}
 \includegraphics[width=\textwidth]{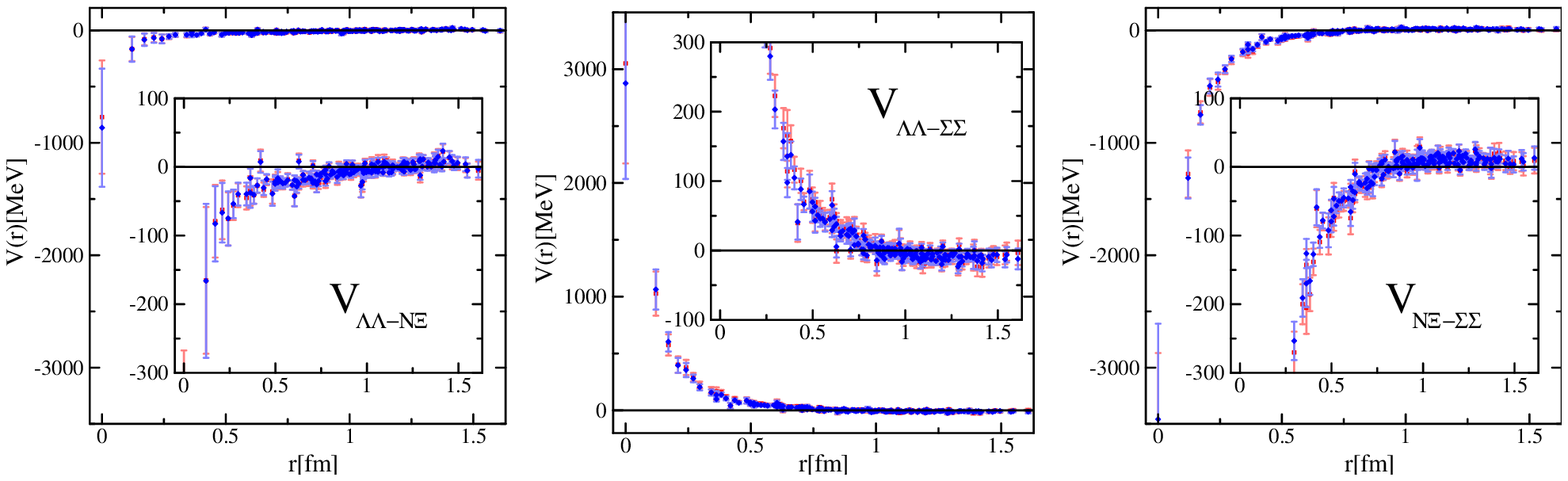}
\end{center}
\caption{The coupled channel potential matrix  from  the NBS wave function for Set 1. The vertical axis is the potential strength in units of [MeV], while the horizontal axis is the relative distance between two baryons in units of [fm]. Taken from Ref.~\protect\cite{Sasaki:2010bh}. }
\label{FIG:PTpbSet1}
\end{figure}
Gauge configurations generated on a $16^3\times 32$ lattice at $a\simeq 0.12$ fm ( therefore $L\simeq 1.9$ fm) in 2+1-flavor full QCD simulation are employed   to calculate the coupled channel potentials at three different values of the light quark mass with the fixed bare strange quark mass\cite{Sasaki:2010bh}.  Quark propagators are calculated with the spatial wall source at $t_0$ with the Dirichlet boundary condition in time at $t=t_0+16$. The wall source is placed at 16 different time slices on each gauge configuration, in order to enhance signals, together with the average over forward and backward propagations in time. 
Corresponding hadron masses and number of gauge configurations are given in table~\ref{tab:hadron_2+1}.

\begin{figure}[tb]
 \begin{center}
 \includegraphics[width=\textwidth]{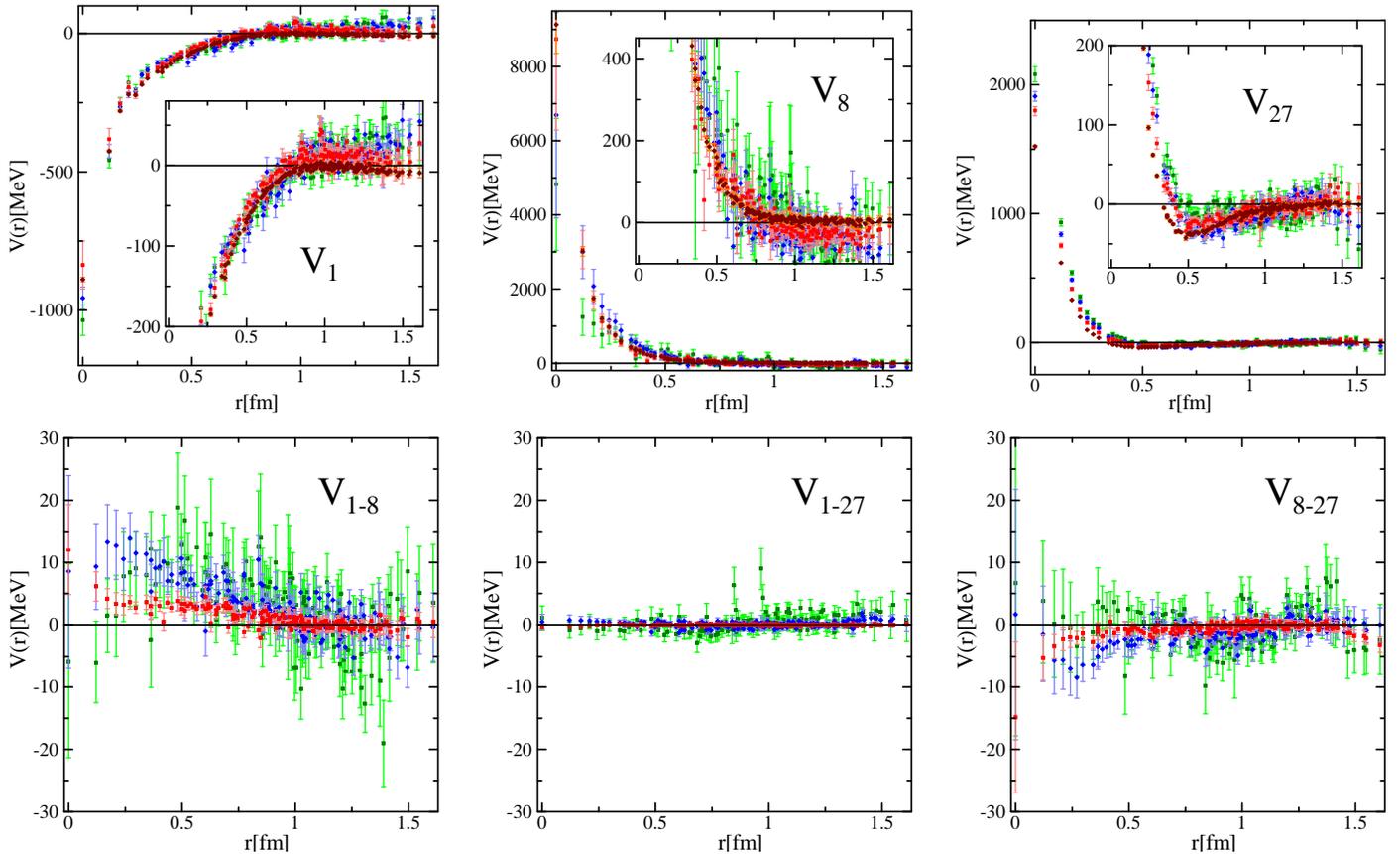}
 \end{center}
\caption{Transition potentials in the flavor SU(3) IR basis. Red, blue and green symbols correspond to results of Set1, Set2 and Set3, respectively. The result of the flavor SU(3) symmetric limit at the same strange quark mass is  also plotted with brown symbols~\protect\cite{Inoue:2010hs}. Taken from Ref.~\protect\cite{Sasaki:2010bh}.}
\label{FIG:potall}
\end{figure}
The coupled channel potential matrix $V_{AB,CD}$ from the NBS wave function for Set 1 is shown in Figure~\ref{FIG:PTpbSet1}. 
The flavor dependence of the height of the repulsive core at short distance region is observed. In particular,
the $\Sigma \Sigma$ potential has the strongest repulsive core of these three channels.
It is interesting to see that  off-diagonal parts of the potential matrix roughly satisfy the hermiticity relation $V_{AB,CD} = V_{CD,AB}$ within statistical errors.
 In addition  the off-diagonal parts are similar in magnitude for  
 $V_{\Lambda \Lambda,\Sigma \Sigma}$ and $V_{N \Xi,\Sigma \Sigma}$ with the diagonal parts, but $V_{\Lambda \Lambda,N \Xi}$ is much smaller. 

In order to compare the results of the potential matrix calculated in three configuration sets, the potentials from the particle basis are transformed to those in the flavor SU(3) irreducible representation (IR) basis as
\begin{eqnarray}
 V^{IR} = U^\dagger V U = 
 \left( \begin{array}{ccc}
 V_{1,1} & V_{1,8} & V_{1,27} \\
 V_{8,1} & V_{8,8} &V_{8,27} \\
 V_{27,1} & V_{27,8} & V_{27,27} \\
 \end{array} \right)
\end{eqnarray}
where $U$ is a unitary transformation matrix whose explicit form is given in Appendix~\ref{app:octet}. The potential matrix in the IR basis is convenient and a good measure of  SU(3) breaking effects by comparing three configuration sets since it is diagonal in the SU(3) symmetric limit.

In Figure~\ref{FIG:potall}, the results of the potential matrix in the IR basis are compared between different configuration sets, together with the one  in the flavor SU(3) symmetric limit. 
As the pion mass decreases, the repulsive core in the $V_{27,27}$ potential increases.
The $V_{1,27}$ and $V_{8,27}$ transition potentials are consistent with zero within statistical errors.
On the other hand, it is noteworthy that the flavor SU(3) symmetry breaking effect becomes manifest in the $V_{1,8}$ transition potential.

\subsection{Time dependent method}
One of the practical difficulties to extract the NBS wave function and the potential from  
the correlation function Eq.(\ref{eq:4-pt}) is to achieve the ground state saturation in numerical simulations at large but finite $t-t_0$ with reasonably small statistical errors.
While the stability of the potential against $t-t_0$ has been confirmed within statistical errors in numerical simulations reviewed in the report, the determination of $W$ for the ground state suffers from systematic errors due to contaminations of possible excited states, which can be seen as follows. There exist 3 different methods to determine $W$. The most well-known method is to determine $W$ from the $t-t_0$ dependence of the correlation function eq.(\ref{eq:4-pt}) summed over $\br$ to pick up the zero momentum state. On the other hand, one can determine $\bk^2$ of $W$ by fitting the $\br$ dependence of the NBS wave function with its expected asymptotic behavior at large $r$ or by reading off the constant shift of the Laplacian part of the potential from zero at large $r$.
Although the latter two methods usually give consistent results within statistical errors, the first method (the $t$ dependence method) sometimes leads to a result different from those determined by the latter two (called the $\br$ dependent method together) at the value of $t-t_0$ usually employed in numerical simulations. Although, in principle,  the increase of $t-t_0$ is needed in order to see an agreement between $t$ and $\br$ dependence methods, it is difficult in practice due to larger statistical errors at larger $t-t_0$ for the two-baryon system.

In order to overcome this practical difficulty, the method to extract the potential from the NBS wave function has been modified as follows.  Let us consider the correlation function Eq.(\ref{eq:4-pt}) again:
\be
F(\br,t) = \sum_{W\le W_{\rm th} } A_W \phi^W(\br ) e^{-W t} + O(e^{-W_{\rm th}t}) .
\ee
If $t$ is large enough so that contributions from $O(e^{-W_{\rm th}t})$ terms can be neglected\footnote{This limitation for $t$ can be loosened if the coupled channel potentials are introduced as in the previous subsections.}, we have
\be
H_0 F(\br, t) \simeq \sum_W A_W\int d^3\br^\prime [E_W\delta^{(3)}(\br-\br^\prime) - U(\br,\br^\prime)]\varphi^W(\br^\prime) e^{-W t}
\ee
where $E_W=k_W^2/(2\mu) = (W^2-4m_N^2)/(4m_N)$ with $\mu=m_N/2$. By using the non-relativistic approximation that $W=2\sqrt{k_W^2+m_N^2} =2m_N + k_W^2/m_N + O(k_W^4/m_N^3)$,
\be
\left[ H_0 +\frac{d}{d t} +2m_N\right] F(\br, t) =- \int d^3\br^\prime U(\br,\br^\prime) F(\br, t)
\simeq -V^{\rm LO}(\br) F(\br, t)
\ee
where the velocity expansion is introduced in the last line and higher other than the leading order terms are then omitted. The leading order potential is therefore given by
\be
V^{\rm LO}(\br) = -\frac{\left[ H_0 +\frac{d}{d t} +2m_N\right] F(\br, t) }{F (\br, t)}
\ee
or
\be
V^{\rm LO}(\br) = -\frac{\left[ H_0 +\frac{d}{d t} \right] R(\br, t) }{R (\br, t)}
\ee
where $R(\br, t) = F (\br, t)/e^{-2 m_N t}$.  Here it is assumed that $O(e^{-W_{\rm th} t})$ contributions can be neglected at large $t$.
The non-relativistic formula for $V^{\rm LO}(r)$ above can be easily generalized to the case that masses of two particles are different, by the replacement that $R(\br, t) = F (\br, t)/e^{- (m_A+m_B) t}$.
Note also that the potential extracted in this method automatically satisfies that $V^{\rm LO}(r) \rightarrow 0$ as $r\rightarrow 0$ without the constant shift. 
This property may be used to check whether this extraction works correctly or not.

On the lattice, the $t$ derivative should be approximated by the $t$ difference. In practice, one may adopt a particular method for the $t$ difference,  
in order to reduce statistical as well as systematic errors for $V^{\rm LO}(\br)$.

The non-relativistic approximation can be removed by using the second order derivative in $t$ as
\be
V^{\rm LO}(\br) =\frac{ \displaystyle \left[ -H_0 +\frac{1}{4m_N}\frac{d^2}{d t^2} -m_N\right] F(\br, t)}{ F (\br, t)} ,
\ee
as long as $O(e^{-W_{\rm th} t})$ contributions are negligibly small. For this method to apply, two particles should have the same mass.
Statistical errors of the second order difference on the lattice must be kept small in numerical simulations. 

One may introduce a more general correlation function as
\be
F(\bx , \by, t) = \int d^3\bx_1d^3\by_1\langle 0 \vert T\{N(\bx_1+\bx,t) N(\bx_1,t)\} T\{\overline{N}(\by_1+\by,0) \overline{N}(\by_1,0)\} \vert 0\rangle .
\ee  
Using this new quantity, we have  
\be
\left[ H_0 -\frac{1}{4m_N}\frac{d^2}{d t^2} +m_N\right] R(\bx,\by, t) = -\int d^3\,\bz U(\bx,\bz) F(\bz, \by,t),
\ee
from which the non-local potential is extracted as
\be
U(\bx,\by) = \int d^3\bz \left[- H_0 +\frac{1}{4m_N}\frac{d^2}{d t^2} -m_N\right] F(\bx,\bz, t) \cdot
\tilde F^{-1}(\bz,\by, t).
\ee
Here $\tilde F^{-1}(\bx,\by, t)$ is  the approximated inverse of the hermitian operator $F (\bx,\by, t)$, defined by
\be
\tilde F^{-1}(\bx,\by, t) =\sum_{\lambda_n\not= 0 } \frac{1}{\lambda_n(t)} v_n(\bx, t) v_n^\dagger(\by,t)
\ee
where $\lambda_n(t)$ and $v(\bx,t)$ are an eigenvalue and a corresponding eigenvector of $F (\bx,\by, t)$, respectively, and  zero eigenvalues are removed in the summation.
Since the modified potential
\be
\hat U(\bx,\by) =  U(\bx,\by) + \sum_{\lambda_n=0} c_n v_n(\bx, t) v_n^\dagger(\by,t)
\ee 
also satisfies the same Schr\"odinger equation $\forall\{c_n \}$, the non-local potential is NOT unique, and $U(\bx,\by)$ is scheme dependent, as discussed before.

\subsection{Bound $H$ dibaryon in flavor SU(3) limit}
As an application of the method in the previous subsection, let us consider the singlet potential in the flavor SU(3) limit in order to investigate whether the bound $H$ dibaryon exists or not in this case.  

At the leading order of the velocity expansion, the central potential is defined in this method by
\be
V_C^{(X)} (r) = -\frac{ \left[H_0 +\displaystyle \frac{d}{d t}\right] R(\br,t-t_0)}{ R(\br,t-t_0)} ,
\ee
which is calculated on $16^3\times 32$, $24^3\times 32$ and $32^3\times 32$ lattices at $a=0.121(2)$ fm and three values of the quark mass, where  the PS meson mass and the octet baryon mass are given by $(m_{\rm PS}, m_B) = (1015(1){\rm MeV}, 2030(2){\rm MeV} )$, $(837(1){\rm MeV}, 1748(1){\rm MeV} )$ and $(673(1){\rm MeV}, 1485(2){\rm MeV} )$ on a $32^3\times 32$ lattice\cite{Inoue:2010es}.

\begin{figure}[tb]
\begin{center}
\includegraphics[width=0.45\textwidth]{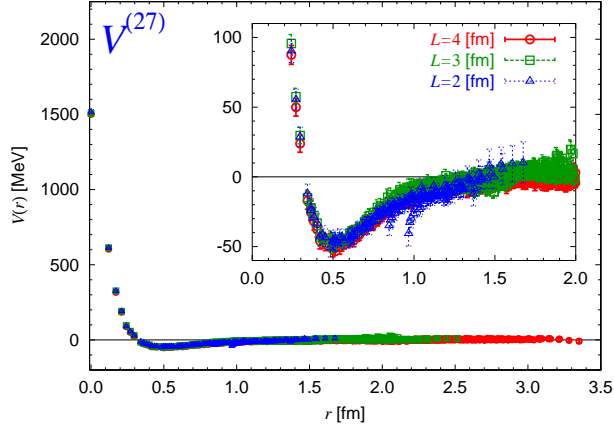}
\end{center}
\vspace{-0.5cm}
\caption{The flavor 27-plet potential $V_C^{(27)}(r)$ obtained for $L=1.94,\, 2.90,\, 3.87$ fm
at $m_{\rm ps}=1015$ MeV and $(t-t_0)/a = 10$. Taken from Ref.~\protect\cite{Inoue:2010es}.
}
\label{fig:27-plet}
\end{figure} 
To check the qualitative consistency with previous results, the central potential in the 27-plet channel is plotted in Fig.~\ref{fig:27-plet} obtained in three different lattice volumes with $L=1.94,\,2.90,\,3.87$ fm at $m_{\rm ps}=1015$ MeV and $(t-t_0)/a=10$. 
This is the case corresponding to the spin-singlet NN potential.
 Compared with statistical errors, the $L$ dependence is found to be negligible. 
 The $t$ dependence is also small as long as $(t-t_0)/a \geq 9$. 
As expected,  the potential approaches zero automatically for large $r$.  The figure shows a repulsive core at short distance  surrounded by an attractive well at medium and long distances, which is  qualitatively consistent with previous results in quenched and  full QCD simulations.

\begin{figure}[tb]
\begin{center}
\includegraphics[width=0.45\textwidth]{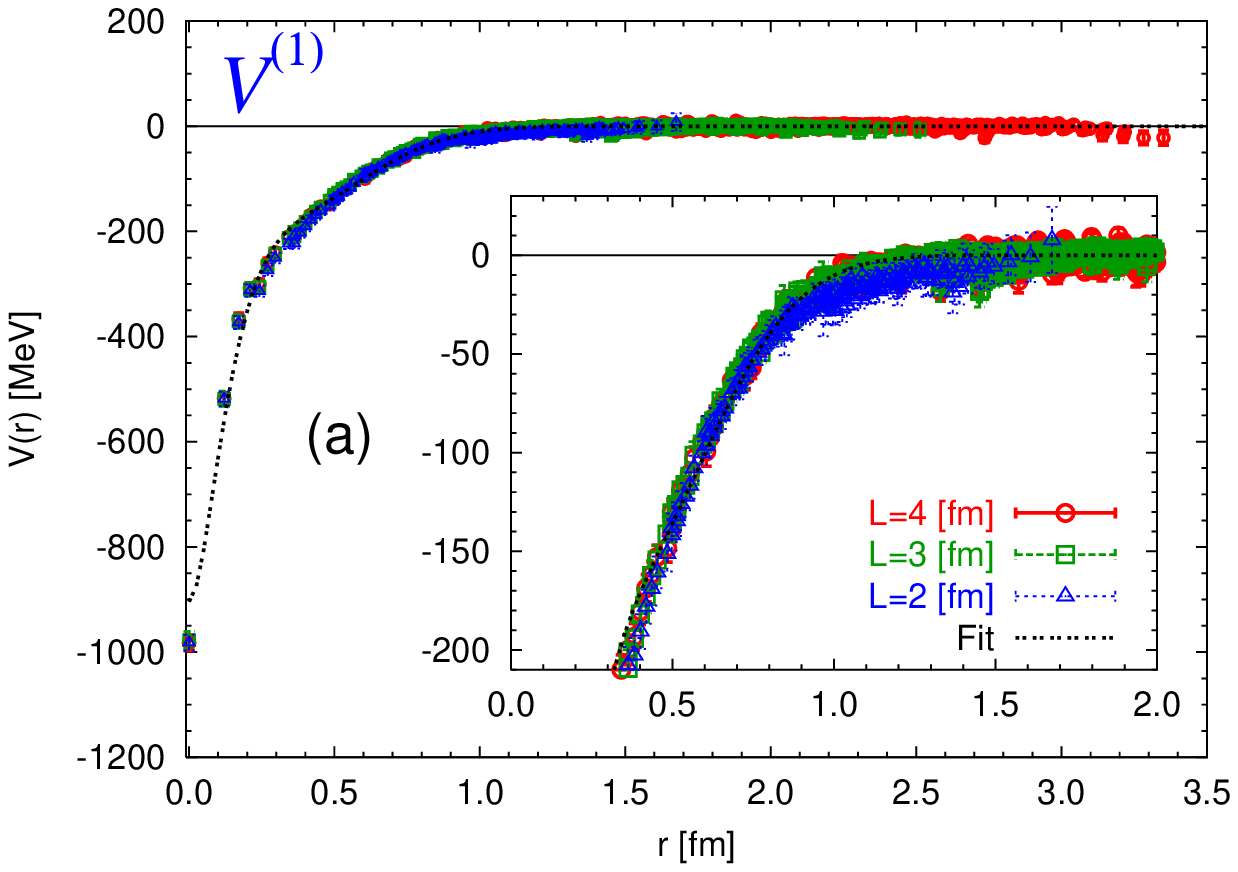}
\includegraphics[width=0.45\textwidth]{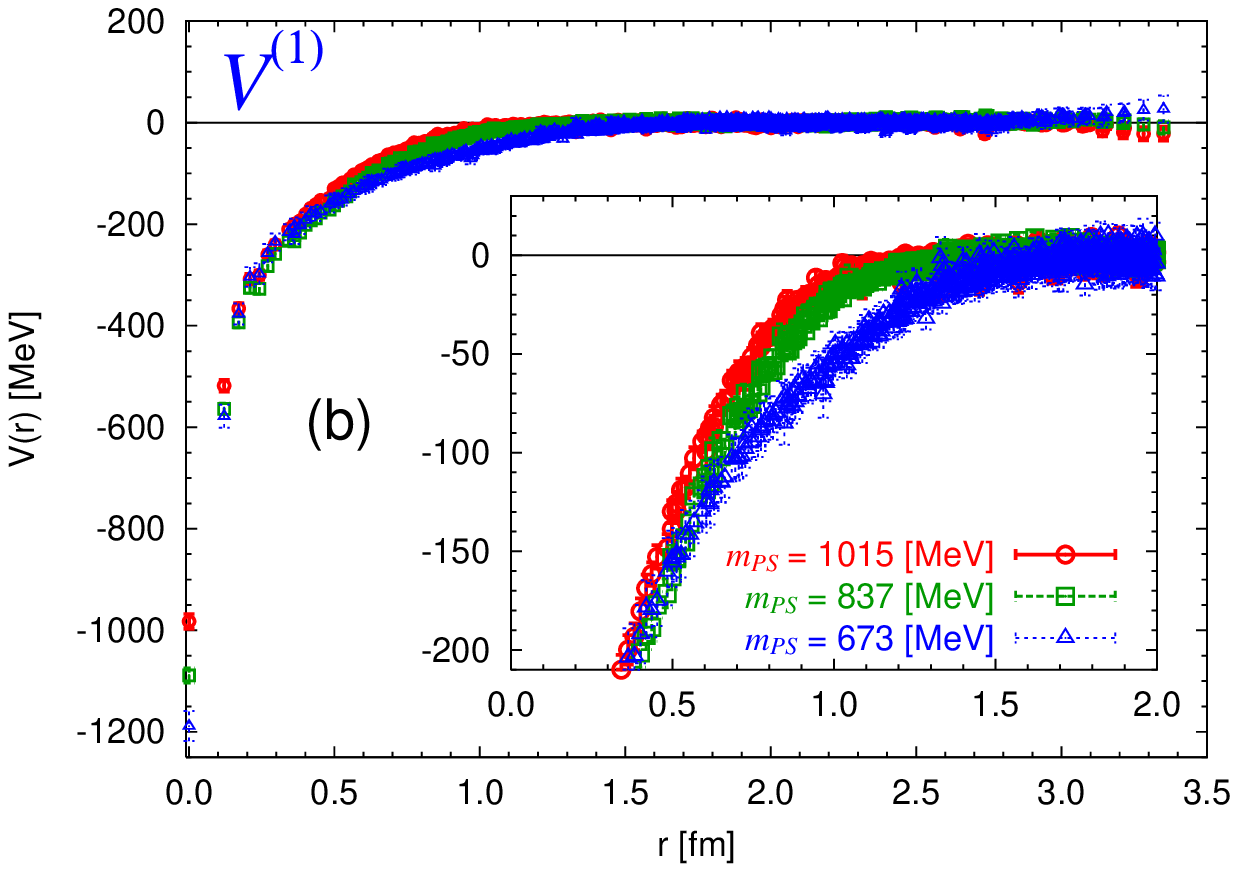}
\end{center}
\vspace{-0.5cm}
\caption{ The flavor-singlet potential $V_C^{(1)}(r)$ at $(t-t_0)/a = 10$. 
 (Left) Results for $L=1.94,\, 2.90,\, 3.87$ fm at $m_{\rm ps}=1015$ MeV.
 (Right) Results for $L=3.87$ fm at  $m_{\rm ps}=1015,\, 837,\, 673$ MeV. 
  Taken from Ref.~\protect\cite{Inoue:2010es}. }
\label{fig:singlet}
\end{figure}
Shown in Fig.~\ref{fig:singlet}(Left) and  Fig.~\ref{fig:singlet}(Right)
 are the volume and the quark mass dependences 
 of the central potential in the flavor-singlet channel $V_C^{(1)}(r)$, respectively, at $(t-t_0)/a=10$
 where the potentials do not have appreciable change with respect to the  choice of $t$.
 The flavor-singlet potential is shown to have an ``attractive core" and to be well localized in space.
 Because of the latter property, no significant volume dependence of the potential is observed within the statistical errors, as seen in Fig.~\ref{fig:singlet}(Left).
 As  the quark mass decreases in Fig.~\ref{fig:singlet}(Right) , the long range part of the attraction tends to increase. 
 
The resultant potential is fitted by the following analytic function composed of
 an attractive Gaussian core plus a long range (Yukawa)$^2$ attraction:
$
  V(r) = b_1 e^{-b_2\,r^2} + b_3(1 - e^{-b_4\,r^2})\left( {e^{-b_5\,r}}/{r} \right)^2 .
$
 With the five parameters, $b_{1,2,3,4,5}$,   the lattice results can be fitted
 reasonably well with $\chi^2/{\rm dof} \simeq 1$.
 The fitted result for $L=3.87$ fm is shown by the dashed line in Fig.~\ref{fig:singlet}(Left). 
 
\begin{figure}[tb]
\begin{center}
\includegraphics[width=0.45\textwidth]{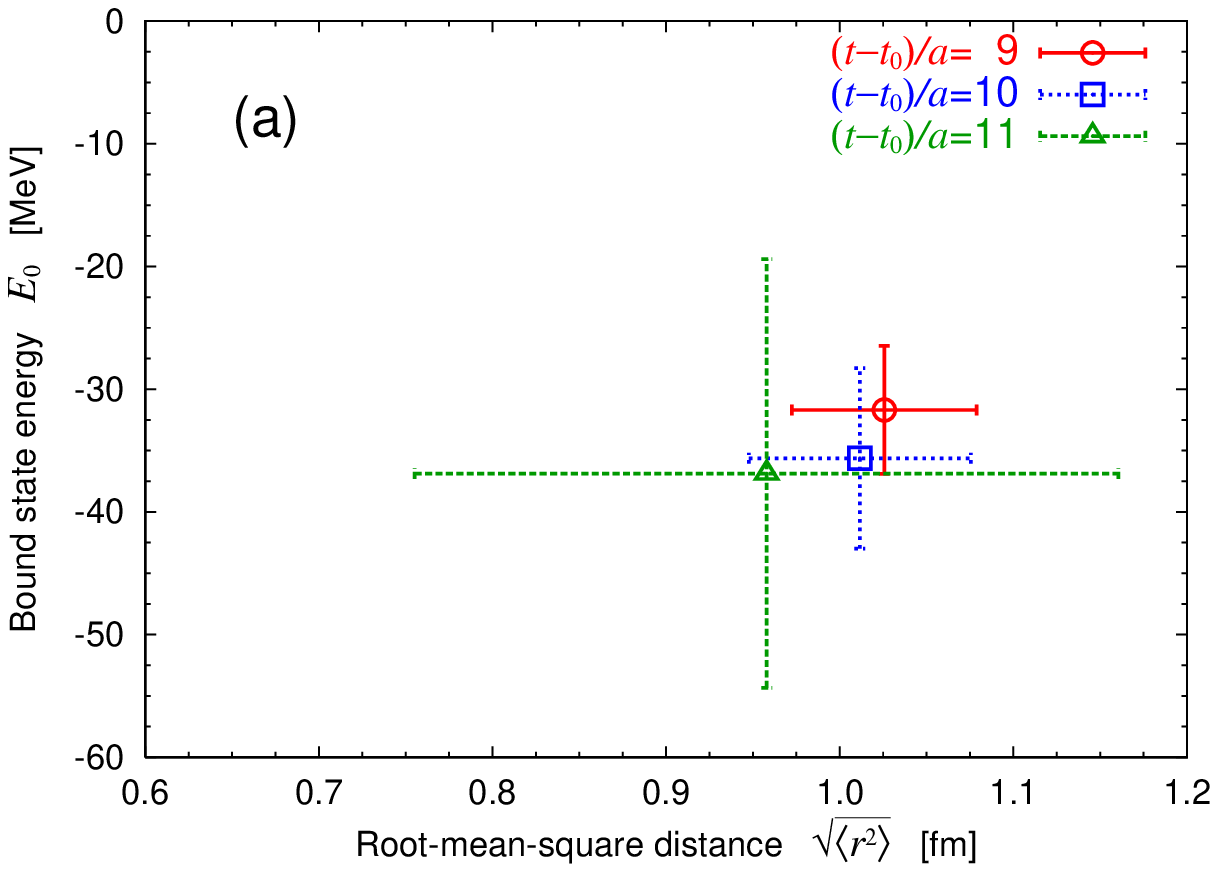}
\vspace{0.3cm}
\includegraphics[width=0.45\textwidth]{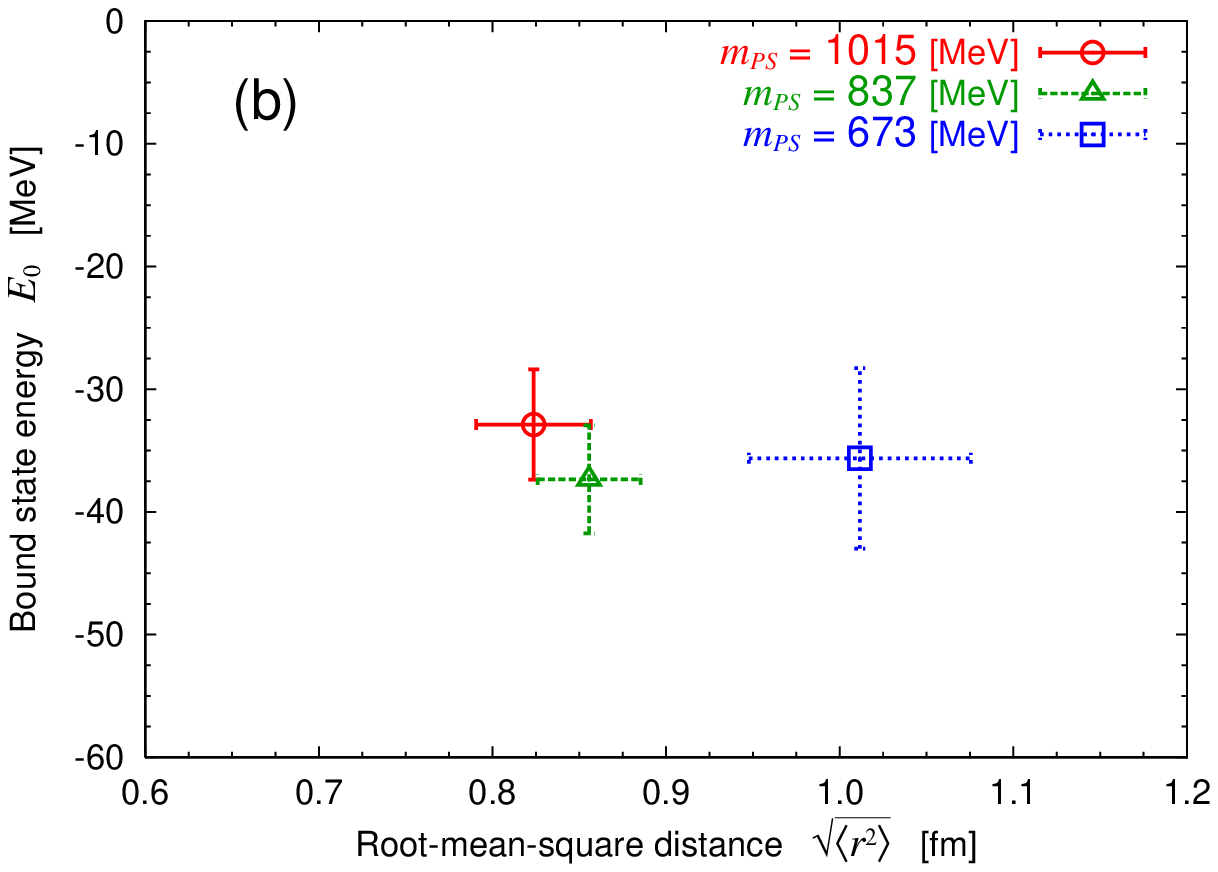}
\end{center}
\vspace{-0.5cm}
\caption{The bound state energy $E_0\equiv -B_H$ and the rms distance $\sqrt{\langle r^2\rangle}$ of the $H$-dibaryon 
  obtained from the potential at $L=3.87$ fm.
(Left) $(t-t_0)/a$ dependence at $m_{\rm ps}=673$ MeV. 
 (Right) Quark mass dependence at $(t-t_0)/a=10$. Taken from Ref.~\protect\cite{Inoue:2010es}.}
\label{fig:H-dibaryon}
\end{figure}

Solving the Schr\"{o}dinger equation with the fitted potential in infinite volume,
the energies and the wave functions are obtained at the present quark masses in the flavor SU(3) limit.
 It turns out that, at each quark mass, there is only one bound state
 with binding energy 30--40 MeV.  
 In Fig.~\ref{fig:H-dibaryon}(Left), 
 the energy and the root-mean-squared (rms) distance
 of the bound state  are plotted in the case of 
  $(t-t_0)/a=9, 10, 11$ at $m_{\rm ps}=673$ MeV and  $L=3.87$ fm, where
  errors are estimated by the jackknife method.  
  Although the statistical error increases as $t$ increases,
 we observe small changes of central values, which are considered as the systematic errors.
 Fig.~\ref{fig:H-dibaryon}(Right) shows  the energy and the rms distance of the bound state at each quark mass obtained from the potential at  $L=3.87$ fm and $(t-t_0)/a=10$. 
 Despite the fact that the potential has quark mass dependence,
 the resultant binding energies of the $H$-dibaryon are insensitive in the present range of the quark masses.
 This is due to the fact that the increase of the  attraction toward the lighter quark
 mass is partially compensated by the increase of the kinetic energy for the lighter baryon mass. 
 It is noted that there appears no bound state for the potential of the 27-plet channel 
 in the present range of the quark masses.
   
The final results of the binding energy $B_H$ and the rms distance $\sqrt{\langle r^2\rangle}$
are summarized below, where the 1st and 2nd parentheses  correspond to statistical errors 
and systematic errors from the $t$-dependence, respectively.
\begin{eqnarray}
m_{ps}=1015 ~\mbox{MeV} :~ \  B_H &=& 32.9 (4.5)(6.6) ~\mbox{MeV} \nonumber \\
~~  \sqrt{\langle r^2 \rangle} &=& 0.823(33)(40)  ~\mbox{fm} \nonumber \\
 m_{ps}=~837 ~\mbox{MeV} :~ \  B_H &=& 37.4 (4.4)(7.3) ~\mbox{MeV} \nonumber \\
~~  \sqrt{\langle r^2 \rangle} &=& 0.855(29)(61)  ~\mbox{fm}  \nonumber \\
 m_{ps}=~673 ~\mbox{MeV} :~ \ B_H &=& 35.6 (7.4)(4.0) ~\mbox{MeV}
 \nonumber \\
  ~~  \sqrt{\langle r^2 \rangle} &=& 1.011(63)(68)  ~\mbox{fm}
 \nonumber
\end{eqnarray} 
 Since the binding energy is insensitive to the quark masses,
 there may be a possibility of weakly bound or resonant $H$-dibaryon even in the real world with lighter quark masses and the flavor SU(3) breaking. To make a definite conclusion on this point,
 the $\Lambda\Lambda-N\Xi-\Sigma\Sigma$ coupled channel analysis is necessary
 for $H$ in the (2+1)-flavor lattice QCD simulations, as discussed in Sec.~\ref{sec:coupled_channel}. 

Recently the existence of $H$-dibaryon is also suggested by a direct calculation of its binding energy in 2+1 full QCD simulations\cite{Beane:2010hg}, where $B_H=16.6(2.1)(4.6)$ MeV is reported in the $L\rightarrow \infty$ extrapolation at $m_\pi \simeq 389$ MeV .

\section{Other applications}
\label{sec:others}
In the last section, some other applications of the potential method are reviewed.

\subsection{Three nucleon force}
Recent precise calculations of  few-nucleon systems clearly indicate that the 2 nucleon force alone is insufficient to understand the nuclei,  which calls for three (and/or more) nucleon forces.
Actually, the three nucleon force (TNF) is supposed to play an important and nontrivial role in various phenomena in nuclear and astrophysics.
For the binding energies of light nuclei,  the attractive TNF is required to reproduce the experimental data. On the other hand, the repulsive TNF is necessary to reproduce the empirical saturation density of  symmetric nuclear matter. For the equation of state(EoS) of asymmetric nuclear matter, a repulsive TNF is required
to explain the observed maximum neutron star mass.

Pioneered by Fujita-Miyazawa~\cite{Fujita:1957zz},
the TNF has been mainly studied from the two-pion exchange picture
with the $\Delta$-excitation.
In addition, the repulsive TNF is often introduced 
phenomenologically~\cite{Pudliner:1995wk}.
Recently, the TNF based on chiral effective field theory is developing~\cite{vanKolck:1994yi},
but the unknown low-energy constants can be obtained 
only by the fitting to the experimental data.
Since the TNF  originates from the fact that the nucleon is not a fundamental particle,
it is essential to study the TNF from fundamental degrees of freedom(DoF) , i.e., quarks and gluons.

In Ref.~\cite{Doi:2010yh,Doi:2011gq} 
such first-principle calculations of the TNF in lattice QCD have been reported.
If the potential method is applied to the three nucleon (3N) system,
a straightforward calculation is impossible due to the significantly enlarged DoF. In Ref.~\cite{Doi:2010yh,Doi:2011gq}, two different approaches have been considered.

The NBS wave function for 3N is defined by
\be
\varphi^W(\br_{12},\br_{123})e^{-Wt} = \langle 0 \vert N(\bx_1,t) N(\bx_2,t) N(\bx_3,t) \vert W \rangle
\ee 
where $\br_{12}\equiv \bx_1-\bx_2$, $\br_{123}\equiv \bx_3-(\bx_1+\bx_2)/2$ are the Jacobi coordinates, and 
$\vert W\rangle $ is the 3N state with energy $W$.
At the leading order of the velocity expansion, the NBS wave function satisfies
\be
\left[ -\frac{1}{2\mu_{12}}\nabla^2_{r_{12}} -\frac{1}{2\mu_{123}}\nabla^2_{r_{123}} + \sum_{i<j}
V_{{\rm 2N},ij}(\bx_i-\bx_j) + V_{\rm TNF}(\br_{12},\br_{123})
\right] \varphi^W(\br_{12},\br_{123}) = E \varphi^W(\br_{12},\br_{123})
\ee
where $V_{{\rm 2N},ij}(\bx_i-\bx_j) $ denotes the potential between $(i,j)$-pair,  $V_{\rm TNF}(\br_{12},\br_{123})$ the TNF, $\mu_{12}=m_N/2$, $\mu_{123}=2m_N/3$ the reduced masses. 
If $\varphi^W(\br_{12},\br_{123})$ is calculated for all $\br_{12},\br_{123}$ and all $V_{{\rm 2N},ij}(\bx_i-\bx_j) $ are available by lattice calculations, $V_{\rm TNF}(\br_{12},\br_{123})$ can be extracted.
Unfortunately, this is not the case:
Since both $\br_{12}$ and $\br_{123}$ have $L^3$ DoF,
the calculation cost is more expensive by a factor of $L^3$
compared to the 2N system.
Furthermore, the number of diagrams 
to be calculated in the Wick contraction
tends to diverge with a factor of $N_u ! \times N_d !$
($N_{u,d}$ are numbers of u,d quarks in the system).
It is also noted that not all 2N potentials are available in lattice QCD at this moment:
Only parity-even 2N potentials have been obtained so far.

The first method in Ref.~\cite{Doi:2010yh} to avoid these problems is to consider the effective 2N potential in the 3N system by taking the summation over the location of the spectator nucleon $N(\bx_3)$,
\be
\varphi^W(\br_{12}) = \sum_{\bx_3}\varphi^W(\br_{12},\br_{123})= \sum_{\br_{123}}\varphi^W(\br_{12},\br_{123}) .
\ee
The effective potential between $N(\bx_1)$ and $N(\bx_2)$ is then defined by
\be
\left[ -\frac{1}{2\mu_{12}}\nabla^2_{r_{12}}  +V_{\rm eff}(\br_{12})
\right] \varphi^W(\br_{12}) = E \varphi^W(\br_{12}) .
\ee
In this calculation, the DoF of $\br_{123}$ is integrated out beforehand, and thus
the calculation cost is reduced by a factor of  $\sim 1/L^3$, compared to the straightforward calculation. The difference $V_{\rm eff} (\vec{r}) - V_{2N}(\vec{r})$ can be considered to be the ``finite density effect'' in the 3N system. Part of this effect is attributed 
to the genuine 2N potential with the nontrivial 3N correlation,
while another originates from the genuine TNF.

The triton channel( $I=1/2$, $J^P= 1/2^+$) is studied as the 3N system. 
Since the spectator nucleon is projected to the S-wave, the possible quantum numbers between the (effective) 2N are only $^{2S+1}L_J =$ $^1S_0$, $^3S_1$, $^3D_1$.
Gauge configurations in 2-flavor QCD on a $16^3\times 32$ lattice
at $a\simeq 0.16$ fm\cite{Aoki:2002uc} are employed for the calculation at $m_\pi \simeq 1.13$ GeV and $m_N\simeq 2.15$ GeV.

\begin{figure}[tb]
\begin{center}
\includegraphics[width=0.32\columnwidth,angle=270]{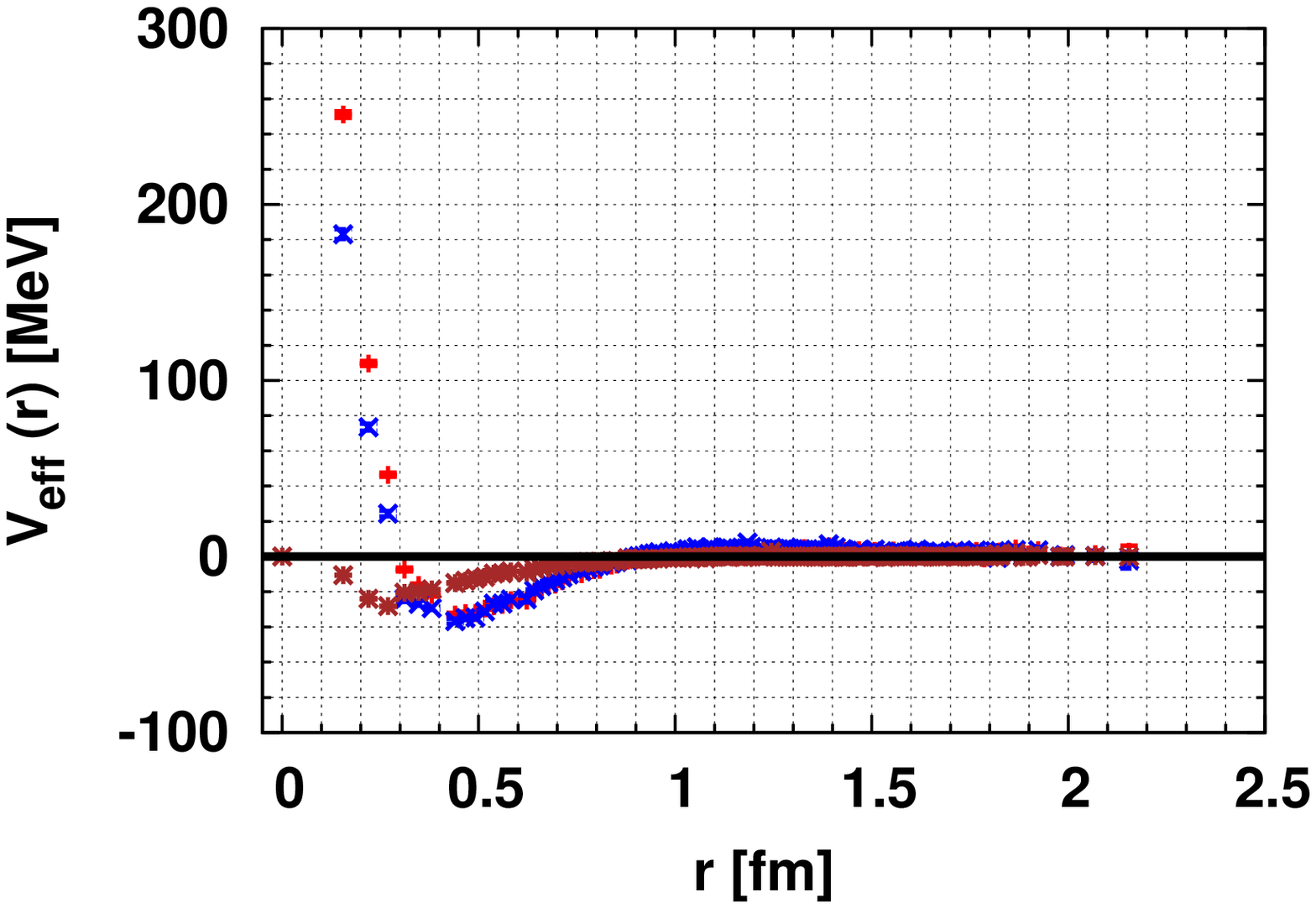}
\includegraphics[width=0.32\columnwidth,angle=270]{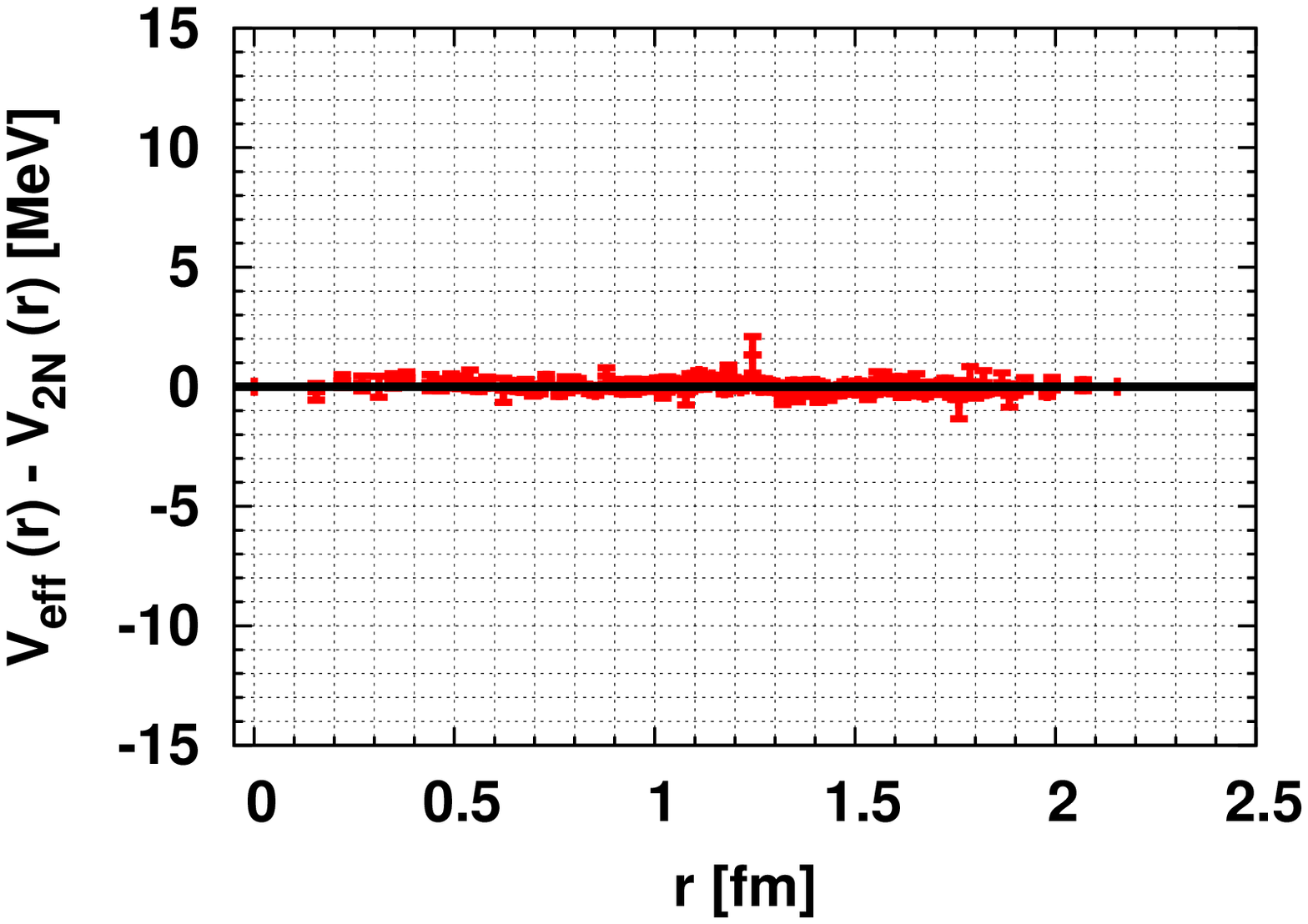}
\end{center}
\caption{(Left) Effective 2N potentials,
where red, blue, brown points correspond to 
$V_{C,{\rm eff}}^{I=1,S=0}$,
$V_{C,{\rm eff}}^{I=0,S=1}$,
$V_{T,{\rm eff}}^{I=0,S=1}$
potential, respectively.
(Right) The difference between the effective 2N and the genuine 2N
for $V_{T}^{I=0,S=1}$ potential.
Taken from Ref.~\protect\cite{Doi:2010yh}. }
\label{fig:pot_eff2N}
\end{figure} 
Fig.~\ref{fig:pot_eff2N}(Left) show  results for $V_{\rm eff}(r)$
in the triton channel at $t-t_0=8$, where the constant shift by energy is not included
for the central potentials. It is noteworthy that $V_{\rm eff}(r)$ is obtained with good precision even though the signal to noise ratio is expected to be worse for more quarks in the system.
Fig.~\ref{fig:pot_eff2N}(Right) gives $V_{\rm eff}(r) - V_{\rm 2N}(r)$ for the tensor potential, which is free from the constant shift. The difference is consistent with zero within a few MeV statistical errors. Similar results are reported for central potentials as well in Ref.\cite{Doi:2010yh}.
There is no indication of a TNF effect. A possible explanation is  that the TNF effect is suppressed at heavy quark mass. Basically similar results are obtained,  however, for lighter pion masses($m_\pi \simeq 0.7$ GeV and 0.57 GeV)\cite{Doi:2010yh}.  Another possibility is that the TNF effect is suppressed by the summation over the location of the spectator nucleon.
While the TNF effect is expected to be enhanced when all three nucleons are close to each other, such 3-dimensional spacial configurations have small contributions in the spectator summations.

In order to assess this possibility, a second method has been investigated   in Ref.\cite{Doi:2010yh,Doi:2011gq},  where the linear setup with $\br_{123}=0$ is used for the 3N wave function.
In this case, the third nucleon is attached to $(1,2)$-nucleon pair with only S-wave.
Considering the total 3N quantum numbers of $I=1/2, J^P=1/2^+$, 
the wave function can be completely spanned by only three bases, which can be labeled
by the quantum numbers of $(1,2)$-pair as $^1S_0$, $^3S_1$, $^3D_1$.
Therefore, the Schr\"odinger equation is  simplified to
the $3\times 3$ coupled channel equations with the bases of 
$\varphi_{^1S_0}$, $\varphi_{^3S_1}$, $\varphi_{^3D_1}$.
Even in this case the subtraction of $V_{\rm 2N}$ remains nontrivial:
the parity-odd potentials, which must be subtracted, are not available in lattice QCD at this moment.
The subtraction problem of parity-odd potentials can be avoided for triton by using the symmetric wave function,
\be
\varphi_S \equiv
\frac{1}{\sqrt{6}}
\Big[
-   \Pu \Nu \Nd + \Pu \Nd \Nu               
                - \Nu \Nd \Pu + \Nd \Nu \Pu 
+   \Nu \Pu \Nd               - \Nd \Pu \Nu
\Big]  .
\label{eq:psi_S}
\ee
Combined with the Pauli principle,
it is automatically guaranteed that any 2N-pairs couple with even parity only,
since this wave function is anti-symmetric in spin/isospin spaces for any 2N-pairs.
Therefore the TNF can be extracted unambiguously in this channel,
without the information of parity-odd 2N potentials.

\begin{figure}[tb]
\begin{center}
\includegraphics[width=0.32\columnwidth,angle=270]{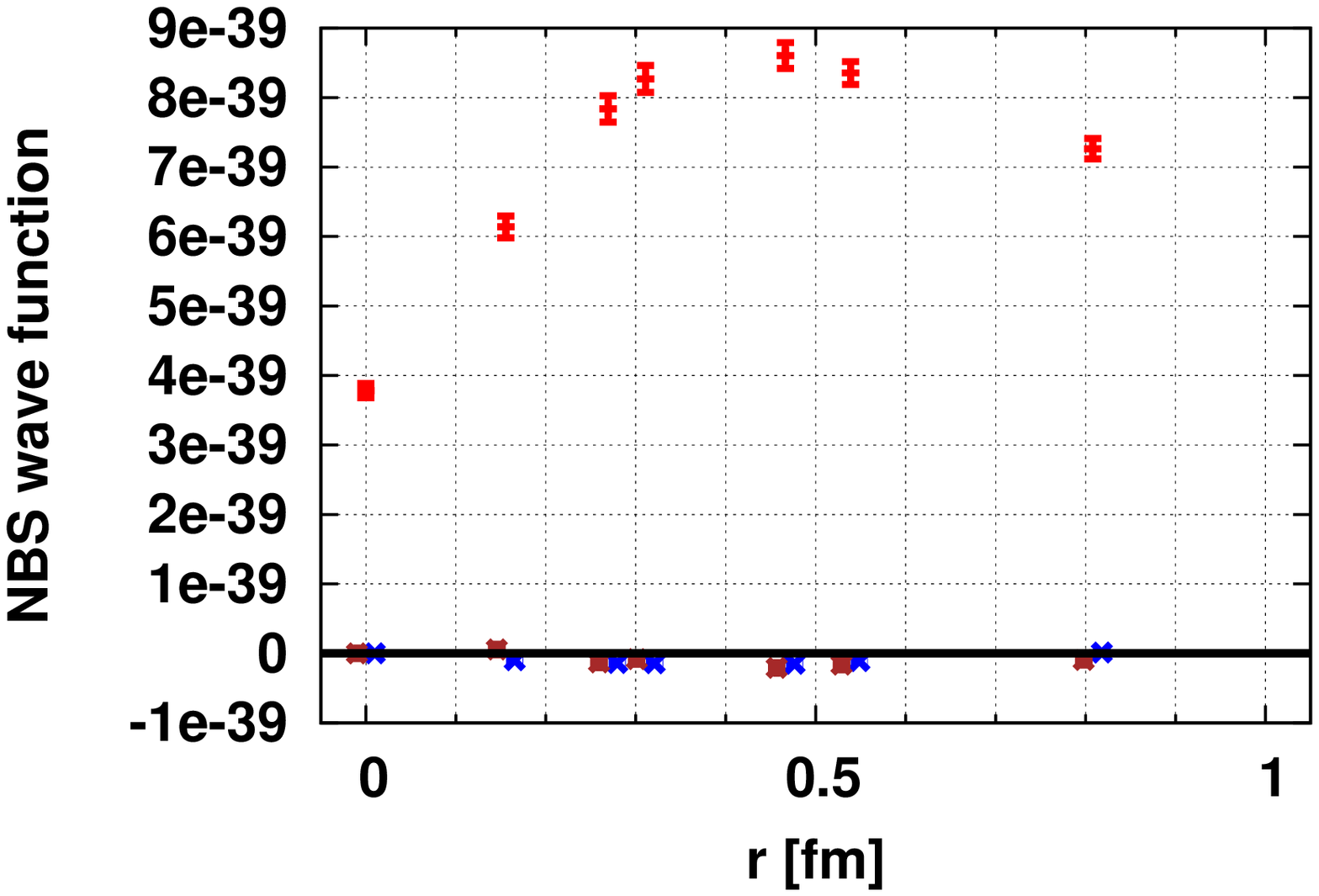}
\includegraphics[width=0.32\columnwidth,angle=270]{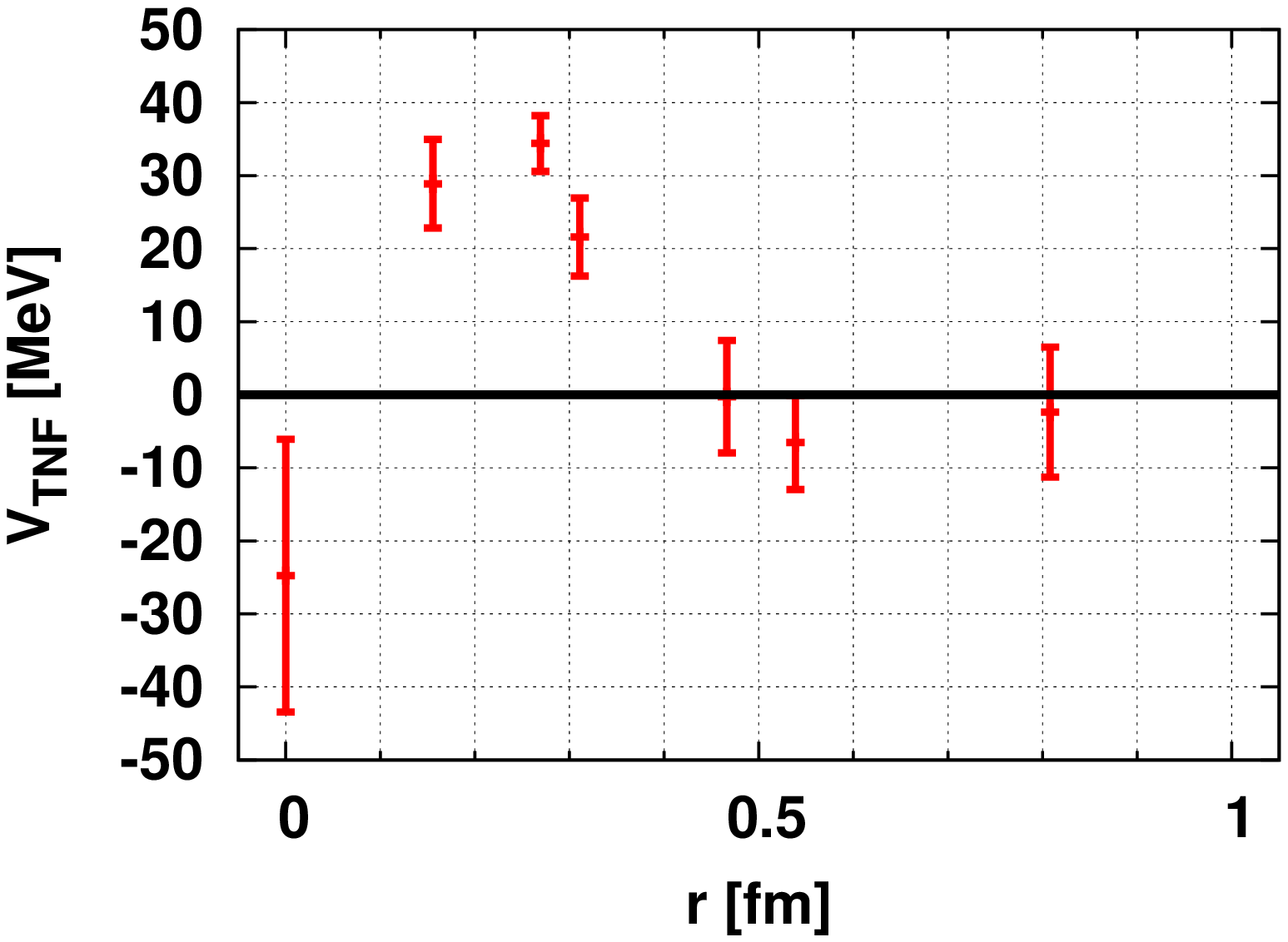}
\end{center}
\caption{(Left) The wave function with linear setup in the triton channel.
Red, blue, brown points correspond to
$\varphi_S$, $\varphi_M$, $\varphi_{^3D_1}$, respectively.
(Right) The scalar/isoscalar TNF in the triton channel,
plotted against the distance $r=\vert \br_{12}/2\vert$ in the linear setup.
Taken from Ref.~\protect\cite{Doi:2010yh}. }
\label{fig:TNR}
\end{figure} 
The same gauge configurations used for the effective 2N potential study are employed in the numerical simulations.
Fig.~\ref{fig:TNR}(Left) gives each wave function of
$\varphi_S= \frac{1}{\sqrt{2}} ( - \psi_{^1S_0} + \psi_{^3S_1} )$, $\varphi_M\equiv \frac{1}{\sqrt{2}} ( + \psi_{^1S_0} + \psi_{^3S_1} )$, $\psi_{^3D_1}$ as a function of $r=\vert \br_{12}/2\vert$ in the triton channel at $t-t_0 = 8$.
Among these three 
$\varphi_S$ dominates the wave function, since 
$\varphi_S$ contains the component for which
all three nucleons are in S-wave.

By subtracting the $V_{2N}$ from the total potentials in the 3N system,
the TNF is detemined.
Fig.~\ref{fig:TNR} (Right) shows results
for the scalar/isoscalar TNF, where the $r$-independent shift of $V_{2N}$ is not included,
and thus about ${\cal O}(10)$ MeV systematic error is understood.
There are various physical implications in Fig.~\ref{fig:TNR} (Right).
At the long distance region of $r$, the TNF is small as is 
expected.
At the short distance region, 
the indication of a repulsive TNF is observed.
Recalling that the repulsive short-range TNF is phenomenologically required 
to explain the saturation density of nuclear matter, etc.,
this is a very encouraging result.
Of course, further study is necessary to confirm this result,
e.g., the study of the ground state saturation,
the evaluation of the constant shift by energies,
the examination of the discretization error.

\subsection{Meson-baryon interactions}
The potential method can be naturally extended to the meson-baryon systems and the meson-meson systems. In this subsection, two applications of the potential method to the meson-baryon system are discussed.

The first application is the study of the $K N$ interaction in the $I(J^P) =0(1/2^-)$ and $1(1/2^-)$ channels by the potential method. These channels may be relevant for the possible exotic state $\Theta^+$, whose existence is still controversial.

The $KN$ potentials in isospin $I=0$ and $I=1$ channels have been calculated in $2+1$ full QCD simulations, employing 700 gauge configurations on a $16^3\times 32$ lattice at $a=0.121(1)$ fm and $(m_\pi, m_K, m_N)=(871(1),912(2),1796(7))$ in unit of MeV\cite{Ikeda:2010sg}.

\begin{figure}[tb]
\begin{center}
\includegraphics[width=0.45\columnwidth]{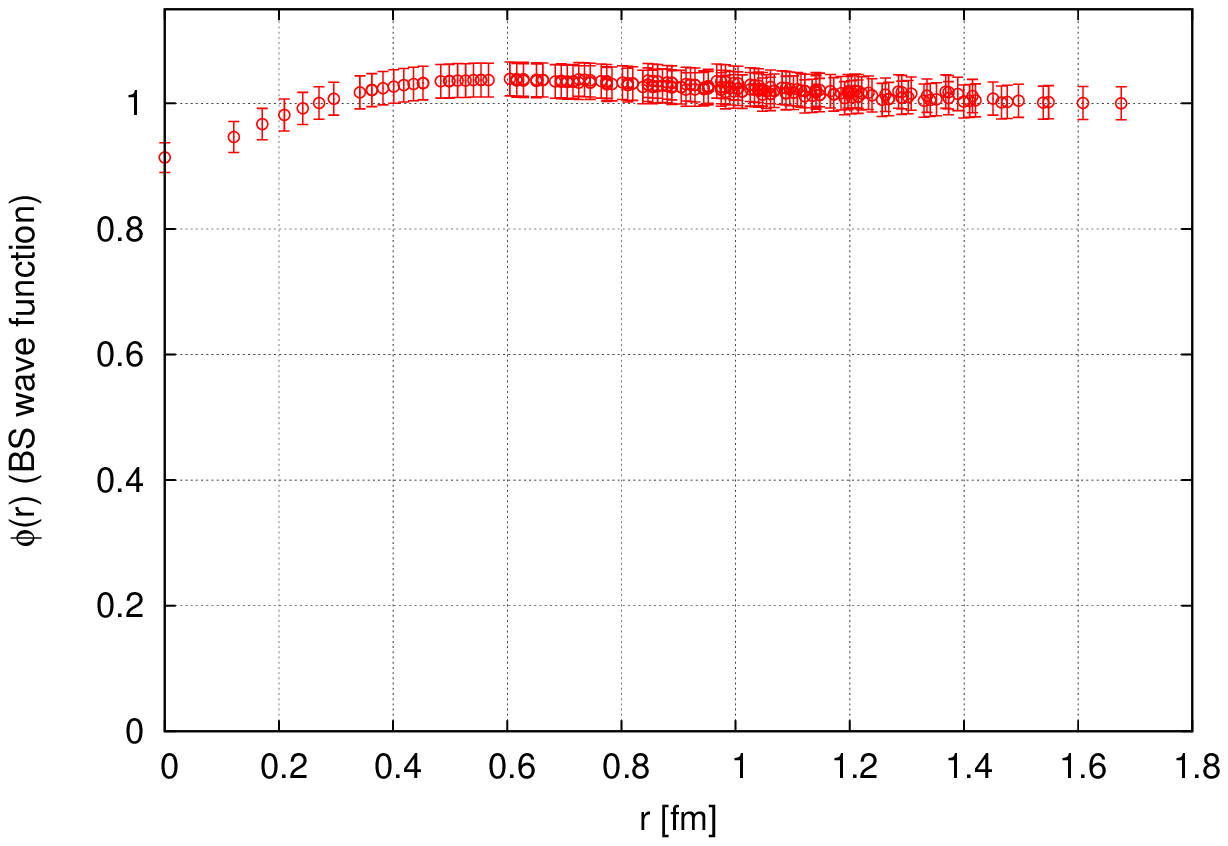}
\includegraphics[width=0.45\columnwidth]{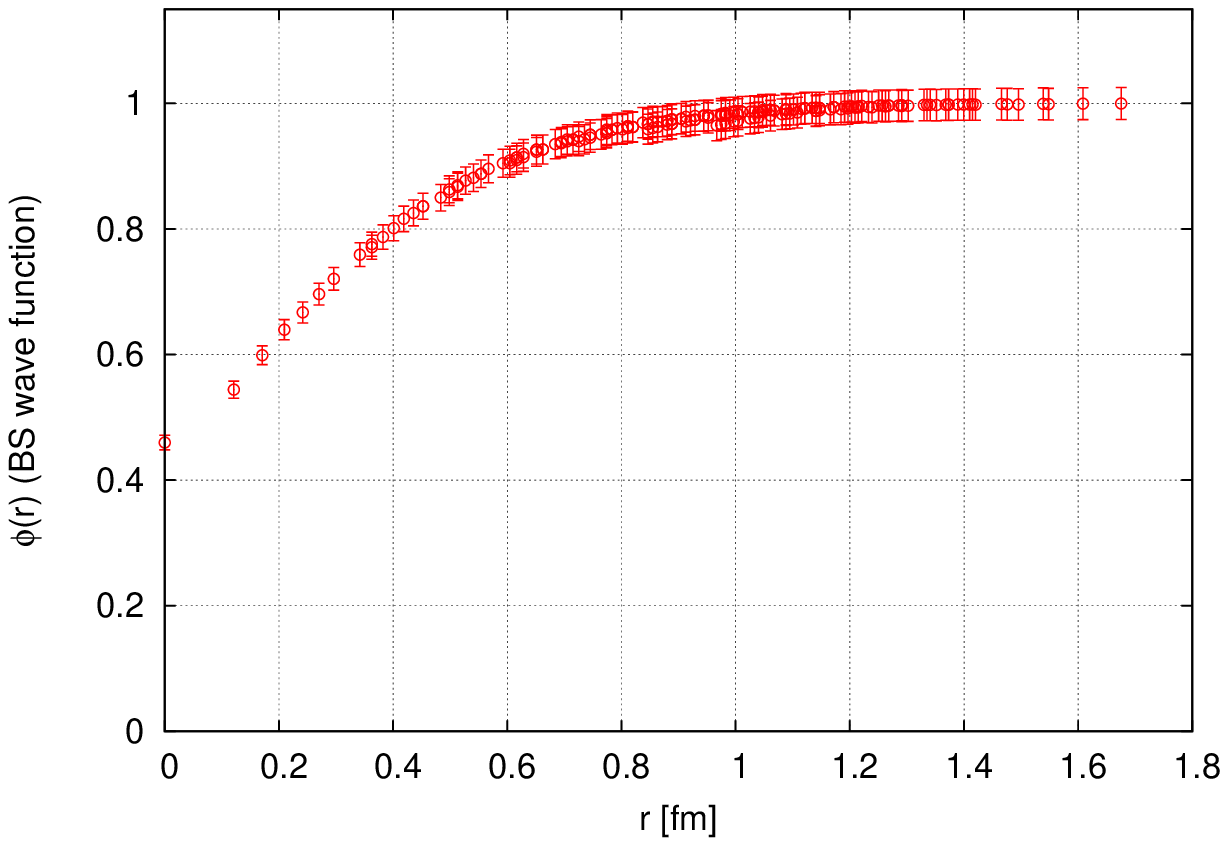}
\end{center}
\caption{The NBS wave function of the $KN$ scattering in the $I=0$ (left) and the $I=1$ (right) channels. Taken from Ref.~\protect\cite{Ikeda:2010sg}. }
\label{fig:NBS_KN}
\end{figure} 

Fig.~\ref{fig:NBS_KN}  shows the NBS wave functions of the $KN$ scatterings 
in the $I=0$ (left) and $I=1$ (right) channels.
The large $r$ behavior of the NBS wave functions in both channels do not show a sign of 
a bound state, though more detailed analysis is needed with larger volumes for a definite conclusion. On the other hand, the small $r$ behavior of the NBS wave functions 
suggests a repulsive interaction at short distance ($r<0.3$ fm).
The repulsion in the $I=1$ channel seems to be stronger than that in the $I=0$ channel.

\begin{figure}[tb]
\begin{center}
\includegraphics[width=0.45\columnwidth]{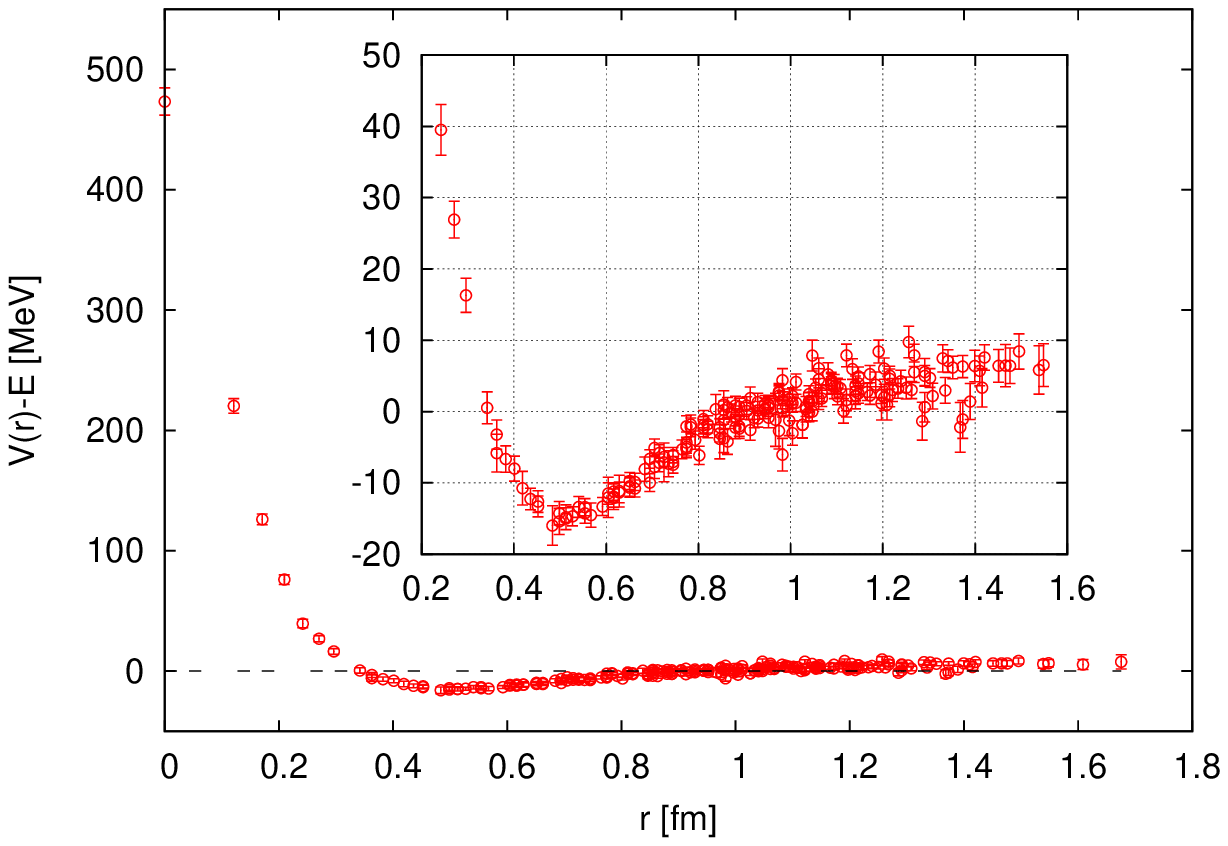}
\includegraphics[width=0.45\columnwidth]{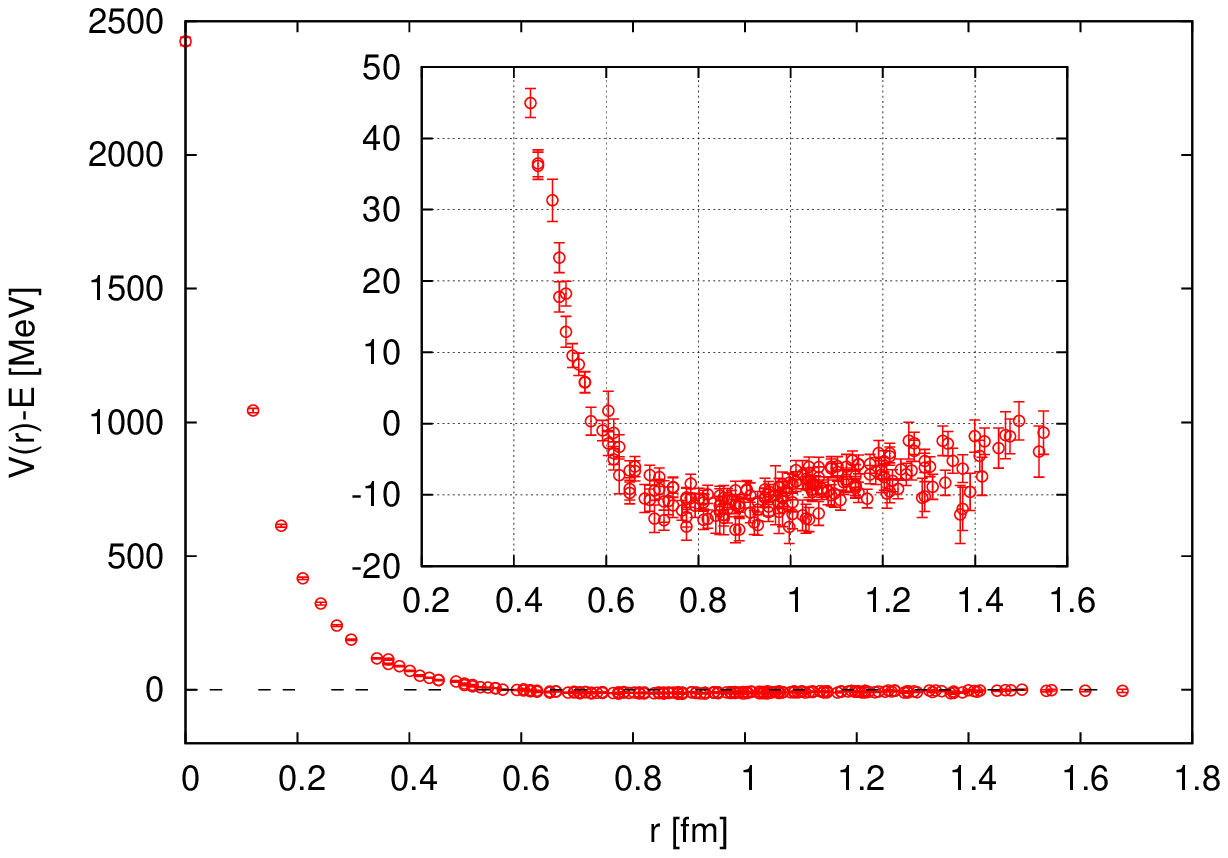}
\end{center}
\caption{The LO potential for the $KN$ state without the energy shift $E$ in the $I=0$ (left)
 and the $I=1$ channels (right). }
\label{fig:pot_NK}
\end{figure} 

The LO potential $V(r)$ for the $KN$ state without the constant energy shift $E$ is shown
 in Fig.~\ref{fig:pot_NK}  in the $I=0$ (left)  and $I=1$ (right).
As expected from the NBS wave functions in Fig.~\ref{fig:NBS_KN},
the repulsive interactions are observed at short distance in both channels,
while the attractive well appears at the medium distance ($0.4<r<0.8$ fm) in the $I=0$ channel.
These results indicate that there are no bound states
in $I(J^{\pi})=0(1/2^-)$ and $1(1/2^-)$ states at $m_{\pi} \simeq 870$ MeV.

The second application is to study the charmonium-nucleon interaction using the potential method.  Since  charmonium does not share the same quark flavor with the nucleon,
the charmonium-nucleon interaction is mainly induced by the genuine QCD effect of multi-gluon exchanges. Theoretical studies based on QCD suggest that the  $c\bar{c}$-$N$ interaction is weakly attractive. It is  argued that the $c\bar{c}$-nucleus
 ($A$) bound system may be realized for the mass number $A\ge 3$ if the
 attraction between the charmonium and the nucleon is sufficiently
 strong~\cite{Brodsky:1989jd,Wasson:1991fb}.  Precise information on the $c\bar{c}$-$N$ potential $V_{c\bar{c}N}(r)$ is therefore  indispensable for exploring nuclear-bound charmonium states such as the $\eta_c$-${}^{3}{\rm He}$ or $J/\psi$-${}^{3}{\rm He}$ bound state in few body calculations~\cite{Belyaev:2006vn}.  
 
 In Ref.~\cite{Takahashi:2009ef}, the charmonium-nucleon potentials are calculated in quenched QCD  on $16^3\times 48$ and $32^3\times 48$ lattices at $a\simeq 0.94$ fm at three different values of the light quark mass corresponding to $(m_\pi, m_N)\simeq (640,1430), (720,1520), (870,1700)$ in unit of MeV and one fixed value of the charm quark mass  corresponding to $m_{\eta_c} \simeq 2920$ MeV and $m_{J/\Psi} \simeq 3000$ MeV.

\begin{figure}[tb]
\begin{center}
\includegraphics[width=0.32\columnwidth,angle=270]{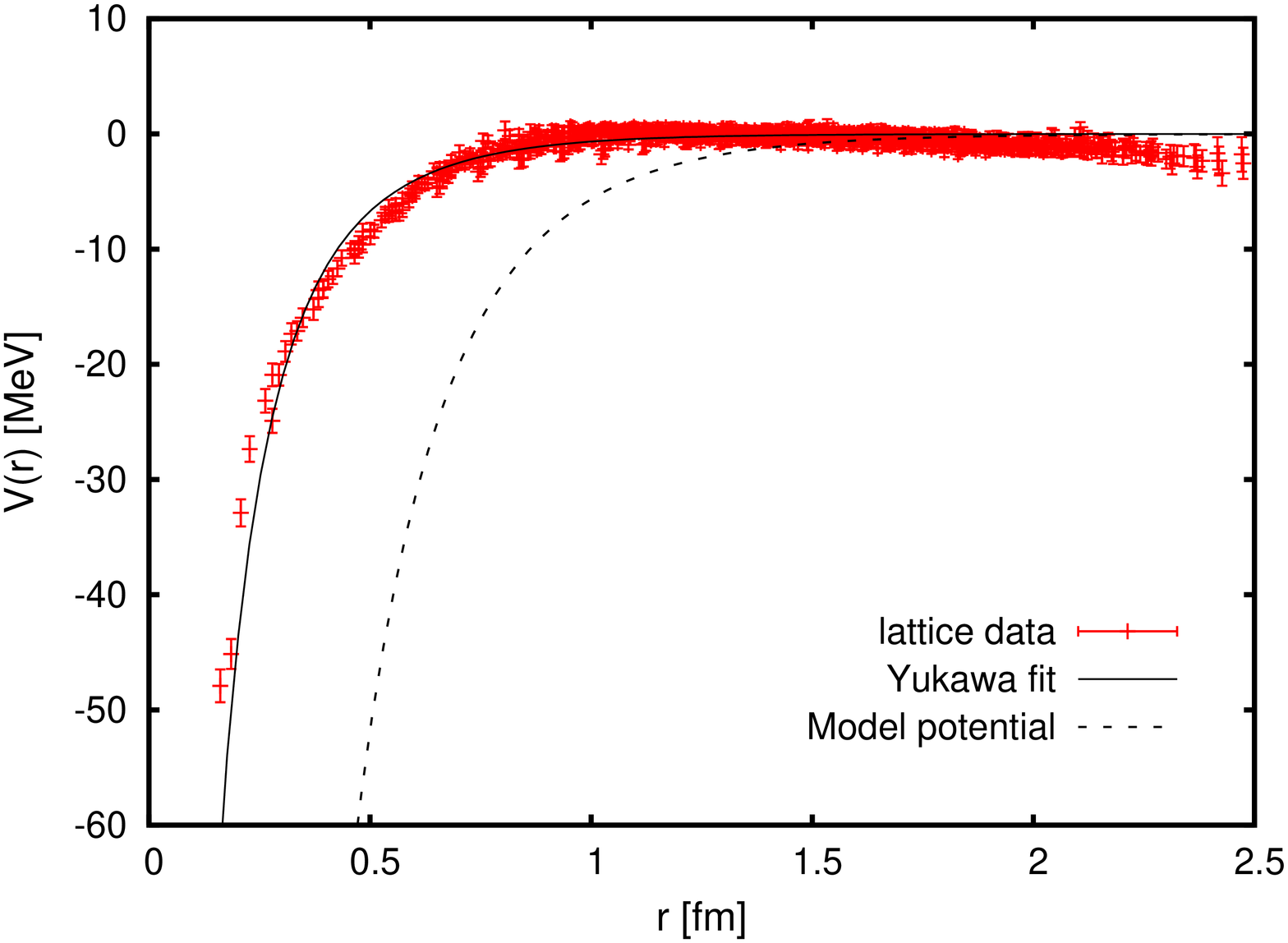}
\includegraphics[width=0.32\columnwidth,angle=270]{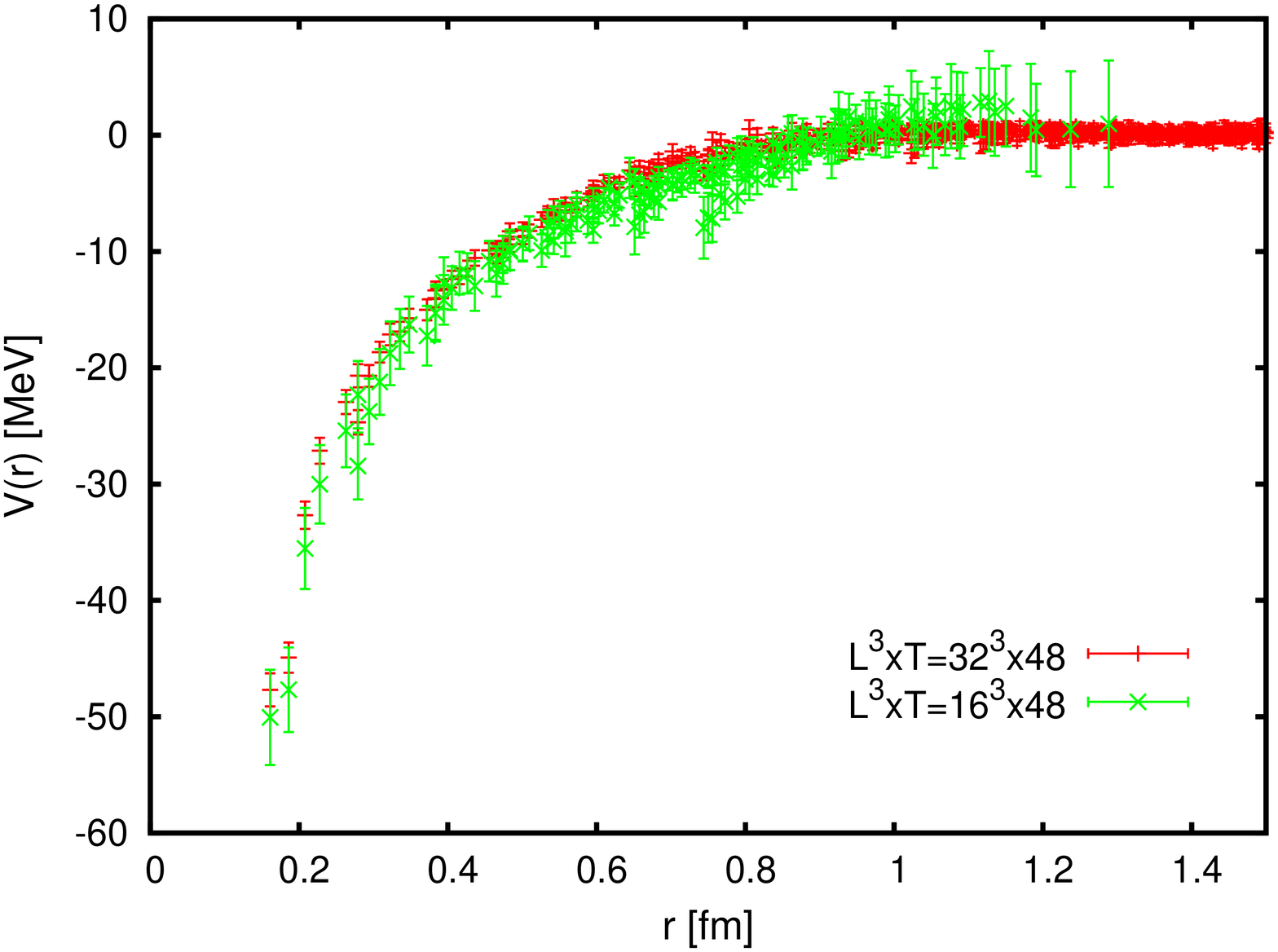}
\end{center}
\caption{(Left) The effective central potential in the $s$-wave $\eta_c$-$N$ system at $m_\pi= 640$ MeV.  The solid line is the fit by  the Yukawa potential while the dashed line is the one by the phenomenological potential adopted in Ref.~\protect\cite{Brodsky:1989jd}.
(Right) The volume dependence of the $\eta_c$-$N$ potential.  Taken from Ref.~\protect\cite{Kawanai:2010ev}.
  }
\label{fig:pot_ccN}
\end{figure} 

The effective central $\eta_c$-$N$ potential, evaluated from the NBS wave function with measured values of $E$ and $\mu$, is shown in Fig.~\ref{fig:pot_ccN}(Left).
 The $\eta_c$-$N$ potential clearly exhibits an entirely
  attractive interaction between charmonium and the nucleon 
  without any repulsion at all distances.
  Absence of the short range repulsion (the repulsive core) may be related to  absence of
  the Pauli exclusion between the heavy quarkonium and the light hadron. 
  The interaction is exponentially screened in the long distance region
  $r\ge 1$ fm.  
  The fit of the potential with  the Yukawa form $-\gamma e^{-\alpha r}/r$ gives $\gamma \sim 0.1$ and $\alpha \sim 0.6$ GeV (solid line in Fig.~\ref{fig:pot_ccN}),  which are compared with the phenomenological values $\gamma=0.6$ and $\alpha=0.6$ GeV in Ref.~\cite{{Brodsky:1989jd}} (dashed line). The strength of the Yukawa potential $\gamma$
  is six times smaller than the phenomenological value, while the Yukawa
  screening parameter $\alpha$ obtained from lattice QCD is
  comparable with the phenomenological one. The $c\bar{c}$-$N$ potential is found to be
rather weak in lattice QCD. 

As shown in Fig.~\ref{fig:pot_ccN}(Right), 
there is no significant difference between potentials at two different spatial sizes ($La\approx 3.0$ and 1.5 fm). The size dependence of the $\eta_c$-$N$ potential seems small since
the $\eta_c$-$N$  potential is quickly screened to zero and turns out to be short
 ranged. 
 
\begin{figure}[tb]
\begin{center}
\includegraphics[width=0.32\columnwidth,angle=270]{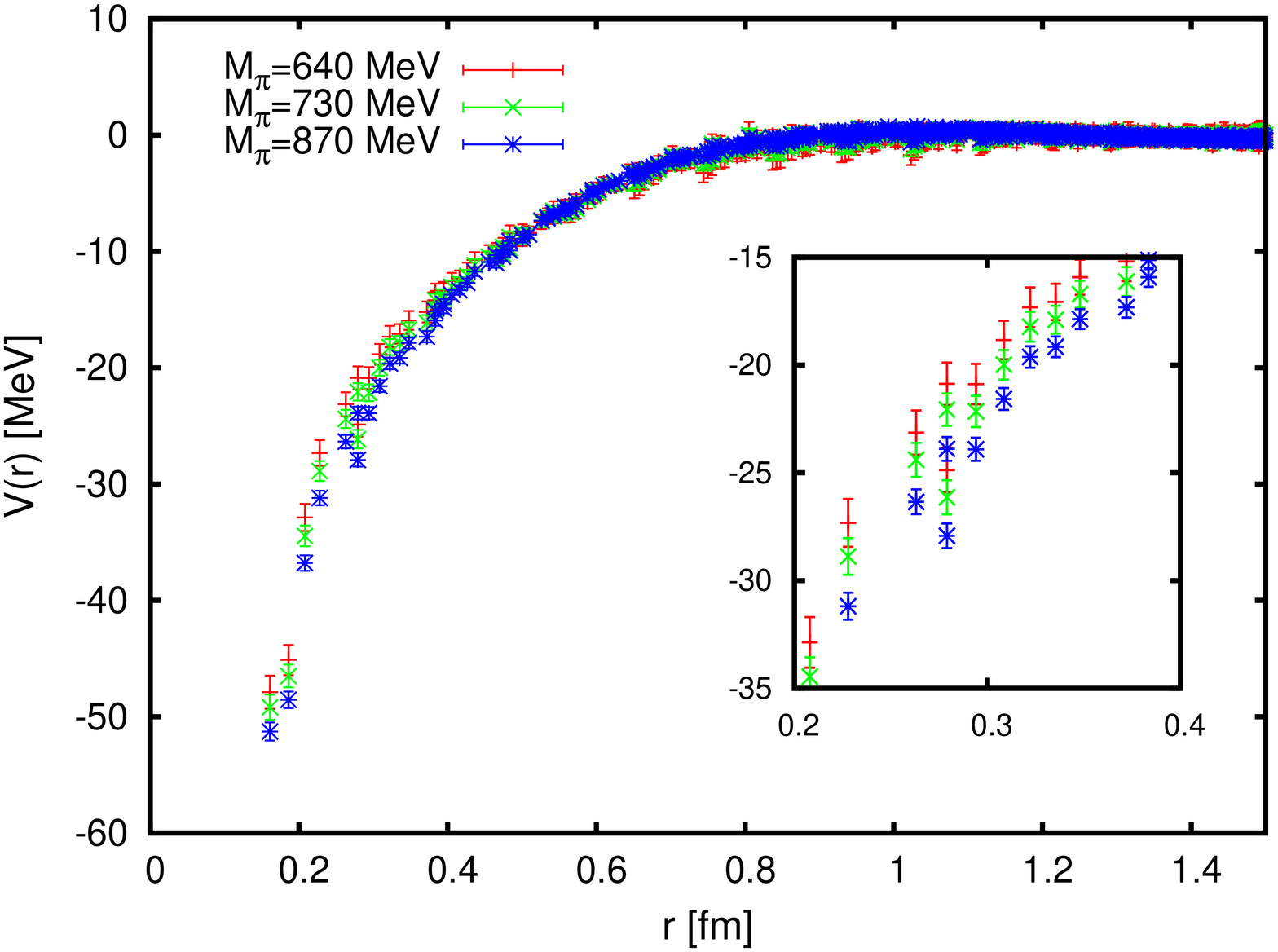}
\includegraphics[width=0.32\columnwidth,angle=270]{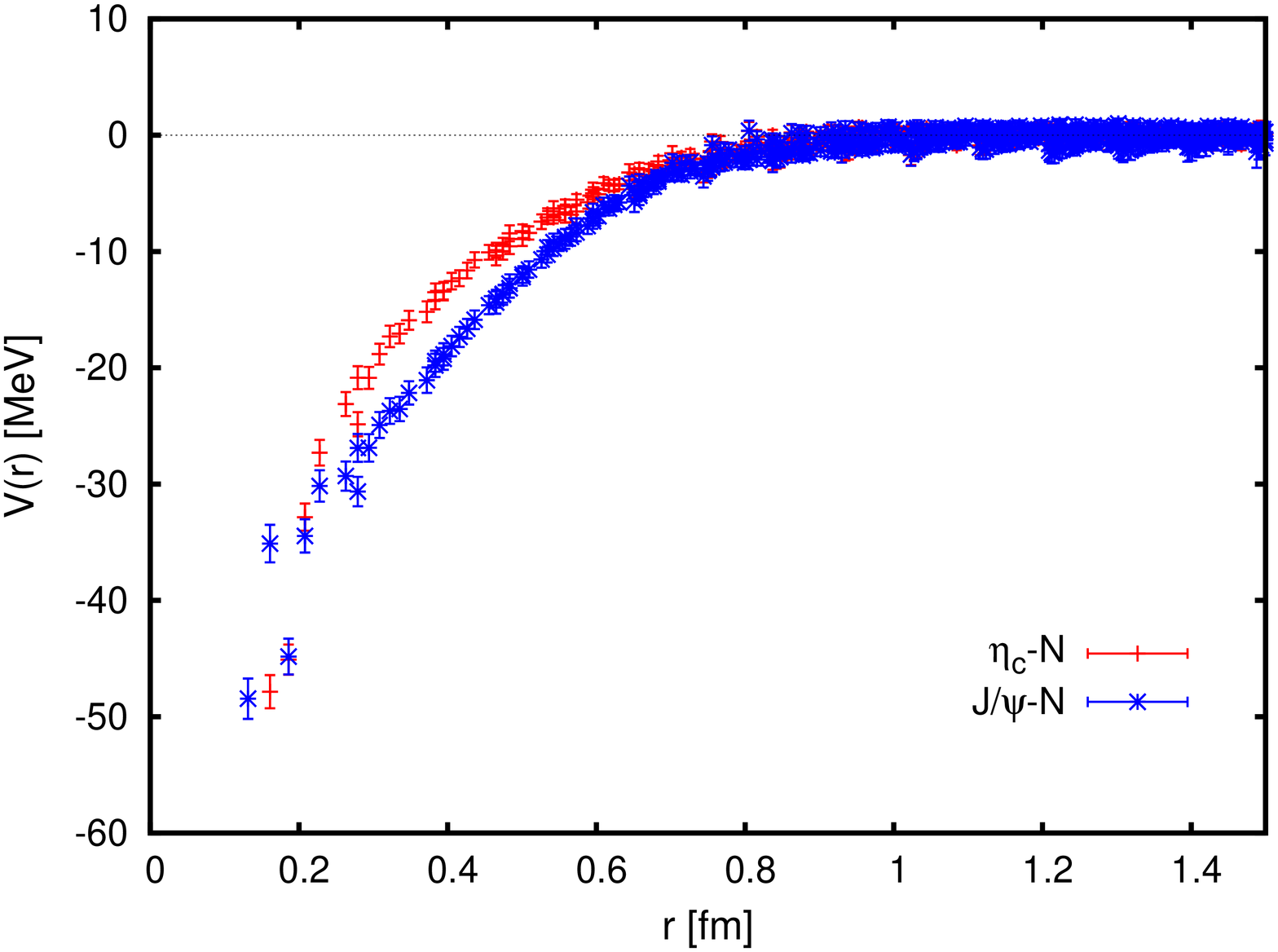}
\end{center}
\caption{(Left) The quark-mass dependence of the $\eta_c$-$N$ potential.
(Right) The  spin-independent part of the central $J/\psi$-$N$
  potential at $m_\pi = 640$ MeV, together with the $\eta_c$-$N$ potential for comparison.
  Taken from Ref.~\protect\cite{Kawanai:2010ev}.
}
\label{fig:pot_JPsiN}
\end{figure} 

No large quark-mass dependence is observed in Fig.~\ref{fig:pot_JPsiN}(Left). 
This may be understood by the argument that
  the $c\bar{c}$-$N$ interaction is mainly governed by multi-gluon exchanges,
  which do not depend explicitly on the quark mass.
  Taking a closer look at the inset of Fig.~\ref{fig:pot_JPsiN}(Left), however, one finds
 that the attractive interaction in the $\eta_c$-$N$ system tends to get slightly weaker as the light
 quark mass decreases. 

The  spin-independent part of the central $J/\psi$-$N$ potential is shown  at $m_\pi = 640$ MeV in Fig.~\ref{fig:pot_JPsiN}(Right), together with the $\eta_c$-$N$ potential for comparison.
While no qualitative difference between the $\eta_c$-$N$ and $J/\psi$-$N$ potentials is observed, the attractive interaction in the $J/\psi$-$N$ potential is a little
  stronger than the $\eta_c$-$N$ potential. 
  The attraction, however, is still not strong enough to form a bound state in the
  $J/\psi$-$N$ channel.
  
\subsection{Two-color QCD and potentials}
In Ref.~\cite{Takahashi:2009ef}, potentials between baryons in two-color (SU(2)) QCD are investigated on the lattice.

In two-color QCD,  two quarks (diquark)  form a "baryon", whose local interpolating operator is given by
\be
D_{ij,\Gamma}(\bx,t)\equiv \varepsilon^{ab}q_i(\bx,t)\Gamma q_j(\bx,t),
\ee
where $\varepsilon^{ab}$ is the $2\times 2$ anti-symmetric tensor, $i,j$ are flavor indices and
$\Gamma=C,C\gamma_5, C\gamma_\mu, C\gamma_\mu\gamma_5$ with the charge conjugation matrix $C$. The NBS wave function is then defined by
\be
\varphi^{W,\Gamma}_{ij,kl}(\br )e^{-Wt} = \langle 0\vert T\{ D_{ij,\Gamma}(\bx+\br,t) D_{kl,\Gamma}
(\bx,t)\} \vert W\rangle
\ee
where $\vert W \rangle $ is an eigenstate of two color QCD with total energy $W$ and  "baryon" number 2. The potential derived from this wave function is denoted as 
$V_{ij,kl}^\Gamma(\br)$.

The potentials have been calculated in SU(2) quenched QCD on a $24^3\times 48$ lattice at $a\simeq 0.1$ fm determined from the string tension with the assumption that $\sqrt{\sigma}=440$ MeV.  The lightest "baryon" corresponds to the scalar diquark state ($\Gamma=C$). Therefore the NBS wave function is evaluated for $\Gamma=C$ at
four values of the quark mass, which give $m_S=1.044(2),0.836(2),0.618(2), 0.377(3)$ in lattice unit with $m_S$ being the scalar diquark mass.  The potential is then extracted from the $A_1$ state with $\Gamma = C$, and is therefore denoted by $V_{ij,kl}(r)$.

\begin{figure}[tb]
\begin{center}
\includegraphics[width=0.45\columnwidth]{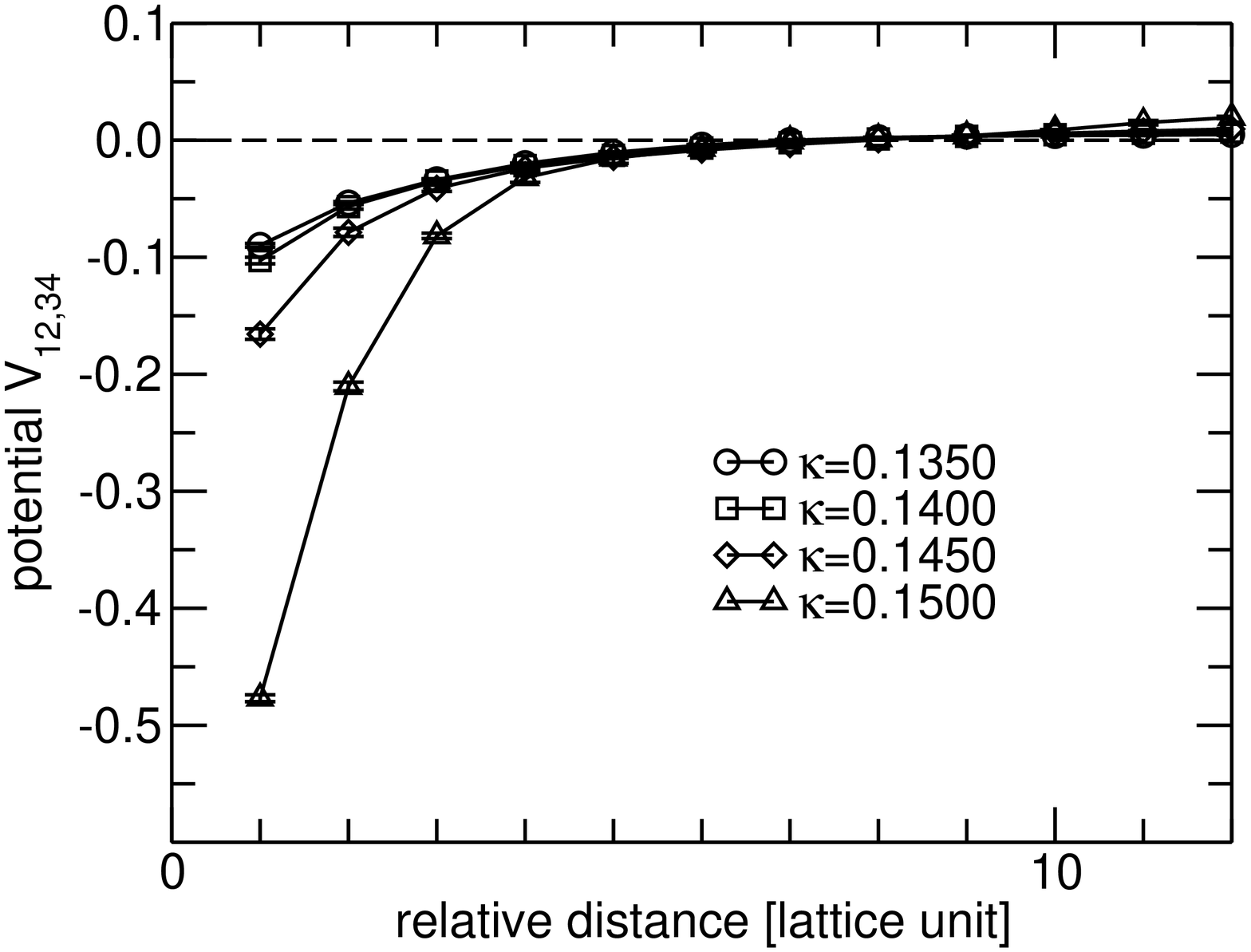}
\includegraphics[width=0.45\columnwidth]{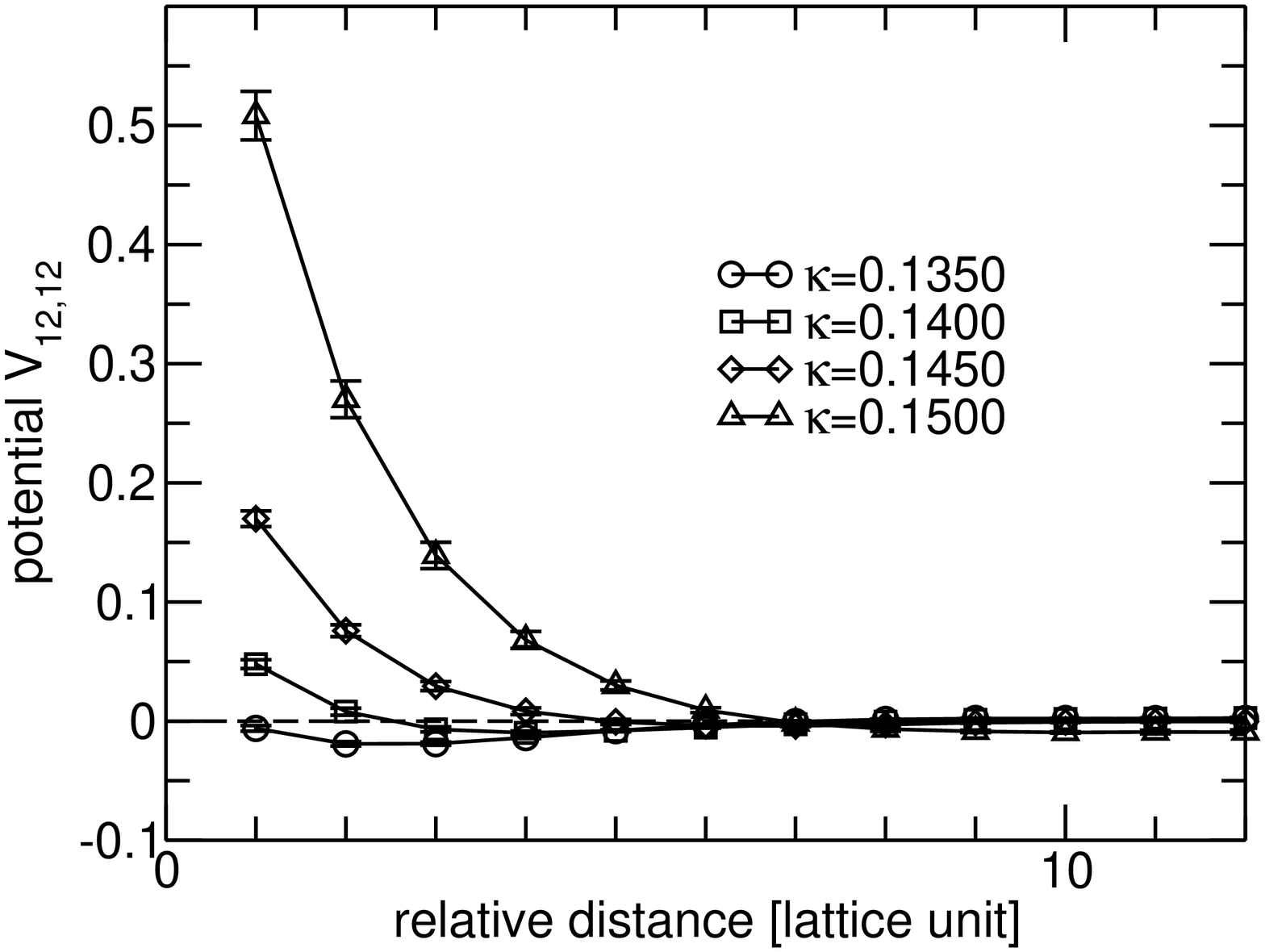}
\end{center}
\caption{
\label{fig:potential_2color}
The LO central potentials $V_{12,34}(r)$ (Left) and $V_{12,12}(r)$ (Right)
as functions of the relative distance $r$ in lattice unit. Here $\kappa = 0.135,14,145$ and $0.15$ correspond to $m_S=1.044,0.836, 0.618$ and $0.377$ in lattice unit, respectively.
Taken from Ref.~\protect\cite{Takahashi:2009ef}.
}
\end{figure}

Fig.~\ref{fig:potential_2color} shows the LO central potentials plotted as functions of $r$ for
$(i,j,k,l)=(1,2,3,4)$(left) and $(1,2,1,2)$(right), where quark-exchange diagrams are absent for the former while they exist for the latter .

The potential $V_{12,34}(r)$ has an attractive interaction at all length scales, which becomes stronger as $m_S$ decreases. The $r$-dependence is monotonic at all values of $m_S$.
At large $r$ ($r\ge 4$ in lattice units), the potential has small $m_S$ dependence, while it has stronger $m_S$ dependence at  small $r$ ($r\le 4$). The magnitude of the potential decreases as $m_S$ increases and finally the potentials at $m_S=0.836$ and $1.044$ coincide with each other. 

For $V_{12,12}$, on the other hand,  strong repulsions appear at short distance.
The $m_S$ dependence is not monotonic:  The potential is a smooth function of $r$ at small $m_S$, while it has a pocket at intermediate distance for $m_S=1.044$.
As $m_S$ decreases, the repulsive core rapidly increases and the attractive pocket disappears.

The qualitative difference between the two cases suggests that contributions from quark exchange diagrams, which more or less represent the Pauli exclusion effect, are responsible for the repulsive core. On the other hand, the attraction  may be explained by gluon exchanges,
which are the main part of the potential $V_{12,34}(r)$.
More details analysis can be found in Ref.~\cite{Takahashi:2009ef}.

\section{Conclusion}
\label{sec:conclusion}

In this report, we have reviewed the progress on the study of hadron interactions  via the potential method in lattice QCD. The key quantity of the method is the NBS wave function of the two particle state. Not only is the asymptotic behavior of the NBS wave related to the scattering phase shift (the phase of the S-matrix) in QCD, but also the non-local but energy-independent potential (or the interaction kernel) can be extracted from the NBS  wave function in the non-asymptotic region.
By construction the potential correctly reproduces the scattering phase shift at all energies below the inelastic threshold. In practice the non-local potential is approximated by the velocity expansion in  terms of local functions, so that physical observables such as the scattering phase shift are approximately calculated.   It is also possible to check the accuracy of the approximation.

Using the local potential approximation at the leading order in the velocity expansion, nucleon-nucleon, hyperon-nucleon and  hyperon-hyperon interactions have already been investigated successfully.  The BB potential in the flavor SU(3) limit shows the attraction in the flavor singlet channel, which is strong enough to form a bound state, $H$-dibaryon in this limit.
The repulsive core is  analyzed in terms of the operator product expansion and the renormalization group.

The potential method is extended to the case that the total energy is above the inelastic threshold.
This method seems to work for the BB interaction with $S=-2$ and $I=0$, where
the coupled channel potentials among $\Lambda\Lambda$, $N\Xi$ and $\Sigma\Sigma$ are obtained.
The potential method is also applied to the three nucleon force, the meson-baryon interactions and two color QCD.

Of course, more careful studies of systematic errors such as the finite volume effect, dynamical quark effect, quark mass dependence and the lattice spacing effect are needed, before applying
potentials obtained in lattice QCD in order to  investigations of nuclei and hyper-nuclei. 

Finally I stress here that the potential derived from the NBS wave function adds a new tool to investigate hadron interactions in lattice QCD, in addition to the standard finite volume method proposed by L\"uscher~\cite{Luscher:1990ux}.
In particular, the hadron scattering above the inelastic threshold  can be treated in lattice QCD by the potential method. Other extensions of this method will also be  looked for.
  
 \section*{Acknowledgement}
 I would like to thank all members of HAL QCD collaboration (Drs. T.~Doi, T.~Hatsuda, Y.~Ikeda, T.~Inoue, N.~Ishii, K.~Murano, H.~Nemura, K.~Sasaki), Dr. J.~Balog and Dr. P.~Weisz for useful discussions. In particular I would like to thank Dr. P.~Weisz  for careful reading of the manuscript and valuable comments.
 I also thank T.~Kawanai for providing me with figures used in this report.
 This work is supported in part by the Grant-in-Aid of the Ministry of Education, Science and Technology, Sports and Culture(MEXT) (No. 20340047) and by the MEXT Grant-in-Aid for Scientific Research on Innovative Areas (No. 2004: 20105001, 20105003).

\appendix
\section{Asymptotic behaviors of the NBS wave function in the spin triplet channel}
\label{app:NBS}
There are three components in the spin triplet channel at given $J$. One component corresponds to the $l=J$ scattering, which is extracted by taking 
$\vert s_1 s_2 \rangle = \frac{1}{\sqrt{2}}\left(\vert +\frac{1}{2},+\frac{1}{2}\rangle -\vert -\frac{1}{2}, -\frac{1}{2}\rangle\right) $.
As $r\rightarrow \infty$, this component becomes
\bea
\varphi^W({\bf r})_{S=1}&\simeq&\sum_{l,l_z} Z^{ll_z}(S=1) Y_{l l_z}(\Omega_{\bf r}) \frac{\sin( k r - l\pi/2 + \delta_{l1}(k))}{k r}  e^{i\delta_{l1}(k)} ,
\eea  
where
\bea
Z^{ll_z}(S)&=& \frac{Z}{2}D^l_{l_z0}(\Omega_{\bf k}) U(\nabla) U(-\nabla) \sum_{s_z}\chi(S,s_z)
\langle l ,S, l_z,s_z\vert l, l_z-s_z\rangle \nonumber \\
&\times&\left(
\langle l,l_z-s_z,l,S\vert l,l_z-s_z,+\frac{1}{2},+\frac{1}{2}\rangle -
\langle l,l_z-s_z,l,S\vert l,l_z-s_z,-\frac{1}{2},-\frac{1}{2}\rangle\right).
\eea

The $l=J\mp 1$ components, which correspond to $\vert s_1 s_2 \rangle = \frac{1}{\sqrt{2}}\left(\vert \frac{1}{2},-\frac{1}{2}\rangle \pm\vert -\frac{1}{2}, +\frac{1}{2}\rangle\right)\equiv \vert \pm\rangle $, 
are mixed with each other, and those asymptotic behavior is given by
\bea
\varphi^W({\bf r})_{S=1,s_1-s_2}&\simeq&\sum_{J,M,l,l_z} Z^{JM}_{ll_z}(S=1) L^{JM}_{ll_z}(S=1,k) 
Y_{l l_z}(\Omega_{\bf r}) \frac{\sin( k r - l\pi/2 + \delta_{l1}(k))}{k r}  e^{i\delta_{l1}(k)} \nonumber \\
&\times& \left(
\begin{array}{cc}
\delta_{l,J-1}& 0 \\
0 & \delta_{l,J+1}\\
\end{array}
\right) R^{JM}_{s_1-s_2}(S=1,k)
\eea
where
\bea
Z^{JM}_{ll_z} (S)&=& Z U(\nabla) U(-\nabla)\sum_{s_z} \chi(S,s_z)\langle l S l_z s_z\vert J M\rangle \\
L^{JM}_{ll_z}(S,k)_{t_1} &=&\sum_{\bar s_1- \bar s_2=\pm 1} \langle JM l S \vert J M \bar s_1 \bar s_2\rangle \left[ X O(k) \right]_{\bar s_1-\bar s_2, t_1} \\
R^{JM}_{s_1-s_2}(S,k)_{t_1} &=&\left[O^{-1}(k) X\right]_{t_1, s_1- s_2} D^J_{M,s_1-s_2}(\Omega_{\bf k}) ,
\eea
and $s_1-s_2$ becomes $\pm 1$ in the above equation.
Here $t_1=\pm 1$ corresponds to $\vert \pm\rangle$ introduced before, and
$X$, which transforms the helicity basis $\{ \vert +\frac{1}{2},-\frac{1}{2}\rangle, \vert-\frac{1}{2},+\frac{1}{2}\rangle\} $ to the new basis $\{ \vert +\rangle, \vert -\rangle$,  is given by 
\be
X =X^{-1}= \frac{1}{\sqrt{2}}
\left(
\begin{array}{cc} 
 1&  1 \\
 1 &  -1\\
\end{array}
\right) .
\ee
The mixing matrix in the new basis has the form
\be
O(k) =\left(
\begin{array}{cc} 
 \cos \theta_J(k) & -\sin\theta_J(k)   \\
 \sin \theta_J(k) &  \cos\theta_J(k)\\
\end{array}
\right) ,
\ee
where $\theta_J(k)$ is the mixing angle between $l=J\pm 1$.

\section{Octet baryons}
\label{app:octet}
Local octet baryon operators are defined by
\bea
p_\alpha(x)&=&\varepsilon_{abc} (u^a(x)C\gamma_5 d^b (x) ) u_\alpha^c(x) \nn\\
n_\alpha(x)&=&\varepsilon_{abc} (u^a(x)C\gamma_5 d^b (x) ) d_\alpha^c(x) \nn\\
\Sigma_\alpha^+(x)&=&-\varepsilon_{abc} (u^a(x)C\gamma_5 s^b (x) ) u_\alpha(x) \nn\\
\Sigma_\alpha^0(x)&=&-\varepsilon_{abc}\frac{1}{\sqrt{2}}[ (d^a(x)C\gamma_5 s^b (x) ) u_\alpha^c(x)+u^a(x)C\gamma_5 s^b (x) ) d_\alpha^c(x)] \nn\\
\Sigma_\alpha^-(x)&=&-\varepsilon_{abc} (d^a(x)C\gamma_5 s^b (x) ) d_\alpha^c(x) \nn\\
\Xi_\alpha^0(x) &=& \varepsilon_{abc} (s^a(x)C\gamma_5 u^b (x) ) s_\alpha^c(x) \nn\\
\Xi_\alpha^-(x) &=& \varepsilon_{abc} (s^a(x)C\gamma_5 d^b (x) ) s_\alpha^c(x) \nn\\
\Lambda_\alpha(x)&=&-\varepsilon_{abc}\frac{1}{\sqrt{6}}[ (d^a(x)C\gamma_5 s^b (x) ) u_\alpha^c(x)+u^a(x)C\gamma_5 s^b (x) ) d_\alpha^c(x)-2u^a(x)C\gamma_5 d^b (x) ) s_\alpha^c(x)] ,
\label{eq:octet}
\eea 
from which the irreducible representations of the flavor SU(3) with baryon number $B=2$ are constructed as
\bea
(BB)^{({\bf 27})}&=& \sqrt{\frac{27}{40}}\Lambda\Lambda -  \sqrt{\frac{1}{40}}\Sigma\Sigma +\sqrt{\frac{12}{40}}N\Xi  \nn \\
(BB)^{({\bf 8}_s)}&=& -\sqrt{\frac{1}{5}}\Lambda\Lambda -  \sqrt{\frac{3}{5}}\Sigma\Sigma +\sqrt{\frac{1}{5}}N\Xi  \nn \\
(BB)^{({\bf 1})}&=& -\sqrt{\frac{1}{8}}\Lambda\Lambda +  \sqrt{\frac{3}{8}}\Sigma\Sigma +\sqrt{\frac{4}{8}}N\Xi  \nn \\
(BB)^{(\overline{\bf 10})} &=&\sqrt{\frac{1}{2}}p n -  \sqrt{\frac{1}{2}}n p \nn \\
(BB)^{({\bf 10})} &=&\sqrt{\frac{1}{2}}p \Xi^+ -  \sqrt{\frac{1}{2}}\Xi^+ p \nn \\
(BB)^{({\bf 8}_a)} &=&\sqrt{\frac{1}{4}}p \Xi^- -  \sqrt{\frac{1}{4}}\Xi^- p -\sqrt{\frac{1}{4}}n \Xi^0 +  \sqrt{\frac{1}{4}}\Xi^0 n 
\label{eq:su3_rep}
\eea
where $\Sigma\Sigma$ and $N \Xi$ are defined 
\bea
\Sigma\Sigma&=& \sqrt{\frac{1}{3}}\Sigma^+\Sigma^- -  \sqrt{\frac{1}{3}}\Sigma^0\Sigma^0 +\sqrt{\frac{1}{3}}\Sigma^-\Sigma^+  \nn \\
N\Xi &=&\sqrt{\frac{1}{4}}p \Xi^- + \sqrt{\frac{1}{4}}\Xi^- p -\sqrt{\frac{1}{4}}n \Xi^0 -  \sqrt{\frac{1}{4}}\Xi^0 n .
\eea

Unitary matrices which rotate the flavor basis to the baryon basis
 are given as follows.
\begin{enumerate}
\item S=$-$1, I=1/2, spin-singlet.
\begin{eqnarray}
 \left(
  \begin{array}{l}
   \ketv N\Lambda> \\
   \ketv N\Sigma>
  \end{array}
  \right)
=
   \left(
    \begin{array}{rr}
       \sqrt{\frac{ 9}{10}} &  -\sqrt{\frac{ 1}{10}} \\
       \sqrt{\frac{ 1}{10}} &   \sqrt{\frac{ 9}{10}} 
    \end{array}
   \right)
   \left(
    \begin{array}{l}
     \ketv{\bf 27}> \\
     \ketv{\bf 8}_s>
    \end{array}
   \right)
\end{eqnarray}

\item S=$-$1, I=1/2, spin-triplet.
\begin{eqnarray}
\hfill
 \left(
  \begin{array}{l}
   \ketv N\Lambda> \\
   \ketv N\Sigma>
  \end{array}
  \right)
=
   \left(
    \begin{array}{rr}
       \sqrt{\frac{ 1}{2}} &  -\sqrt{\frac{ 1}{2}} \\
       \sqrt{\frac{ 1}{2}} &   \sqrt{\frac{ 1}{2}} 
    \end{array}
   \right)
   \left(
    \begin{array}{l}
     \ketv{\overline{\bf 10}}> \\
     \ketv{\bf 8}_a>
    \end{array}
   \right)
\end{eqnarray}

\item S=$-$2, I=0, spin-singlet.
\begin{equation}
 \left(
  \begin{array}{l}
   \ketv \Lambda\Lambda> \\
   \ketv \Sigma\Sigma> \\
   \ketv N\Xi>
  \end{array}
  \right)
=
   \left(
    \begin{array}{rrr}
       \sqrt{\frac{27}{40}} &  -\sqrt{\frac{ 8}{40}} & -\sqrt{\frac{ 5}{40}}  \\
      -\sqrt{\frac{ 1}{40}} &  -\sqrt{\frac{24}{40}} &  \sqrt{\frac{15}{40}}  \\
       \sqrt{\frac{12}{40}} &   \sqrt{\frac{ 8}{40}} &  \sqrt{\frac{20}{40}}
    \end{array}
   \right)
   \left(
    \begin{array}{l}
     \ketv{\bf 27}> \\
     \ketv{\bf 8}_s> \\
     \ketv{\bf 1}>
    \end{array}
   \right)
\end{equation}

\item S=$-$2, I=1, spin-singlet.
\begin{eqnarray}
 \left(
  \begin{array}{l}
   \ketv N\Xi> \\
   \ketv \Sigma\Lambda>
  \end{array}
  \right)
=
   \left(
    \begin{array}{rr}
       \sqrt{\frac{ 2}{5}} &  -\sqrt{\frac{ 3}{5}} \\
       \sqrt{\frac{ 3}{5}} &   \sqrt{\frac{ 2}{5}} 
    \end{array}
   \right)
   \left(
    \begin{array}{l}
     \ketv{\bf 27}> \\
     \ketv{\bf 8}_s>
    \end{array}
   \right)
\end{eqnarray}

\item S=$-$2, I=1, spin-triplet.
\begin{eqnarray}
 \left(
  \begin{array}{l}
   \ketv N\Xi> \\
   \ketv \Sigma\Lambda> \\
   \ketv \Sigma\Sigma>
  \end{array}
  \right)
=
   \left(
    \begin{array}{rrr}
      -\sqrt{\frac{ 1}{3}} &  -\sqrt{\frac{ 1}{3}} & \sqrt{\frac{ 1}{3}} \\
      -\sqrt{\frac{ 1}{2}} &   \sqrt{\frac{ 1}{2}} & 0                   \\
       \sqrt{\frac{ 1}{6}} &   \sqrt{\frac{ 1}{6}} & \sqrt{\frac{ 4}{6}} 
    \end{array}
   \right)
   \left(
    \begin{array}{l}
     \ketv{\overline{\bf 10}}> \\
     \ketv{\bf 10}> \\
     \ketv{\bf 8}_a>
    \end{array}
   \right)
\end{eqnarray}

\item S=$-$3, I=1/2, spin-singlet.
\begin{eqnarray}
 \left(
  \begin{array}{l}
   \ketv \Lambda\Xi> \\
   \ketv \Sigma\Xi>
  \end{array}
  \right)
=
   \left(
    \begin{array}{rr}
      \sqrt{\frac{ 9}{10}} &  -\sqrt{\frac{ 1}{10}}  \\
      \sqrt{\frac{ 1}{10}} &   \sqrt{\frac{ 9}{10}} 
    \end{array}
   \right)
   \left(
    \begin{array}{l}
     \ketv{\bf 27}> \\
     \ketv{\bf 8}_s>
    \end{array}
   \right)
\end{eqnarray}

\item S=$-$3, I=1/2, spin-triplet.
\begin{eqnarray}
 \left(
  \begin{array}{l}
   \ketv \Lambda\Xi> \\
   \ketv \Sigma\Xi>
  \end{array}
  \right)
=
   \left(
    \begin{array}{rr}
      \sqrt{\frac{ 1}{2}} &  -\sqrt{\frac{ 1}{2}}  \\
      \sqrt{\frac{ 1}{2}} &   \sqrt{\frac{ 1}{2}} 
    \end{array}
   \right)
   \left(
    \begin{array}{l}
     \ketv{\bf 10}> \\
     \ketv{\bf 8}_a>
    \end{array}
   \right)
\end{eqnarray}
\end{enumerate}

\end{document}